\documentclass[%
 reprint,
 nofootinbib,
 amsmath,amssymb,
 aps,
 prd,
]{revtex4-2}
\usepackage[hidelinks]{hyperref}
\usepackage{graphicx}
\usepackage{tabularx}
\usepackage{color}

\font\ec=ecrm0800 at 11pt
\def\thorn{\hbox{\ec\char'336}}
\def\edth{\eth}

\newcommand{\CC}{C\nolinebreak\hspace{-.05em}\raisebox{.4ex}{\tiny\bf +}\nolinebreak\hspace{-.10em}\raisebox{.4ex}{\tiny\bf +}}

\begin{document}

\preprint{APS/123-QED}

\title{Metric reconstruction and the Hamiltonian for eccentric, precessing binaries in the small-mass-ratio limit}% Force line breaks with \\
%\thetaanks{A footnote to the article title}%

\author{Zachary Nasipak}
\affiliation{School of Mathematical Sciences and STAG Research Centre,
University of Southampton, Southampton, SO17 1BJ, United Kingdom}
\email{z.nasipak@soton.ac.uk}
% \affiliation{Center for Space Sciences and Technology, University of Maryland Baltimore County, Baltimore, MD, 21250, USA}
% \affiliation{NASA Goddard Space Flight Center, 8800 Greenbelt Road, Greenbelt, Maryland, 20771, USA}

\date{\today}% It is always \today, today,

\begin{abstract}
We calculate the first-order (in the mass-ratio) metric perturbation produced by a small body on an eccentric, precessing bound orbit about a Kerr black hole. We reconstruct the metric perturbation from the maximal spin-weight Weyl scalars, $\psi_0$ and $\psi_4$, using four different methods. The first two follow the work of Chrzanowski, Cohen, Kegeles, and Wald and reconstruct the metric perturbation from either $\psi_0$ or $\psi_4$, leading to perturbations in the ingoing or outgoing radiation gauges. The other two methods build upon the work of Aksteiner, Andersson, and B{\"a}ckdahl and reconstruct the metric perturbation from both $\psi_0$ and $\psi_4$. We compare the local and asymptotic behaviors of the metric across different gauges. We also calculate the generalized redshift invariant along eccentric, precessing orbits in Kerr spacetime for the first time and make the numerical methods employed in these calculations openly available through the Python library \texttt{pybhpt}. Building off recent work by Lewis \emph{et al.}, we also relate our redshift data to the Hamiltonian of the system. Combining our numerical data with this Hamiltonian formulation provides a method for generating waveforms that include post-adiabatic conservative effects and acts as a useful bridge between results in self-force, effective-one-body, and post-Newtonian theory.
\end{abstract}

\maketitle

% Main body with filler text
\section{Introduction}
\label{sec:intro}

Gravitational self-force (GSF) theory is a powerful tool for approximating the gravitational radiation produced by compact object binaries \cite{BaraPoun19, PounWard20}. Within this framework, a binary's mass ratio $\epsilon = m_2/m_1$ is treated as a small parameter, and the full metric $\mathrm{g}_{\mu\nu}$ is linearly separated,
\begin{align}
    \mathrm{g}_{\mu\nu} = {g}_{\mu\nu} + h_{\mu\nu},
\end{align}
into a background field ${g}_{\mu\nu}$ defined by the larger mass $m_1$ and a metric perturbation $h_{\mu\nu}=O(\epsilon)$ sourced by the smaller mass $m_2$. By expanding
\begin{align}
    h_{\mu\nu} = \epsilon h^{(1)}_{\mu\nu} + \epsilon^2 h^{(2)}_{\mu\nu} + \cdots,
\end{align}
the field equations can be solved iteratively at each order in $\epsilon$ to obtain the perturbations $h^{(n)}_{\mu\nu}$. The binary's dynamics and gravitational wave emission are then derived from the metric perturbation $h_{\mu\nu}$ up to some order in $\epsilon$.
% At zeroth-order, the small body follows a geodesic in the background $g_{\mu\nu}$. At first-order, the motion of this small mass sources the perturbation $h^{(1)}_{\mu\nu}$, which encodes the binary's gravitational wave emission. The backreaction of $h^{(1)}_{\mu\nu}$ on the small body also leads to a GSF that drives the small body's gradual inspiral into the more massive black hole. At second-order, nonlinear couplings of the first-order corrections source the second-order metric perturbation $h_{\alpha\beta}^{(2)}$, which further alters the binary's dynamics and gravitational wave signal. 

This GSF framework provides a natural approach for modeling compact object binaries with highly disparate masses, in particular extreme-mass-ratio inspirals (EMRIs). EMRIs, which are characterized by mass ratios $10^{-7} \lesssim \epsilon \lesssim 10^{-4}$, are key targets for future space-based milliHertz gravitational wave detectors \cite{BerrETC19, BakeETC19} Observations of EMRIs have the potential to yield high-precision measurements of source properties \cite{BabaETC17, BerrETC19}, but realizing this potential requires waveform models with first post-adiabatic (1PA) accuracy—i.e., phase errors that scale with $\epsilon^{1/2}$ or $\epsilon$ \cite{HindFlan08, PounWard20, MillPoun21}.\footnote{The post-adiabatic approximation comes from multiscale expansions of the field equations \cite{HindFlan08, PounWard20, MillPoun21}. It describes the evolution of the gravitational wave phase as an expansion in $\epsilon$, with $n$PA effects contributing to the phase at $O(\epsilon^{n-1})$. Thus, 1PA waveforms can have phase errors of $O(\epsilon)$, because they neglect 2PA effects that contribute at this order. However, this scaling breaks down when the system passes through transient $r\theta$-resonances \cite{FlanHind12, Vand14a}. Dissipative effects can accumulate over the resonant timescale $\sim M \epsilon^{-1/2}$, effectively promoting $n$PA errors by a half-order in $\epsilon$. Therefore, if the motion is significantly perturbed by a resonance, 2PA terms contribute at $O(\epsilon^{1/2})$.} Achieving this level of accuracy will necessitate both first- and second-order metric perturbation data (e.g.,  $h^{(1)}_{\mu\nu}$ and $h^{(2)}_{\mu\nu}$) across the astrophysical parameter space.\footnote{Metric perturbations $h^{(n)}_{\mu\nu}$ can be divided into dissipative and conservative pieces. The dissipative piece drives the secular decay of the orbit and contributes at the $(n-1)$PA order, while the remaining conservative corrections contribute at $n$PA order.}

Recent progress also suggests that GSF models possess a domain of validity that is much broader than the extreme mass-ratio regime. For example, 1PA GSF models---when working in a Schwarzschild background $g_{\mu\nu}^\mathrm{Schw}$---can generate inspiral waveforms for precessing binaries with mass ratios as large as $\epsilon \sim 10^{-1}$, while also maintaining subradian phase accuracy \cite{PounETC20, BurkETC24}. Such developments indicate that GSF methods may effectively capture the strong-field dynamics of {intermediate mass-ratio binaries} ($10^{-4} \lesssim \epsilon \lesssim 10^{-1}$) \cite{VandPfei20}. In this intermediate domain, post-Newtonian methods lose accuracy and full numerical relativity becomes computationally prohibitive, making GSF approaches a promising and possibly essential tool for waveform modeling across both ground- and space-based observations.

% However, to unlock high-precision measurements of EMRI properties, GSF models must produce waveforms with first post-adiabatic (1PA) accuracy, where phase errors scale with $\epsilon$ \cite{HindFlan08, PounWard20, MillPoun21}. These so-called 1PA waveforms will require calculating both first- and second-order perturbations.

Despite their promise, GSF models still remain limited in scope. Formation models predict that small mass-ratio binaries will exhibit significant eccentricity and spin-induced precession \cite{BabaETC17, BerrETC19}, yet current GSF implementations are limited in their ability to model one or both of these effects. To date, published first-order self-force models---which include perturbations up to $h^{(1)}_{\alpha\beta}$---either neglect spin-induced precession \cite{BaraSago10, ShahETC11, ShahFrieKeid12, AkcaWarbBara13, VandShah15, OsbuWarbEvan16, DolaETC23}, only include the spin of the smaller secondary mass \cite{WarbOsbuEvan17, AkcaDempDola17}, or are restricted to low eccentricities ($e \leq 0.1$) \cite{Vand18}. Second-order self-force models---which include perturbations up to $h^{(2)}_{\alpha\beta}$---are even more restricted, only modeling quasi-circular inspirals with a slowly-spinning primary and spinning secondary \cite{WardETC21, BurkETC24, AkcaETC20}. A comprehensive effort to compute and tabulate first- and second-order data across a broader range of the parameter space will be essential for fully realizing the scientific potential of self-force models in gravitational wave astronomy.

% Additionally, --- make some comments about how other fields have made use of hamiltonian formalism to understand binary dynamics. self-force has found some success in translating perturbations into hamiltonian quantities, but this has only been restricted to equatorial orbits. Self-force should extend results for generic orbits to better match results to EOB and PN.

To support this endeavor, in this work we provide multiple methods and numerical tools for calculating the first-order metric perturbation $h_{\alpha\beta}^{(1)}$ for a small non-spinning body on an eccentric and precessing (or inclined)\footnote{We work in a frame in which the position and spin angular momentum of the more massive black hole are fixed at leading-order. Thus, inclination and precession are often used interchangeably: one can describe the small body's motion as ``inclined" with respect to the primary black hole's equatorial plane, or one can describe the orbital plane as``precessing" about the more massive body's spin axis.} bound orbit about a more massive, spinning black hole. We make these new tools openly-available through the Python library \texttt{pybhpt} \cite{PYBHPT, PYBHPTv092} to better facilitate the generation of first-order results  across the parameter space and to improve the accessibility of first-order data for the construction of second-order perturbations. We highlight that, with \texttt{pybhpt}, we extend first-order metric perturbation calculations to binaries involving (dimensionless) black hole spins as high as $0.999$, orbital eccentricities up to $0.6$, and inclination (or precession) angles up to $49\pi/100$ radians. 

With these new data, we also calculate a quasi-invariant quantity known as the generalized redshift. The redshift along eccentric orbits has been extensively explored in the literature and has been a useful tool for cross-validation of various metric perturbation calculations and facilitated comparisons between the self-force, effective-one-body, and post-Newtonian frameworks \cite{AkcaETC15, VandShah15, BiniDamoGera16b, BiniDamoGera16c, BiniDamoGera18, BiniGera19a}. Since its introduction, further works have highlighted the increasing importance of the redshift in understanding conservative dynamics. Le Tiec and collaborators connected the local redshift invariant to the binding energy and angular momentum of binary via the \emph{first law of black hole mechanics} \cite{LetiBlanWhit12, Leti15}, while Fujita \emph{et al.} and Lewis \emph{et al.} have connected the redshift to the two-body Hamiltonian \cite{FujiETC17, LewiETC25}. We use \texttt{pybhpt} to compute the generalized redshift to first-order in the mass-ratio along eccentric, precessing orbits in Kerr spacetime for the first time.

We also emphasize that \texttt{pybhpt} allows one to compute metric perturbations using multiple approaches and consequently in multiple gauges. This is beneficial for constructing $h^{(2)}_{\alpha\beta}$, which is sourced by first-order perturbations. To obtain well-defined second-order results, the first-order perturbations must be expressed in gauges that are regular both at the black hole horizon (e.g., ingoing radiation gauge) and at infinity (e.g., outgoing radiation gauge), as well as in gauges like Lorenz gauge that allow for regularization near the worldline to extract physically-relevant information. This work presents metric perturbation data for eccentric, precessing systems across multiple gauges for the first time.

In the following section, we overview the key methods employed both in this work and by \texttt{pybhpt}. In particular, we discuss our procedures for obtaining first-order metric perturbations for eccentric, precessing bound (nonplunging) geodesics in Kerr spacetime. We also introduce the generalized redshift quantity and summarize its connection to the self-force Hamiltonian that governs the conservative dynamics of the perturbed system. All redshift data produced by this work are provided in the Supplemental Material and online \cite{Nasipak_metric-reconstruction-paper_2026, Nasipak2026-st}.

\subsection{Paper overview}

We highlight four main contributions from this work:
\begin{enumerate}
    \item we implement {four} different methods for reconstructing the metric perturbation for eccentric precessing binaries in Kerr spacetime;
    \item we analyze the large-$r$ and near-horizon asymptotics of the perturbations, along with their singular structures near the worldline of the small body;
    \item we compute the generalized redshift invariant for eccentric, precessing binaries \emph{for the first time} and connect these data to the Hamiltonian capturing the conservative dynamics;
    \item we make all of the above methods, tools, and computations openly available through the Python library \texttt{pybhpt}.
\end{enumerate}

These methods leverage a procedure known as \emph{metric reconstruction} \cite{Chrz75, KegeCohe79, CoheKege79, Wald78, AkstAndeBack16} to derive $h_{\alpha\beta}^{(1)}$ from the Weyl scalars $\psi_0$ and $\psi_4$, bypassing the need to solve the perturbed Einstein equations directly. In Kerr spacetime, this is advantageous because solving the separable Teukolsky equation \cite{Teuk73}---which describes $\psi_0$ and $\psi_4$---is typically simpler than tackling the linearized Einstein equations, which are not known to be separable in Kerr.

Two of the metric reconstruction procedures that we employ are based on the work of Chrzanowski \cite{Chrz75}, Cohen and Kegeles (CCK) \cite{CoheKege79, KegeCohe79}, and Wald \cite{Wald78}, and are prevalent in the self-force literature (e.g., \cite{Ori03, ShahFrieKeid12, VandShah15, Vand16, Vand18, BereGravLups24, HollToom24}). These procedures lead to metric perturbations in either the ingoing radiation gauge (IRG) or outgoing radiation gauge (ORG). The other two procedures are based on the work of Aksteiner, Andersson, and B{\"a}ckdahl (AAB) \cite{AkstAndeBack16}. The AAB procedures have been leveraged as an intermediate step for reconstructing metric perturbations in Lorenz gauge \cite{DolaKavaWard22, WardKavaDola24}, though this method has only been applied to circular, nonprecessing (equatorial) orbits thus far \cite{DolaETC23}. We extend the AAB results to bound non-circular orbits in Kerr for the first time. Unlike the CCK procedure, the metric perturbations produced by the AAB procedures do not satisfy known gauge constraint equations. Thus, for reasons that will be apparent later, we will simply refer to the $h^{(1)}_{\mu\nu}$ produced by the two AAB procedures as either the symmetric radiation gauge (SRG) or antisymmetric radiation gauge (ARG) metric perturbations, with symmetric and antisymmetric referring to the nature of the reconstruction procedure rather than the underlying physics of the radiation.

We then use our $h_{\alpha\beta}^{(1)}$ results to calculate the generalized redshift correction \cite{Detw08, BaraSago11, AkcaETC15, VandShah15},
\begin{align}
    \langle \tilde{z}_1 \rangle_t = -\frac{1}{2}\langle \tilde{z}_0 h_{\alpha\beta}^{\mathrm{R}(1)} u^\alpha u^\beta\rangle_t,
\end{align}
where $u^\alpha$ is the geodesic four-velocity of $m_2$ in the Kerr background, $\tilde{z}_0 = (u^t)^{-1}$ is the geodesic redshift in Kerr, and $\langle \cdot \rangle_t$ represents an infinite time average. 
We calculate $\langle \tilde{z}_1 \rangle_t$ because it is (pseudo- or quasi-)invariant for a certain class of ``physically-reasonable'' gauges and thus provides a way of validating our calculations with those previously performed in the literature. (See Sec.~\ref{sec:redshift} or Ref.~\cite{BaraPoun19} for further details on pseudo-invariants in self-force theory.)

To extract $\langle \tilde{z}_1 \rangle_t$ from the metric perturbation, we must separate contributions to the redshift from the regular and singular fields, $h_{\alpha\beta}^{(1)} = h_{\alpha\beta}^{\mathrm{R}(1)} + h_{\alpha\beta}^{\mathrm{S}(1)}$ \cite{DetwWhit03}. The regular perturbation $h_{\alpha\beta}^{\mathrm{R}(1)}$ is finite and smooth along the small mass' worldline and is responsible for its self-forced motion, which is geodesic in the \emph{effective spacetime},
\begin{align} \label{eqn:gtilde}
    \tilde{\mathrm{g}}_{\mu\nu} = g_{\mu\nu} + \epsilon h^{\mathrm{R}(1)}_{\mu\nu} + O(\epsilon^2).
\end{align}
On the other hand, the singular field $h_{\alpha\beta}^{\mathrm{S}(1)}$ diverges along the worldline and does not contribute to the dynamics---it is an artifact from ``skeletonizing'' the smaller body and choosing to represent it as a point mass rather than an extended compact object \cite{BaraPoun19}. In this work, we use mode-sum regularization methods \cite{Heff21} to extract $h^{\mathrm{R}(1)}_{uu}=h^{\mathrm{R}(1)}_{\mu\nu}u^\mu u^\nu$ rather than computing the full regular field $h^{\mathrm{R}(1)}_{\alpha\beta}$.

The averaged redshift invariant $\langle \tilde{z}_1 \rangle_t$ not only serves as a valuable benchmark across self-force calculations but also plays a central role in computing self-force dynamics. Lewis \emph{et al.} \cite{LewiETC25} recently demonstrated that, to first order in $\epsilon$, the conservative dynamics of a binary system are described by a six-dimensional phase-space Hamiltonian (with respect to action-angle variables), with the interaction Hamiltonian determined by $\langle \tilde{z}_1 \rangle_t$. Thus, one can evolve the binary dynamics and gravitational wave signal using the redshift and Hamilton’s equations without directly computing the self-force. Thus $\langle \tilde{z}_1 \rangle_t$ provides all of the conservative self-force information that is necessary for building 1PA waveform models \cite{FujiETC17, LewiETC25}. Consequently, \texttt{pybhpt} provides a platform for constructing 1PA waveform models across the eccentric, precessing parameter space, while also facilitating comparisons between self-force, post-Newtonian and effective-one-body Hamiltonian formulations.

\subsection{Paper outline}

The paper is laid out as follows. In Sec.~\ref{sec:background}, we outline a general procedure for constructing first-order metric perturbations in IRG, ORG, SRG, and ARG in Kerr spacetime. We also summarize how one can then extract the redshift and, consequently, the self-force Hamiltonian from the metric perturbation. In Sec.~\ref{sec:mode-sum-full-sec}, we provide a detailed approach for calculating the metric perturbation for generic bound orbits using a frequency-domain approach. Then, in Sec.~\ref{sec:mode-sum-redshift-full-sec}, we outline a practical algorithm for constructing and regularizing the generalized redshift invariant from our metric perturbation results for eccentric and precessing orbits in Kerr. 

In Sec.~\ref{sec:gauge}, we compute the first-order metric perturbation for a series of different sources and compare these perturbations across the different gauges. In particular, we highlight their asymptotic behaviors near the horizon and infinity, along with their singular behavior along the worldline of the point-particle source. In Sec.~\ref{sec:redshift-results}, we present numerical results for the the generalized redshift, extending these data to precessing (inclined) orbits for the first time. For nonprecessing sources, we validate our results by comparing with those previously published in the literature. For precessing sources, we perform a self-consistency check by computing and comparing redshift values from the ORG, IRG, and SRG metric perturbations. Finally, we summarize these results in Sec.~\ref{sec:conclusion}. 
% We provide the scripts for generating this data in the Supplemental Material.

\section{Background}
\label{sec:background}

\subsection{Gravitational self-force framework}

We consider a binary composed of a small non-spinning compact object of mass $m_2$ bound to more massive black hole with mass $m_1 \gg m_2$ and angular momentum $J$. The background metric $g_{\mu\nu}$ is taken to be the Kerr metric with mass and spin parameters $M=m_1$ and $a = J/M$. We also introduce the dimensionless spin parameter $\hat{a} = a/M$ for later convenience. In Boyer-Lindquist coordinates $(t,r,z=\cos\theta,\phi)$, the Kerr line element is given by,
\begin{multline*}
    ds^2 = -\left(1 - \frac{2Mr}{\Sigma} \right) dt^2 - \frac{4Ma r (1-z^2)}{\Sigma} dtd\phi + \frac{\Sigma}{\Delta} dr^2 
    \\
    + \frac{\Sigma}{(1-z^2)} dz^2 + (1-z^2)^2\left(\frac{r^2+a^2}{1-z^2} + \frac{2Ma^2r}{\Sigma} \right) d\phi^2,
\end{multline*}
where
\begin{align*}
    \Delta &= r^2 - 2Mr + a^2, 
    &
    \Sigma &= r^2+a^2z^2.
\end{align*}
At zeroth-order in the mass-ratio $\epsilon$, the small mass $m_2$ is treated as a point-particle following a Kerr geodesic $x_p$.

To construct higher-order dynamics, one solves for the metric perturbation---sourced by $m_2$---by expanding both $h_{\mu\nu}$ and Einstein's equations order by order in $\epsilon$. This leads to a hierarchical set of linearized equations for $h_{\mu\nu}^{(n)}$,
\begin{align} \label{eqn:EFEs}
    (\mathcal{E}h^{(n)})_{\alpha\beta} = T^{(n)}_{\alpha\beta},
\end{align}
where $\mathcal{E}$ is the linearized Einstein operator and $T^{(n)}_{\alpha\beta}$ is the stress-energy source associated with the $n$th order perturbation $h_{\mu\nu}^{(n)}$. At first-order, $T^{(1)}_{\alpha\beta}$ reduces to the stress-energy of a point-particle source, while at second-order (and higher) the source depends on nonlinear couplings of $h_{\mu\nu}^{(1)}$ and the higher multipole structure of the source. (For further discussion, see Secs. 2 and 5 in Ref.~\cite{PounWard20} and references therein.)

The equations of motion then take the form
\begin{align} \label{eqn:EOMs}
    u^\alpha \nabla_\alpha u^\beta = \epsilon F^{(1)}_\beta[h^{\mathrm{R}(1)}] + \epsilon^2 F^{(2)}_\beta[h^{\mathrm{R}(2)}] + \cdots, 
\end{align}
where $F^{(n)}_\beta$ is the $n$th-order GSF that drives the inspiral of the small body. Thus, at first and higher orders, the mass and motion of the small secondary source the perturbations $h^{(n)}_{\mu\nu}$. The secondary interacts with these metric corrections, giving rise to the GSF that drives its inspiral into the more massive primary. Crucially, the GSF does not depend on the physical retarded metric perturbation, which diverges along the small body's worldline, but instead depends on the {regular} components $h^{\mathrm{R}(n)}_{\alpha \beta}$, which are finite along the worldline.

\subsection{Background Kerr geodesic}

The motion of the small mass $m_2$ is initially treated as a timelike Kerr geodesic
\begin{align}
   x_p^\alpha(\tau) \doteq [t_p(\tau), r_p(\tau), z_p(\tau), \phi_p(\tau)],
\end{align}
with proper time $\tau$ and four-velocity $u^\alpha = dx_p^\alpha/d\tau$. The geodesic possesses three constants of motion related to the three Killing symmetries---$\xi^{\mu}_{(t)}$, $\xi^{\mu}_{(\phi)}$, $\mathcal{Q}_{\mu\nu}$---of Kerr spacetime: the orbital energy $\mathcal{E} = - \xi^{\mu}_{(t)}u_\mu = -u_t$, the $z$-component of the orbital angular momentum $\mathcal{L}_z = \xi^{\mu}_{(\phi)} u_\mu = u_\phi$, and the Carter constant $\mathcal{Q} = \mathcal{Q}_{\mu\nu} u^\mu u^\nu$ (see Eqs.~(7)-(8) in Ref.~\cite{SteiWarb20}). 

Due to these first integrals of motion, and by introducing the Carter-Mino time parameter $\lambda$ defined by $\Sigma d\lambda = d\tau$, the geodesic equations reduce to a separable system of first-order equations. These equations have analytic solutions in terms of elliptic functions \cite{FujiHiki09, Vand20, DysoVand23}. For bound motion, they can also be solved via spectral integration methods \cite{HoppETC15, NasiOsbuEvan19}. From here on we restrict ourselves to bound, nonplunging geodesics, and, thus, the spectral methods are used for this work, as described in Appendix \ref{app:pybhpt}.

The radial motion librates between the turning points $r_\mathrm{min}$ and $r_\mathrm{max}$ with Mino frequency $\Upsilon_r$, while the polar motion oscillates between the turning points $z_\mathrm{max}$ and $z_\mathrm{min} = -z_\mathrm{max}$ with Mino frequency $\Upsilon_z$. Defining the phase variables $q_r = \Upsilon_r \lambda$ and $q_z = \Upsilon_z \lambda$, solutions take the form,
\begin{subequations} \label{eqn:geo}
    \begin{align}\label{eqn:geoTSol}
    t_p(\lambda) &= \Upsilon_t \lambda + \Delta \hat{t}_{(r)}(q_r + q_{r0}) - \Delta\hat{t}_{(r)}(q_{r0}) 
    \\ \notag
    & \qquad + \Delta \hat{t}_{(z)}(q_z + q_{z 0}) - \Delta \hat{t}_{(z)}(q_{z 0}) + q_{t0},
    \\
    r_p(\lambda) &= \Delta \hat{r}_{(r)}(q_r + q_{r0}),
    \\
    z_p(\lambda) &= \Delta \hat{z}_{(z)}(q_z + q_{z 0}),
    \\ \label{eqn:geoPhiSol}
    \phi_p(\lambda) &= \Upsilon_\phi \lambda + \Delta \hat{\phi}_{(r)}(q_r + q_{r0}) - \Delta \hat{\phi}_{(r)}(q_{r0})
    \\ \notag
    & \qquad + \Delta \hat{\phi}_{(z)}(q_z + q_{z 0}) - \Delta \hat{\phi}_{(z)}(q_{z 0}) + q_{\phi 0},
\end{align}
\end{subequations}
where the functions $\Delta \hat{x}^\alpha_{(r)}$ and $\Delta \hat{x}^\alpha_{(r)}$ are $2\pi$-periodic [e.g., $\Delta\hat{t}_{(r)}(q) = \Delta\hat{t}_{(r)}(q + 2\pi)$], $\Upsilon_t$ and $\Upsilon_\phi$ are the rates at which $t_p$ and $\phi_p$ accumulate with respect to Mino time, and the initial phases $q_0^\alpha\doteq [q_{t0},q_{r0},q_{z0},q_{\phi 0}]$ define the initial conditions of the orbit at $\lambda = 0$. (See Refs.~\cite{DrasFlanHugh05, Vand20, NasiEvan21} for further discussion.) The fundamental coordinate-time frequencies of the orbit are then given by
\begin{align} \label{eqn:frequencies}
    \Omega_r &= \frac{\Upsilon_r}{\Upsilon_t},
    &
    \Omega_z &= \frac{\Upsilon_z}{\Upsilon_t},
    &
    \Omega_\phi &= \frac{\Upsilon_\phi}{\Upsilon_t}.
\end{align}

% Rather than solving \eqref{eqn:EFEs} and \eqref{eqn:EOMs} simultaneously to obtain a self-consistent evolution of the binary, it is also advantageous to perform a two-timescale (or post-adiabatic) expansion of the field equations and equations of motion.

% For a Schwarzschild background, the lefthand-side of \eqref{eqn:EFEs} is separable due to spherical symmetry. In Kerr spacetime, however, 

\subsection{Teukolsky formalism}

Given a Kerr background and the geodesic motion of the point-particle source, we can then construct the resulting first-order metric perturbation $h^{(1)}_{\alpha\beta}$. However, since there is no known separable form of the linearized Einstein equations, it is much less convenient to solve for $h^{(n)}_{\alpha \beta}$ or $h^{\mathrm{R}(n)}_{\alpha \beta}$ directly. Instead, one can solve for perturbations in terms of the Weyl scalars $\{\psi_0,\psi_1,\psi_2,\psi_3,\psi_4\}$ \cite{NewmPenr62, NewmPenr63}. These scalars are constructed by projecting the Weyl tensor $C_{\alpha\beta\gamma\delta}$ onto a null tetrad $e^\alpha_\mathbf{a} \doteq (l^\alpha, n^\alpha, m^\alpha, \bar{m}^\alpha)$, where $l^\alpha$ and $n^\alpha$ are aligned with the principal null directions of the spacetime, and the tetrad is normalized such that $l^\alpha n_\alpha = -1 = - m_\alpha \bar{m}^\alpha$. In this work, we make use of the Kinnersley tetrad, which is explicitly defined in Appendix \ref{app:kinnersley}.
%Specifically, we have that
% \begin{subequations} \label{eqn:WeylFull}
%     \begin{align}
%     \psi_0 &= C_{\alpha\beta\gamma\delta} l^\alpha m^\beta l^\gamma m^\delta,
%     \\
%     \psi_4 &= C_{\alpha\beta\gamma\delta} n^\alpha \bar{m}^\beta n^\gamma \bar{m}^\delta.
% \end{align}
% \end{subequations}

With this choice of tetrad, $\psi_2$ is the only nonvanishing Weyl scalar in Kerr. For linear perturbations of Kerr, nearly all gauge-invariant information is encoded in the perturbed (maximal spin-weight) scalars $\psi_0$ and $\psi_4$ \cite{Wald73}.\footnote{The remaining gauge-invariant information is captured by certain \emph{completion} terms related to stationary and axisymmetric perturbations within Kerr. These completion terms are further discussed in Sec.~\ref{sec:completion}.} They are related to the metric perturbation via
\begin{subequations} \label{eqn:Weyl}
    \begin{align}
        \psi_0 &= \psi_0^{(1)} + O(\epsilon^2) = (\mathcal{T}_0)^{\alpha\beta} h^{(1)}_{\alpha\beta} + O(\epsilon^2),
        \\
        \psi_4 &= \psi_4^{(1)} + O(\epsilon^2) = (\mathcal{T}_4)^{\alpha\beta} h^{(1)}_{\alpha\beta} + O(\epsilon^2),
    \end{align}
\end{subequations}
where $\psi_0^{(1)} \sim \psi_4^{(1)} \sim \mathcal{O}(\epsilon)$ and the operators $\mathcal{T}_0$ and $\mathcal{T}_4$ are defined in Appendix \ref{app:def-operators}. Defining $\zeta = -(\psi_2/M)^{-1/3}$ and dropping all higher-order (nonlinear) corrections, these scalars then satisfy the Teukolsky equation(s) \cite{Teuk73},
\begin{subequations} \label{eqn:teuk}
    \begin{align}
    {\mathcal{O}_0} \psi_{0} &= {\mathcal{O}} \psi_{0} = T_{0},
    \\
    {\mathcal{O}}_4 \psi_4 &= {\mathcal{O}'} \psi_{4}  =\zeta^{-4}\mathcal{O} (\zeta^{4} \psi_{4}) = \zeta^{-4} T_{4},
\end{align}
\end{subequations}
where 
\begin{align} \label{eqn:source}
    T_{0,4} = (\mathcal{S}_{0,4})^{\alpha\beta} T_{\alpha\beta}^{(1)},
\end{align}
and the operators $\mathcal{S}_0$, $\mathcal{S}_4$, $\mathcal{O}$ and $\mathcal{O}'$ are also defined in Appendix \ref{app:def-operators}. For a point-particle source, the stress-energy reduces to
\begin{align} \label{eqn:Tmunu-pp}
    T_{(1)}^{\alpha\beta} = {8\pi} m_2 \int u^\alpha u^\beta \frac{\delta^{(4)}(x-x_p)}{\sqrt{-g}} d\tau,
\end{align} 
where $g$ is the determinant of $g_{\mu\nu}$. Crucially, the Teukolsky operator $\mathcal{O}$ is separable in Boyer-Lindquist coordinates, providing a much more convenient avenue for solving perturbations of Kerr. Because we will only focus on linear perturbations for the remainder of this work, we will drop the superscript ``${(1)}$" and simply denote the linear metric perturbation as $h_{\mu\nu}$ hereafter.

% The goal then is to relate the perturbed scalars $\psi_0$ and $\psi_4$ to the metric perturbation $h^{(1)}_{\alpha\beta}$.

\subsection{Metric reconstruction for vacuum perturbations of Kerr spacetime}
\label{sec:metricrecon}

The metric perturbation $h_{\mu\nu}$ can then be ``reconstructed'' from the Weyl scalars $\psi_0$ and $\psi_4$ using the procedures originally proposed by CCK and Wald \cite{Chrz75, CoheKege79, KegeCohe79, Wald78}, or those more recently developed by AAB \cite{AkstAndeBack16}. Following Wald's notation \cite{Wald78}, one begins with the operator identity
\begin{align}
    (\mathcal{S}_{0,4})^{\alpha\beta} (\mathcal{E} h')_{\alpha\beta} = \mathcal{O}_{0,4} (\mathcal{T}_{0,4})^{\alpha\beta}h'_{\alpha\beta},
\end{align}
which is valid for some solution $h'$ to the linearized Einstein equations. Taking the adjoint of the operators, as defined by Wald (see Eq.~(10) in Ref.~\cite{Wald78}), we then have
\begin{align} \label{eqn:adjoint}
    (\mathcal{E} \mathcal{S}_{0,4}^\dagger \phi)_{\alpha\beta} = (\mathcal{T}^\dagger_{0,4}\mathcal{O}^\dagger_{0,4}\phi)_{\alpha\beta},
\end{align}
where $\phi$ is some test field of spin-weight $\pm 2$, the adjoint operators $\mathcal{S}_{0,4}^\dagger$, $\mathcal{T}_{0,4}^\dagger$, and $\mathcal{O}_{0,4}^\dagger$ are defined in Appendix \ref{app:def-operators}, and we have made use of the fact that $\mathcal{E}^\dagger = \mathcal{E}$. Inserting the ansatz
\begin{align} \label{eqn:hrec-ansatz}
    h^\mathrm{rec}_{\alpha\beta} = c_0(\mathcal{S}_{0}^\dagger \phi_0)_{\alpha\beta} + c_4 (\mathcal{S}_{4}^\dagger \phi_4)_{\alpha\beta},
\end{align}
into \eqref{eqn:adjoint}, we then see that, for constant parameters $c_0$ and $c_4$, $h^\mathrm{rec}_{\alpha\beta}$ is a valid vacuum solution to the linearized Einstein equations if the potentials $\phi_{0,4}$, also known as \emph{Hertz potentials}, satisfy the vacuum adjoint Teukolsky equations, $\mathcal{O}^\dagger_{0,4}\phi_{0,4} = 0$. [Recall Eq.~\eqref{eqn:teuk}.] Furthermore, one can ensure that $h^\mathrm{rec}_{\alpha\beta}$ is the vacuum perturbation representing the Weyl scalars $\psi_{0,4}$ by enforcing Eq.~\eqref{eqn:Weyl}, leading to differential relations between $\psi_{0,4}$ and $\phi_{0,4}$.

Our choice of $c_0$ and $c_4$ then affects the relationship between the Hertz potentials and Weyl scalars, as well as the gauge of the reconstructed metric perturbation. The CCK procedure amounts to setting $c_4 =0$ or $c_0 =0$, resulting in a metric perturbation in either IRG or ORG,
\begin{subequations}
    \begin{align}
        l^\alpha h^\mathrm{IRG}_{\alpha\beta} &= 0 = g^{\alpha\beta}h^\mathrm{IRG}_{\alpha\beta},
        \\
        n^\alpha h^\mathrm{ORG}_{\alpha\beta} &= 0 = g^{\alpha\beta}h^\mathrm{ORG}_{\alpha\beta},
    \end{align}
\end{subequations}
respectively. Meanwhile the AAB procedures are constructed by the choices $c_4 = - c_0$ and $c_0 = c_4$, which result in metric perturbations in gauges that we refer to as either ARG or SRG, respectively. Both are also traceless gauges.

Therefore, we explicitly express our reconstructed metric perturbations as
\begin{subequations} \label{eqn:reconstruct}
    \begin{align}
        h^\mathrm{ORG,rec}_{\alpha\beta} &= 4\mathrm{Re}(\mathcal{S}_4^\dagger \zeta^4 \Phi^\mathrm{O}_0)_{\alpha\beta},
        \\
        h^\mathrm{IRG,rec}_{\alpha\beta} &= 4\mathrm{Re}(\mathcal{S}_0^\dagger \zeta^4 \Phi^\mathrm{I}_4)_{\alpha\beta},
        \\
        h^\mathrm{SRG,rec}_{\alpha\beta} &= 4\mathrm{Re}\left[(\mathcal{S}_4^\dagger \zeta^4 \Phi^\mathrm{S}_0)_{\alpha\beta} + (\mathcal{S}_0^\dagger \zeta^4 \Phi^\mathrm{S}_4)_{\alpha\beta}\right],
        \\ \label{eqn:reconstruct-ARG}
        h^\mathrm{ARG,rec}_{\alpha\beta} &= 4\mathrm{Re}\left[(\mathcal{S}_4^\dagger \zeta^4 \Phi^\mathrm{A}_0)_{\alpha\beta} - (\mathcal{S}_0^\dagger \zeta^4 \Phi^\mathrm{A}_4)_{\alpha\beta}\right],
    \end{align}
\end{subequations}
where $\Phi_0$ and $\Phi_4$ are our (rescaled) Hertz potentials, and we introduce the factors of $2$ and $\zeta^4$ to simplify expressions later on. We relate $\Phi_0$ and $\Phi_4$ to $\psi_0$ and $\psi_4$ by inserting \eqref{eqn:reconstruct} into \eqref{eqn:Weyl}, resulting in
\begin{subequations} \label{eqn:hertzInversion}
    \begin{align} \label{eqn:hertzInversionORG}
        2\psi_4 &= \thorn'^4 \overline{\zeta^4 \Phi^\mathrm{O}_0},
        &
        2\psi_0 &= \edth^4 \overline{\zeta^4 \Phi^\mathrm{O}_0} + 3 M \pounds_\xi \Phi^\mathrm{O}_0,
        \\ \label{eqn:hertzInversionIRG}
        2\psi_0 &= \thorn^4 \overline{\zeta^4\Phi^\mathrm{I}_4},
        &
        2\psi_4 &= \edth'^4 \overline{\zeta^4\Phi^\mathrm{I}_4} - 3 M \pounds_\xi \Phi^\mathrm{I}_4,
        \\ \label{eqn:hertzInversion+RG}
        \psi_0 &= \edth^4 \overline{\zeta^4 \Phi^\mathrm{S}_0},
        &
        \psi_4 &= \edth'^4 \overline{\zeta^4 \Phi^\mathrm{S}_4},
        \\ \label{eqn:hertzInversion-RG}
        \psi_0 &= 3 M \pounds_\xi \Phi^\mathrm{A}_0,
        &
        \psi_4 &= 3 M \pounds_\xi \Phi^\mathrm{A}_4,
    \end{align}
\end{subequations}
with $\xi^\alpha = \zeta(\rho' l^\alpha - \rho n^\alpha - \tau' m^\alpha + \tau \bar{m}^\alpha)$ and the coefficients $\rho$, $\rho'$, $\tau$, and $\tau'$ defined in Appendix \ref{app:GHP}. Conveniently, $\pounds_\xi$ commutes with $\thorn$, $\edth$, $\thorn'$, and $\edth'$ \cite{PounWard20}. Leveraging this fact, we can combine \eqref{eqn:reconstruct-ARG} and \eqref{eqn:hertzInversion-RG},
\begin{align} \label{eqn:reconstruct-RG2}
    M\pounds_\xi h^\mathrm{SRG,rec}_{\alpha\beta} &= \frac{4}{3}\mathrm{Re}\left[(\mathcal{S}_4^\dagger \zeta^4 \psi_0)_{\alpha\beta} - (\mathcal{S}_0^\dagger \zeta^4 \psi_4)_{\alpha\beta}\right],
\end{align}
yielding a direct relationship between the Weyl scalars and the metric perturbation. Combining this result with \eqref{eqn:Weyl}, one then obtains the AAB operator identity
\begin{align}
    M\pounds_\xi = \frac{2}{3}\mathcal{S}_4^\dagger \zeta^4 \mathcal{T}_0 - \frac{2}{3}\mathcal{S}_0^\dagger \zeta^4 \mathcal{T}_4 + \mathrm{c.c.}
\end{align}
discussed in other works \cite{AkstAndeBack16, DolaKavaWard22}. Unlike the CCK reconstruction, this AAB operator identity can also be extended to include sourced perturbations as investigated in Ref.~\cite{WardKavaDola24}. However, $\xi^\alpha = \delta^\alpha_t$ in Boyer-Lindquist coordinates, and thus $\pounds_\xi = \partial_t $, indicating that the ansatz \eqref{eqn:reconstruct-ARG} can only reconstruct time-dependent pieces of the metric perturbation from $\psi_0$ and $\psi_4$ without introducing linear-in-time modes to $h_{\alpha\beta}$. Static mode contributions to $h_{\alpha\beta}$ must be handled separately, e.g., Ref.~\cite{DolaETC23}. Thus, in this work we will only focus on radiative modes when calculating $h^\mathrm{ARG,rec}_{\alpha\beta}$.
% Note that we will also want the relations
% \begin{subequations}
%     \begin{align}
%         2\psi_0&= \thorn^4 \overline{\zeta^4\Phi^-_4} - \edth^4 \overline{\zeta^4\Phi^-_0},
%         \\
%         2\psi_4 &= \thorn'^4 \overline{\zeta^4\Phi^-_0} - \edth'^4 \overline{\zeta^4\Phi^-_4},
%     \end{align}
% \end{subequations}
% for later when dealing with static contributions to these scalars.

%\subsection{Expressions for the Kinnersley tetrad}

\subsection{Metric reconstruction and completion for point-particle sources}
\label{sec:completion}

While the metric reconstruction procedures described above are primarily valid for \emph{vacuum} perturbations of Kerr, Ori \cite{Ori03} demonstrated that one could extend these methods to perturbations excited by point-particle sources by leveraging the fact that the metric perturbation satisfies the homogeneous linearized field equations everywhere except along the worldline of the particle. Therefore, one can reconstruct the full metric perturbation by taking the limit of the solutions in the two vacuum regions, leading to
\begin{align} \label{eqn:reconOri}
    h^\mathrm{\mathcal{G}RG,rec}_{\alpha\beta} = \begin{cases}
        h^{\mathrm{\mathcal{G}RG,rec}-}_{\alpha\beta}, & r \leq r_p(t),
        \\
        h^{\mathrm{\mathcal{G}RG,rec}+}_{\alpha\beta}, & r \geq r_p(t),
    \end{cases}
\end{align}
where $r_p(t)$ is the radial position of the particle, and $\mathcal{G}$RG refers to the gauge of the perturbation with $\mathcal{G}=$ \{A,I,O,S\}. However, these reconstructed perturbations suffer from string singularities \cite{Ori03, PounMerlBara14}, along which $h^\mathrm{\mathcal{G}RG,rec}_{\alpha\beta}$ fails to satisfy the field equations for a point-particle stress-energy source [Eqs.~\eqref{eqn:EFEs} and \eqref{eqn:Tmunu-pp}]. These singularities give rise to either half-string solutions, with singularities extending from the worldline along null rays toward either the horizon or infinity, or full-string solutions, with singularities extending in both directions \cite{PounMerlBara14}. Alternatively, one can ``glue" together two half-string solutions, thus eliminating string singularities in the vacuum region but at the expense of introducing a gauge singularity at $r_p(t)$. These so-called \emph{no-string solutions} \cite{MerlETC16} have been employed in previous metric reconstruction calculations (e.g., \cite{Vand17, Vand18}) and are the solutions that we construct in this work.

To solve for the inhomogeneous metric perturbation, one also needs to take into account certain \emph{completion pieces} which are not produced by the reconstruction procedure. As originally observed by Wald \cite{Wald73, Wald78} (and noted in other metric reconstruction procedures \cite{MerlETC16, Vand17, Vand18, GreeHollZimm20, ToomETC22, HollToom24}) $\psi_0$ and $\psi_4$ do not contain all of the physical information of the metric perturbation. For the systems we consider, the missing information amounts to residual perturbations within the Kerr family,
\begin{align} \label{eqn:hcomp}
    h^\mathrm{comp \pm}_{\alpha\beta} = c_M^\pm \frac{\partial g_{\alpha\beta}}{\partial M} + c_J^\pm \frac{\partial g_{\alpha\beta}}{\partial J}.
\end{align}
The $\pm$ notation has the same meaning as Eq.~\eqref{eqn:reconOri}---it refers to the radial domain in which each completion piece contributes---while the constants $c_M^\pm$ and $c_J^\pm$ are determined by the nature of the source. For point-particles following Kerr geodesics, we have $c_J^- = 0 = c_M^-$, $c_M^+ = M\mathcal{E}$, and $c_J^+ = M^2\mathcal{L}_z$ \cite{MerlETC16, Vand17}. Our exact expressions for the completion piece are provided in Appendix \ref{app:completion}.

Alternatively, as first proposed by Green, Hollands, and Zimmerman (GHZ) \cite{GreeHollZimm20}, one can ``complete" the solution by introducing a \emph{corrector tensor} $x_{\alpha\beta}$ so that the full solution, $h^\mathrm{\mathcal{G}RG,rec}_{\alpha\beta}+x_{\alpha\beta}$, satisfies the point-particle-sourced field equations. Toomani \emph{et al.}~\cite{ToomETC22} later demonstrated that this corrector tensor includes the completion term $h^\mathrm{comp\pm}_{\alpha\beta}$, along with singular terms that either contribute additional structure to the string singularities (in the half- and full-string cases) or to the gauge singularity (in the no-strings case), ensuring that the field equations are globally satisfied. A key advantage of the GHZ method is that, unlike the CCK-Ori procedures, it can be applied to extended sources \cite{ToomETC22, HollToom24}, as demonstrated in Ref.~\cite{BourETC24}. In this work, however, we restrict our focus to point-particle sources.

Combining all of these terms, the total metric perturbation in the two vacuum regions (and in the gauge $G'$)
\begin{align} \label{eqn:h1decomp}
    h^{G'}_{\alpha\beta} &= \begin{cases}
        h^{G'-}_{\alpha\beta}, & r \leq r_p(t),
        \\
        h^{G'+}_{\alpha\beta}, & r \geq r_p(t),
    \end{cases}
\end{align} 
is then generically given by
\begin{align} \label{eqn:h1pm}
    h^{{G'\pm}}_{\alpha\beta} &= h^{G,\mathrm{rec}\pm}_{\alpha\beta} + x_{\alpha\beta} + \pounds_{\xi^\pm} g_{\alpha\beta},
\end{align}
where ${G}=\mathrm{\mathcal{G}RG} = \{\mathrm{ARG}, \mathrm{IRG}, \mathrm{ORG}, \mathrm{SRG}\}$ represents the gauge of the reconstructed metric, $x_{\alpha\beta}$ is the corrector tensor, and $\pounds_{\xi^\pm} g_{\alpha\beta} = -\nabla_\alpha \xi^\pm_\beta-\nabla_\beta \xi^\pm_\alpha$ is a pure gauge contribution resulting from any additional transformations $x^\alpha \rightarrow x^\alpha + \epsilon \xi^\alpha_\pm$ within the two domains that takes the perturbation from gauge ${G}\rightarrow {G}'$. One can leverage this additional gauge freedom to add certain \emph{gauge-smoothing} or \emph{gauge-fixing} terms that ensure a particular level of continuity of the metric perturbation across the worldline. This may be necessary when computing local quantities, such as the GSF. 

While a general prescription for calculating these gauge-smoothing corrections was given in Ref.~\cite{BiniGera19d}, we do not explicitly compute these terms for the full metric perturbation. Instead, when extracting the redshift, as we will describe in the following section, we only need to impose the condition that $h_{\alpha\beta}u^\alpha u^\beta$ is continuous across the worldline \cite{MerlETC16, BiniGera19d}. Taking into account $h^\mathrm{\mathcal{G}RG,rec+}_{\alpha\beta}u^\alpha u^\beta = h^\mathrm{\mathcal{G}RG,rec-}_{\alpha\beta}u^\alpha u^\beta$ at $x_p$ for radiation gauges\footnote{As we demonstrate in Sec.~\ref{sec:gauge}, this also extends to SRG but most likely not to ARG, which possesses a different singular structure.} \cite{PounMerlBara14}, we then have
\begin{align} \label{eqn:gauge-fix}
    \left.\pounds_{\xi^{-}} g_{\alpha\beta}u^\alpha u^\beta \right\vert_{x=x_p} = \left.h^\mathrm{comp+}_{\alpha\beta}u^\alpha u^\beta\right\vert_{x=x_p} ,
\end{align}
and $\pounds_{\xi^{+}} g_{\alpha\beta}u^\alpha u^\beta = 0$. We leave the explicit derivation of the full gauge-fixing term for generic orbits for future work. Additionally, as done in previous works \cite{VandShah15, Vand17, Vand18}, we do not complete the solution by constructing the full corrector tensor $x_{\alpha\beta}$, but just the contribution coming from $h^\mathrm{comp \pm}_{\alpha\beta}$, since this is the only completion information we need to construct the redshift.

\subsection{Redshift}
\label{sec:redshift}

As previously mentioned, the small body follows a geodesic in the {effective spacetime} $\tilde{g}_{\mu\nu}$ [see Eq.~\eqref{eqn:gtilde}] with four-velocity $\tilde{u}^\alpha = d x^\alpha_p/d\tilde{\tau}$ and proper time $\tilde{\tau}$. One can then define a redshift within the effective spacetime as\footnote{Equation \eqref{eqn:redshift-full} is referred to as the redshift, because it relates redshifted measurements between an asymptotic observer and an observer along the worldline $x_p(\tilde{\tau})$. However, this redshift is not physical since $z$ is defined with respect to the \emph{unphysical} effective spacetime $\tilde{g}_{\mu\nu}$.}
\begin{subequations}\label{eqn:redshift-full}
    \begin{align}
    \tilde{z} &= \frac{1}{\tilde{u}^t},
    \\ 
    &= \frac{1}{{u}^t}\left[ 1- \frac{\epsilon}{2}h_{\mu\nu}^{\mathrm{R}(1)} u^\mu u^\nu  + O(\epsilon^2)\right],
    \\
    &= \tilde{z}_0 + \epsilon \tilde{z}_1 + O(\epsilon^2),
\end{align}
\end{subequations}
where $\tilde{z}_0 = (u^t)^{-1}$ denotes the redshift in the background spacetime and
\begin{align} \label{eqn:redshift}
    \tilde{z}_1 = -\frac{1}{2} \tilde{z}_0 h_{uu}^{\mathrm{R}(1)}
\end{align}
is its first-order correction with $h_{uu}^{\mathrm{R}(1)} = h_{\mu\nu}^{\mathrm{R}(1)} u^\mu u^\nu$.\footnote{Note that other works (e.g., \cite{Detw08, BaraSago10, VandShah15}) define the redshift as the inverse of \eqref{eqn:z}: $U = \langle \tilde{z}^{-1}\rangle_{\tilde{\tau}} = \langle \tilde{z}\rangle_t^{-1}$. Consequently $\langle \tilde{z}_1 \rangle_t = - \langle \tilde{z}_{(0)} \rangle_t^{2} U_1$, where $U_1$ is the first-order correction to the averaged inverse redshift.}

In the case of circular equatorial (nonprecessing, noneccentric) orbits, Detweiler first recognized that the redshift correction $\tilde{z}_1$ presents a (quasi-)invariant measure of conservative perturbations \cite{Detw08}. Barack and Sago later generalized the ``invariance" of this redshift measure by taking the proper-time average of $z$ over one radial period \cite{BaraSago10}. In this work, we follow Ref.~\cite{FujiETC17} and present the natural extension of the redshift invariant to eccentric, precessing orbits. It is given by the infinite coordinate-time average of \eqref{eqn:redshift-full} at the fixed frequencies $\{\Omega_r, \Omega_\theta,\Omega_\phi\}$, masses $\{m_1, m_2\}$ and primary spin $a$,
\begin{align} \label{eqn:z}
    \langle \tilde{z}\rangle_t (\Omega_r, \Omega_\theta, \Omega_\phi; m_1,m_2, a) &= \left\langle \frac{d\tilde{\tau}}{dt} \right\rangle_t,
\end{align}
where
\begin{align}
    \langle f \rangle_{t'} = \lim_{T'\rightarrow\infty} \frac{1}{2T'} \int_{-T'}^{T'} f(t') dt'.
\end{align}
Expanding in the mass-ratio, we then have
\begin{align}
    \langle \tilde{z}\rangle_t(\Omega_i) = \langle \tilde{z}_{(0)}\rangle_t(\Omega_i) + \epsilon \langle \tilde{z}_{(1)}\rangle_t(\Omega_i) + O(\epsilon^2),
\end{align}
where $\langle \tilde{z}_{(0)}\rangle_t(\Omega_i)$ is the redshift in the background spacetime evaluated along a \emph{geodesic} with frequencies $\Omega_i\doteq(\Omega_r,\Omega_\theta,\Omega_\phi)$ (as opposed to $\langle \tilde{z}_{0}\rangle_t(\Omega_i)$, which is the redshift in the background spacetime evaluated along a (self-)forced orbit with frequencies $\Omega_i$). Similarly, $\langle \tilde{z}_{(1)}\rangle_t(\Omega_i) = \langle \tilde{z}_{1}\rangle_t(\Omega_i) + O(\epsilon)$ is given by Eq.~\eqref{eqn:redshift}, but with the right-hand side once again evaluated along a geodesic with frequencies $\Omega_i$.\footnote{As derived in \cite{LewiETC25}, the fact that we can evaluate the zeroth- and first-order terms along geodesic motion is a special consequence of the fixed-frequency expansion. Other parametrizations lead to additional $O(\epsilon)$ corrections due to differences in the redshift along a geodesic and a (self-)accelerated orbit, as measured in the background spacetime. (See Appendix D in \cite{LewiETC25}.)}

Like previous definitions of the redshift, Eq.~\eqref{eqn:z} is a quasi-invariant, by which we mean that $\langle \tilde{z} \rangle_t$ is invariant across a class of {physically-reasonable} gauges that asymptotically preserve the values of the orbital frequencies $\Omega_i$, are continuous across the worldline, and respect the periodic nature of the orbit.\footnote{It is mentioned in the literature that these conditions require a gauge that is {asymptotically} flat, but this is too strict of a requirement. One needs a gauge in which the static \emph{monopole} and \emph{dipole} contributions are asymptotically flat. Thus, Lorenz gauge perturbations must be corrected to calculate the "correct" redshift invariant, while IRG perturbations, which are singular at infinity, do not need to be corrected.} (See \cite{BaraPoun19} for further discussion of quasi-invariance.) Our reconstructed metric perturbations, with the inclusion of the gauge-fixing term in \eqref{eqn:gauge-fix}, satisfy these conditions, enabling comparisons with previous results in the literature.

We note that, because the zeroth-order correction $\langle \tilde{z}_{(0)}\rangle_t$ is known analytically and straightforward to calculate, most of the computational effort in calculating $\langle \tilde{z}\rangle_t$ comes from computing $\langle \tilde{z}_{1}\rangle_t$. Thus, the first-order redshift correction $\langle \tilde{z}_{1}\rangle_t$ is often referred to simply as the redshift, even though the redshift is technically the full quantity $\langle \tilde{z}\rangle_t$. We also adopt this nomenclature and will refer to $\langle \tilde{z}_{1}\rangle_t$ as the redshift and redshift correction throughout this work. 

\subsection{Hamiltonian}

As previously mentioned, the redshift also plays a crucial role in understanding the conservative evolution of the binary. The conservative dynamics of the binary are captured by the six-dimensional Hamiltonian \cite{LewiETC25},
\begin{align} \label{eqn:H6D}
    H(J_i, \varphi^i)  = E_{(0)}(J_i) + \frac{\epsilon}{2} \langle \mathcal{H}_{(1)}\rangle_t(J_i) + O(\epsilon^2),
\end{align}
where $(J_i, \varphi^i)$ represent the action-angle variables of the six-dimensional phase space, $E_{(0)}(J_i)$ is the orbital energy of a geodesic parametrized by the actions $J_i = \{J_r, J_\theta, J_\phi\}$, and the interaction Hamiltonian $\langle \mathcal{H}_{(1)}\rangle_t$ is simply related to the redshift via
\begin{align}
    \langle \mathcal{H}_{(1)}\rangle_t = m_2\left[ \langle \tilde{z}_1 \rangle_t + O(\epsilon) \right].
\end{align}
Thus, by computing $\langle \tilde{z}_1 \rangle_t$ across this phase space, one can use Hamilton's equations,
\begin{align}
    \frac{d\varphi^i}{dt} &= \frac{\partial H}{\partial J_i}
    &
    \frac{dJ_i}{dt} &= -\frac{\partial H}{\partial \varphi^i},
\end{align}
to solve for the conservative dynamics of the system (to first-order in the mass-ratio).

\section{Mode-sum reconstruction for a point-particle source}
\label{sec:mode-sum-full-sec}

We now describe our frequency-domain procedures for reconstructing the metric perturbation using the methods outlined in Sec.~\ref{sec:background}. In particular, in Secs.~\ref{sec:mode-sum-weyl} and \ref{sec:mode-sum-hertz} we describe our methods for calculating the Weyl scalars and the Hertz potential(s), respectively, while in Sec.~\ref{sec:mode-sum-recon} we discuss the reconstruction of the metric perturbation. Note that some results are tetrad-dependent and, therefore, only valid when using the Kinnersley tetrad (see Appendix \ref{app:kinnersley}), as we do in this work.

\subsection{Mode-sum expressions for Weyl scalars}
\label{sec:mode-sum-weyl}

As outlined in Sec.~\ref{sec:background}, to solve for $h_{\mu\nu}$ we first construct the Weyl scalars $\psi_0$ and $\psi_4$, which satisfy the Teukolsky equation with spin-weights $s=+2$ and $s=-2$, respectively. By transforming to the frequency-domain, the Teukolsky equation is amenable to separation of variables in Boyer-Lindquist coordinates. For a point-particle source, we express the Weyl scalars in terms of the mode-decomposed extended homogeneous solutions \cite{BaraOriSago08}
\begin{subequations} \label{eqn:Weylmodesum}
    \begin{align}
    \psi_{\pm 2} &= \psi_{\pm 2}^\mathcal{I} \Theta(r-r_p(t)) + \psi_{\pm 2}^\mathcal{H} \Theta(r_p(t)-r),
    \\
    \psi_{\pm 2}^\mathcal{J} &= \sum_{jm\omega} \psi^\mathcal{J}_{\pm 2jm\omega}(r)S_{\pm 2jm\omega}(z) e^{i(m\phi - \omega t)},
\end{align}
\end{subequations}
with $\mathcal{J} = \{\mathcal{H}, \mathcal{I}\}$, $\psi_{+2} = \psi_0$ and $\psi_{-2} = \zeta^{4}\psi_4$. Above we have introduced the shorthand notation
\begin{align}
    \sum_{jm\omega} = \sum_{j=2}^\infty \sum_{m=-j}^j \sum_{k=-\infty}^\infty \sum_{n=-\infty}^\infty,
\end{align}
with $\omega \rightarrow \omega_{mkn} = m\Omega_\phi+k\Omega_\theta+n\Omega_r$ in the summand. Note that for noneccentric (circular and spherical) orbits all $n\neq 0$ modes vanish, while all $k\neq 0$ modes vanish for nonprecessing (equatorial) orbits.

The angular eigenfunctions $S_{\pm 2jm\omega}(z)$ are known as spin-weighted spheroidal harmonics, and satisfy
\begin{align}
    \left[\frac{d}{dz}\left((1-z^2)\frac{d}{dz}\right) - V^{(\theta)}_{sjm\omega}(z)\right]S_{sjm\omega}(z)= 0,
\end{align}
with potential
\begin{align}
    V^{(\theta)}_{sjm\omega}(z) &= a^2\omega^2(1-z^2)+\frac{(m+sz)^2}{1-z^2}
	\\ \notag
    & \qquad \qquad \qquad -2a\omega (m-s z)-s-\lambda_{sjm\omega},
\end{align}
spheroidicity $\gamma=a\omega$ and eigenvalue $\lambda_{sjm\omega}$. In the limit $a\omega=0$, the spheroidal harmonics reduce to spin-weighted spherical harmonics $Y_{s\ell m}(z)$ with $\lambda_{s\ell m0}=\ell(\ell+1)-s(s+1)$. (We will use $\ell$ for harmonic mode numbers and reserve $l$ for the tetrad leg $l^\alpha$.) In this work, we construct spheroidal harmonics by expanding them as series in spherical harmonics,
\begin{equation} \label{eqn:mixingCoeffs}
    S_{sjm\omega}(z) = \sum_{l=l_\mathrm{min}}^\infty b^{\ell}_{sjm\omega}\, Y_{s\ell m}(z),
\end{equation}
with $\ell_\mathrm{min} = \mathrm{max}[|m|,|s|]$. The coupling coefficients $b^{\ell}_{sjm\omega}$ then satisfy a five-term recurrence relation, which can be solved as an eigenvalue problem \cite{Hugh00b}. In practice, the coupling coefficients decay rapidly with $|\ell-j|$; therefore, even for large values of the spheroidicity (e.g., $a\omega > 5$), one only needs 10 to 30 terms in \eqref{eqn:mixingCoeffs} to compute $S_{sjm\omega}(z)$ to machine precision. (See Appendix \ref{app:pybhpt} for further details on our numerical algorithm.)

The extended homogeneous radial functions $\psi^\mathcal{J}_{\pm 2jm\omega}(r)$ are built from the retarded inhomogeneous solutions $\psi^\mathrm{ret}_{\pm 2jm\omega}(r)$, which satisfy the sourced radial Teukolsky equation,
\begin{align} \label{eqn:teuk-radial}
    \left[\Delta^{-s} \frac{d}{dr}\left( \Delta^{s+1} \frac{d}{dr}\right) + V_{sjm\omega}(r)\right] \psi^\mathrm{ret}_{sjm\omega}(r) &= T_{sjm\omega}(r),
\end{align}
with potential
\begin{align*}
    V_{sjm\omega}(r) = \left(\frac{K^2-2is(r-M)K}{\Delta}+4is\omega r - \lambda_{sjm\omega} \right),
\end{align*}
and $K=(r^2+a^2)\omega-ma$. The sources $T_{\pm 2jm\omega}(r)$ for point-particles on bound geodesics are given in Appendix \ref{app:source}. 

We solve for $\psi^\mathrm{ret}_{\pm 2jm\omega}(r)$ by first constructing the retarded Green's function from the homogeneous solutions
\begin{subequations} \label{eqn:Rasymp}
    \begin{align} \label{eqn:RinHor}
    R^\mathcal{H}_{sjm\omega} (r \rightarrow r_+) &\sim \left(\frac{M^2}{\Delta}\right)^{s} e^{-i k r_*},
    \\ \label{eqn:RupInf}
    R^\mathcal{I}_{sjm\omega} (r \rightarrow \infty) &\sim 
    \left(\frac{M}{r}\right)^{2s+1} e^{i\omega r_*},
\end{align}
\end{subequations}
where we have introduced the shifted frequency $k=\omega - m\Omega_+$, the horizon frequency $\Omega_+ = m\hat{a}/(2r_+)$, outer horizon $r_+ = M(1+\kappa)$, and $\kappa = \sqrt{1-\hat{a}^2}$. (Recall $\hat{a}=a/M$.) We have also introduced the tortoise coordinate
\begin{align} \label{eqn:rstar}
    r_* = r + \frac{r_+}{\kappa} \ln \frac{r-r_+}{2M} 
    - \frac{r_-}{\kappa} \ln \frac{r-r_-}{2M}.
\end{align}
The resulting inhomogeneous solutions then satisfy retarded boundary conditions at the horizon and infinity,
\begin{subequations} \label{eqn:psiAsymp}
    \begin{align} \label{eqn:psiMinus}
    \psi^\mathrm{ret}_{\pm2jm\omega} (r \leq r_\mathrm{min}) &= Z^\mathcal{H}_{\pm 2jm\omega} R^\mathcal{H}_{\pm 2jm\omega}(r),
    \\ \label{eqn:psiPlus}
    \psi^\mathrm{ret}_{\pm2jm\omega} (r \geq r_\mathrm{max}) &= Z^\mathcal{I}_{\pm 2jm\omega} R^\mathcal{I}_{\pm 2jm\omega}(r),
\end{align}
\end{subequations}
with amplitudes
\begin{align} \label{eqn:Zteuk}
    Z^\mathcal{H/I}_{\pm 2jm\omega} = \int_{r_\mathrm{min}}^{r_\mathrm{max}} \frac{\Delta^{\pm2} R^\mathcal{I/H}_{\pm2jm\omega}(r) T_{\pm2jm\omega}(r)}{{\mathcal{W}_{\pm2jm\omega}}} dr,
\end{align}
and constant Wronskian 
\begin{align*}
    \mathcal{W}_{sjm\omega} = \Delta^{s+1}\left[R^\mathcal{H}_{sjm\omega}\frac{d}{dr} R^{\mathcal{I}}_{sjm\omega} - R^\mathcal{I}_{sjm\omega}\frac{d}{dr} R^{\mathcal{H}}_{sjm\omega}\right].
\end{align*}
Recall that $r_\mathrm{min}$ and $r_\mathrm{max}$ are the radial turning points of the point-particle's motion. 

Extending \eqref{eqn:psiMinus} and \eqref{eqn:psiPlus} to be valid for all $r_+ \leq r \leq \infty$ then gives us the mode contributions to the {extended homogeneous solutions} \cite{BaraOriSago08},
\begin{align} \label{eqn:EHSmode}
    \psi^\mathcal{J}_{\pm 2jm\omega}(r) &= Z^\mathcal{J}_{\pm 2jm\omega} R^\mathcal{J}_{\pm 2jm\omega}(r),
\end{align}
which are used in Eq.~\eqref{eqn:Weylmodesum}. For eccentric or precessing sources, the solutions in \eqref{eqn:EHSmode} do not satisfy the inhomogeneous wave equation mode-by-mode, but the full mode-sum in \eqref{eqn:Weylmodesum} satisfies the four-dimensional inhomogeneous Teukolsky equation with retarded boundary conditions \cite{BaraOriSago08}. Our numerical methods for producing $Z^\mathcal{J}_{\pm 2jm\omega} $ and $R^\mathcal{J}_{\pm 2jm\omega}(r)$ are outlined in Appendix \ref{app:pybhpt}.

Connecting these extended homogeneous solutions to the form of our reconstructed metric perturbation in Eqs.~\eqref{eqn:reconOri}-\eqref{eqn:h1pm}, we see that the horizon-side solutions contribute to the metric perturbation $h^-_{\alpha\beta}$ in the domain $r \leq r_p(t)$, while the infinity-side solutions contribute to $h^+_{\alpha\beta}$ in $r \geq r_p(t)$. From $h^-_{\alpha\beta}$, one obtains the regular part of a half-string solution with the string extending to infinity; from $h^+_{\alpha\beta}$, a half-string solution with the string extending to the horizon; and from $h^\pm_{\alpha\beta}$, a no-string solution valid throughout the entire vacuum region.

\subsection{Mode-sum expressions for Hertz potentials}
\label{sec:mode-sum-hertz}

If we restrict ourselves to Boyer-Lindquist coordinates and the Kinnersley tetrad, the inversion relations in \eqref{eqn:hertzEqns} reduce to,
\begin{subequations} \label{eqn:hertzEqns}
    \begin{align}
        \psi_0 &= \frac{1}{2}\hat{D}_0 \overline{\zeta^{4}\Phi^\mathrm{I}_4},
        &
        \psi_4 &= \frac{1}{8}\left[\hat{L}_4\overline{\zeta^{4}\Phi^\mathrm{I}_4} - 12 M\partial_t {\Phi^\mathrm{I}_4} \right],
        \\
        \psi_4 &= \frac{1}{2}\hat{D}_4 \overline{\Phi^\mathrm{O}_0},
        &
        \psi_0 &=\frac{1}{8}\left[ \hat{L}_0\overline{\Phi^\mathrm{O}_0} + 12 M\partial_t \Phi^\mathrm{O}_0\right],
        \\
        \psi_0 &=\frac{1}{4}\hat{L}_0\overline{\Phi^{+}_0},
        &
        \psi_4 &=\frac{1}{4}\hat{L}_4\overline{\zeta^{4}\Phi^{+}_4},
        \\
        \psi_0 &= 3M\partial_t {\Phi^{-}_0},
        &
        \psi_4 &= 3M\partial_t {\Phi^{-}_4},
    \end{align}
\end{subequations}
where, to condense notation, we have defined the operators,
\begin{subequations} \label{eqn:chandraOps}
\begin{align}
    \hat{L}_0 &= \mathcal{L}^4,
    &
    \hat{D}_0 &= \mathcal{D}^4,
    \\
    \hat{L}_4 &= \zeta^{-4} (\mathcal{L}')^4,
    &
    \hat{D}_4 &= \zeta^{-4} \Big(\frac{\Delta}{2}\Big)^2 (\mathcal{D}')^4 \Big(\frac{\Delta}{2}\Big)^2,
\end{align}
\end{subequations}
with
\begin{subequations}
\begin{align}
    \mathcal{L} &= \eth_+ - ia\sqrt{1-z^2} \partial_t,
    \\
    \mathcal{L}' &= \eth_- + ia\sqrt{1-z^2} \partial_t,
    \\
    \mathcal{D} &= \partial_r + \frac{r^2+a^2}{\Delta}\partial_t + \frac{a}{\Delta} \partial_\phi,
    \\
    \mathcal{D}' &= \partial_r - \frac{r^2+a^2}{\Delta}\partial_t - \frac{a}{\Delta} \partial_\phi.
\end{align}
\end{subequations}
We have also introduced the spin-raising and -lowering operators,
\begin{subequations} \label{eqn:spinOps}
    \begin{align}
        \eth_+ &= (1-z^2)^{-1/2}[(1-z^2)\partial_z - i\partial_\phi + s z],
        \\
        \eth_- &=  (1-z^2)^{-1/2}[(1-z^2)\partial_z + i\partial_\phi - s z],
    \end{align}
\end{subequations}
which raise or lower the spin-weight of harmonics, 
\begin{subequations}
    \begin{align}
        \eth_+ Y_{s\ell m} = \mu^{+s}_l Y_{(s+1)\ell m},
        \\
        \eth_- Y_{s\ell m} = \mu^{-s}_l Y_{(s-1)\ell m},
    \end{align}
\end{subequations}
with
\begin{align}
    \mu^n_{\ell} = -\mathrm{sgn}(n)\sqrt{(\ell - n)(\ell + n + 1)}.
\end{align}
Note that, in Schwarzschild spacetime, these raising and lowering operators are related to the Geroch-Held-Penrose (GHP) operators by $\eth_+ = \sqrt{2}r\eth$ and $\eth_- = \sqrt{2}r\eth'$.
% Therefore, ${}_s Y_{\ell m}$ can be related to the scalar harmonics $Y_{\ell m}$ via
% \begin{align} \label{eqn:sYlmRaisingAndLowering}
%     {}_s Y_{\ell m} = \begin{cases}
%     \sqrt{\frac{(l-s)!}{(l+s)!}} \eth^s Y_{\ell m} & 0 \leq s \leq l,
%     \\
%     (-1)^s \sqrt{\frac{(l+s)!}{(l-s)!}} (\eth^\dagger)^{-s} Y_{\ell m} & -l \leq s \leq 0.
%     \end{cases}
% \end{align}
% Explicitly, acting this operator on the $s=\pm 2$ harmonics, we find
% \begin{align}
%     \eth^{\dagger} {}_{+2} Y_{\ell m} &= - \sqrt{(l+2)(l-1)}\;{}_{+1}Y_{\ell m},
%     \\
%     \left[\eth^{\dagger}\right]^2 {}_{+2} Y_{\ell m} &= +\sqrt{(l+2)(l-1)\ell(\ell+1)}\;Y_{\ell m},
%     \\
%     \eth {}_{-2} Y_{\ell m} &= +\sqrt{(l+2)(l-1)}\;{}_{-1}Y_{\ell m},
%     \\
%     \left[\eth\right]^2 {}_{-2} Y_{\ell m} &= +\sqrt{(l+2)(l-1)\ell(\ell+1)}\;Y_{\ell m}.
% \end{align}
% Additionally, in anticipation of calculating the self-force, we consider three applications of the spin operators,
% \begin{align}
%     \left[\eth^{\dagger}\right]^3 {}_{+2} Y_{\ell m} &= -\ell(\ell+1)\sqrt{(l+2)(l-1)}\;{}_{-1}Y_{\ell m},
%     \\
%     \left[\eth\right]^3 {}_{-2} Y_{\ell m} &= +\ell(\ell+1)\sqrt{(l+2)(l-1)}\;{}_{+1}Y_{\ell m}.
% \end{align}

% \subsection{Mode sum expressions for the Hertz potential}

Because the rescaled Hertz potentials satisfy the Teukolsky equations and their adjoints, e.g.,
\begin{align}
    \mathcal{O}^\dagger_4 \zeta^{4}\Phi_0 &= \zeta^{4} \mathcal{O}_{0} \Phi_0,
    &
    \mathcal{O}^\dagger_0 \zeta^{4}\Phi_4 &= \zeta^{4} \mathcal{O}_{4}\Phi_4,
\end{align}
(see Appendix \ref{app:GHP}),
they can be expressed in mode-sum forms comparable to \eqref{eqn:Weylmodesum},
\begin{subequations} \label{eqn:Hertzmodesum}
    \begin{align}
    \Phi_{\pm 2}^\mathcal{G} &= \Phi_{\pm 2}^{\mathcal{G},\mathcal{I}} \Theta(r-r_p(t)) + \Phi_{\pm 2}^{\mathcal{G},\mathcal{H}} \Theta(r_p(t)-r),
    \\
    \Phi_{\pm 2}^{\mathcal{G},\mathcal{J}} &= \sum_{jm\omega} \Phi^{\mathcal{G},\mathcal{J}}_{\pm 2jm\omega}(r)S_{\pm 2jm\omega}(z) e^{i(m\phi - \omega t)},
\end{align}
\end{subequations}
where $\mathcal{G} = \{\mathrm{I}, \mathrm{O}, \mathrm{S}, \mathrm{A}\}$ as before,
\begin{subequations}
\begin{align}
    \Phi_{-2}^{\mathrm{I},\mathcal{J}} &= \zeta^4 \Phi_4^\mathrm{I},
    &
    \Phi_{+2}^{\mathrm{O},\mathcal{J}} &= \Phi_0^\mathrm{O},
    \\
    \Phi_{-2}^{\mathrm{S},\mathcal{J}} &= \zeta^4 \Phi_4^\mathrm{S},
    &
    \Phi_{+2}^{\mathrm{S},\mathcal{J}} &= \Phi_0^\mathrm{S},
    \\
    \Phi_{-2}^{\mathrm{A},\mathcal{J}} &= \zeta^4 \Phi_4^\mathrm{A},
    &
    \Phi_{-2}^{\mathrm{A},\mathcal{J}} &= \Phi_0^\mathrm{A},
\end{align}
\end{subequations}
and $\Phi^{\mathcal{G},\mathcal{J}}_{\pm 2jm\omega}(r) = \mathcal{Z}^{\mathcal{G},\mathcal{J}}_{\pm 2jm\omega} {R}^\mathcal{J}_{\pm 2 jm \omega}(r)$ are solutions to the radial Teukolsky equation with complex amplitudes $\mathcal{Z}^{\mathcal{G},\mathcal{J}}_{\pm 2jm\omega}$.

To determine the Hertz amplitudes $\mathcal{Z}^{\mathrm{G},\mathcal{J}}_{\pm 2jm\omega}$, we insert \eqref{eqn:Hertzmodesum} into \eqref{eqn:hertzEqns}. Using the symmetries of the mode functions (see Appendix \ref{app:symmetries}) and the Teukolsky-Starobinsky identities (see Appendix \ref{app:TS}), the Hertz amplitudes can be expressed in terms of the asymptotic amplitudes of $\psi_0$ and $\psi_4$,
\begin{subequations} \label{eqn:hertzAmpRG1}
    \begin{align}
        \mathcal{Z}^{\mathrm{I},\mathcal{J}}_{-2jm\omega} &= (-1)^{j+m+k}  \frac{2{Z}^\mathcal{J}_{+2jm\omega}}{{\alpha}^{\mathrm{TS},\mathcal{J}}_{+2jm\omega}} 
        \\
        &= (-1)^{j+m+k}  \frac{8{Z}^\mathcal{J}_{-2jm\omega}}{{C}^\mathrm{TS}_{jm\omega}},
        \\
        \mathcal{Z}^{\mathrm{O},\mathcal{J}}_{+2jm\omega} &= (-1)^{j+m+k}  \frac{32{Z}^\mathcal{J}_{-2jm\omega}}{{\alpha}^{\mathrm{TS},\mathcal{J}}_{-2jm\omega}} 
        \\
        &= (-1)^{j+m+k}  \frac{8{Z}^\mathcal{J}_{+2jm\omega}}{\overline{{C}^\mathrm{TS}_{jm\omega}}},
    \end{align}
\end{subequations}
where the Teukolsky-Starobinsky constant ${C}^\mathrm{TS}_{jm\omega}$ is defined in Eq.~\eqref{eqn:CTS}.
% where we have made use of the fact that $\overline{{\alpha}^\mathcal{J}_{\pm 2j-m-\omega}} = {\alpha}^\mathcal{J}_{\pm 2jm\omega}$ and ${\alpha}^\mathcal{J}_{-2jm\omega} {\alpha}^\mathcal{J}_{+2jm\omega} = |{C}_{jm\omega}|^2$.
% 
For the symmetric and antisymmetric reconstruction procedures, we arrive at similarly simple forms,
\begin{subequations} \label{eqn:hertzAmpRG2}
    \begin{align}
        \mathcal{Z}^{S,\mathcal{J}}_{\pm 2jm\omega} &= (-1)^{j+m+k}  \frac{4{Z}^\mathcal{J}_{\pm 2jm\omega}}{\mathrm{Re} \left[C^\mathrm{TS}_{jm\omega} \right]},
        \\ \label{eqn:minuSRGamp}
        \mathcal{Z}^{A,\mathcal{J}}_{\pm 2jm\omega} &= -\frac{{Z}^\mathcal{J}_{\pm 2jm\omega}}{3iM\omega}.
    \end{align}
\end{subequations}
Note that, as mentioned in Sec.~\ref{sec:intro}, \eqref{eqn:minuSRGamp} breaks down for static ($\omega=0$) modes.

Making use of \eqref{eqn:mixingCoeffs}, we can further re-expand onto a basis of spin-weighted spherical harmonics,
\begin{subequations} \label{eqn:HertzYslm}
\begin{align}
    \Phi_{\pm 2}^{\mathcal{G},\mathcal{J}} &= \sum_{\ell m\omega} \check{\Phi}^{\mathcal{G},\mathcal{J}}_{\pm 2\ell m\omega}(r)Y_{\pm 2\ell m} e^{i(m\phi - \omega t)},
    \\
    \check{\Phi}^{\mathcal{G},\mathcal{J}}_{\pm 2\ell m\omega}(r) &= \sum_j b^{\ell}_{sjm\omega}\Phi^{\mathcal{G},\mathcal{J}}_{\pm 2jm\omega}(r),
\end{align}
\end{subequations}
which will be useful for leveraging the properties of the spin-raising and -lowering operators [i.e., \eqref{eqn:spinOps}] in the proceeding calculations.

\begin{widetext}

\subsection{Mode-sum expressions for metric perturbations}
\label{sec:mode-sum-recon}

To assemble mode-sum expressions for the metric perturbations, we first calculate the intermediate quantities $h^{\mathcal{G}, \mathcal{J}}_{(-2)\alpha\beta} = (\mathcal{S}_0^\dagger \Phi^{\mathcal{G}, \mathcal{J}}_{-2})_{\alpha\beta}$ and $h^{\mathcal{G}, \mathcal{J}}_{(+2)\alpha\beta} = (\mathcal{S}_4^\dagger \zeta^4 \Phi^{\mathcal{G}, \mathcal{J}}_{+2})_{\alpha\beta}$ using \eqref{eqn:HertzYslm}, leading to
\begin{subequations}
\begin{align}
    h^{\mathcal{G}\mathrm{RG}}_{(\pm2)\alpha\beta} &= h^{\mathcal{G}, \mathcal{I}}_{(\pm 2)\alpha\beta}\Theta(r-r_p(t)) + h^{\mathcal{G}, \mathcal{H}}_{(\pm 2)\alpha\beta}\Theta(r_p(t)-r),
    \\
    h^{\mathcal{G}, \mathcal{J}}_{(\pm 2)\alpha\beta} &= \sum_{n_i}h_{(\pm2)\alpha\beta}^{(n_t,n_r,n_s,n_\phi)}(r,z) \partial_t^{n_t}\partial_r^{n_r} \eth^{n_s}_\mp \partial_\phi^{n_\phi}  \sum_{\ell m\omega} \check{\Phi}^{\mathcal{G},\mathcal{J}}_{\pm 2jm\omega}(r) Y_{\pm 2\ell m}(z) e^{i(m\phi - \omega t)} + \mathrm{c.c.},
\end{align}
\end{subequations}
where we have introduced the notation
\begin{align}
    \sum_{n_i} = \sum_{n_t = 0}^2 \sum_{n_r = 0}^2 \sum_{n_s=0}^2 \sum_{n_\phi=0}^2.
\end{align} 
Recall that $\mathcal{G}\mathrm{RG}$ refers to the resulting gauge of $h_{(\pm2)\alpha\beta}$, with $\mathcal{G}=$ \{I,O,S,A\}. The functions $h_{(\pm2)\alpha\beta}^{(n_t,n_r,n_s,n_\phi)}(r,z) $ are independent of the mode-numbers and vanish unless $n_t+n_r+n_s+n_\phi\leq 2$ and $n_t,n_r,n_s,n_\phi\leq 2$. Thus, one has up to 15 independent coefficients for each component of the metric perturbation. For completeness, the nonvanishing coefficients are given in Boyer-Lindquist coordinates in Appendix \ref{app:metricCoeffs}. We find this particular decomposition of the mode-sum numerically advantageous, because the coefficients only need to be calculated once, and then can be applied to each mode. 

Acting with the derivatives on the mode functions then leads to
\begin{subequations} \label{eqn:modeSumHABpm}
\begin{align}
    h^{\mathcal{G}, \mathcal{J}}_{(\pm 2)\alpha\beta}(t,r,z,\phi) &= \sum_{n_i}h_{(\pm2)\alpha\beta}^{(n_t,n_r,n_s,n_\phi)}(r,z) \sum_{\ell m\omega} \mathcal{M}^{(n_t,n_s,n_\phi)}_{\pm 2\ell m\omega} \partial_r^{n_r} \check{\Psi}^{\mathcal{G},\mathcal{J}}_{\pm 2\ell m\omega}(r) Y_{\pm (2-n_s)\ell m}(z) e^{i(m\phi - \omega t)} + \mathrm{c.c.},
    \\
    &= \sum_{\ell m\omega} \sum_{s_n=0}^2 h^{(s_n)\mathcal{G}, \mathcal{J}}_{(\pm 2\ell m\omega)\alpha\beta}(r,z) Y_{\pm s_n\ell m}(z) e^{i(m\phi - \omega t)} + \mathrm{c.c.},
\end{align}   
\end{subequations}
\end{widetext}
with $s_n = 2-n_s$ and scalar coefficients
\begin{align}
    \mathcal{M}^{(n_t,n_s,n_\phi)}_{\pm 2\ell m\omega} &= (-i\omega)^{n_t}(im)^{n_\phi} \prod_{s=0}^{n_s} \mu^{\mp(2-s)}_{\ell}.
\end{align}
Finally, to assemble full mode-sum expressions of the metric perturbation, we can insert \eqref{eqn:modeSumHABpm} into \eqref{eqn:reconstruct}. Combining these expressions with the completion pieces in \eqref{eqn:hcomp}, we arrive at full expressions for the metric perturbation $h_{\alpha\beta}$ in Eq.~\eqref{eqn:h1decomp} with
\begin{subequations}
\begin{align}
    h^{\mathcal{G}\mathrm{RG}-}_{\alpha\beta} &= h^{\mathcal{G}\mathrm{RG}, \mathcal{H}}_{(\pm 2)\alpha\beta},
    \\
    h^{\mathcal{G}\mathrm{RG}+}_{\alpha\beta} &= h^{\mathcal{G}\mathrm{RG}, \mathcal{I}}_{(\pm 2)\alpha\beta} + h^\mathrm{comp+}_{\alpha\beta},
\end{align}
\end{subequations}
and similarly for the gauge-fixed $h_{uu}$,
\begin{subequations} \label{eqn:huuModes}
\begin{align} 
    h^{\mathcal{G}\mathrm{RG}-}_{uu} &= h^{\mathcal{G}\mathrm{RG}, \mathcal{H}}_{(\pm 2)uu} + \pounds_{\xi^-}g_{uu},
    \\
    h^{\mathcal{G}\mathrm{RG}+}_{uu} &= h^{\mathcal{G}\mathrm{RG}, \mathcal{I}}_{(\pm 2)uu} + h^\mathrm{comp+}_{uu}.
\end{align}
\end{subequations}

\section{Constructing the redshift correction}
\label{sec:mode-sum-redshift-full-sec}

As discussed in Sec.~\ref{sec:redshift}, computing the redshift correction $\langle \tilde{z}_1\rangle_t$ involves extending $h_{uu}$ off the worldline, evaluating and regularizing it to obtain $h^\mathrm{R}_{uu}$, and taking its infinite time average. The following subsections detail our procedure for carrying out these steps to extract $\langle \tilde{z}_1\rangle_t$ for eccentric, precessing bound orbits in Kerr spacetime.

\subsection{Mode-sum regularization}
\label{sec:regularization}

We make use of mode-sum regularization to extract $h^\mathrm{R}_{uu} = h_{uu} - h^\mathrm{S}_{uu}$. (Note that, for brevity, we drop any notation referring to the gauge of the perturbations in this subsection.) This procedure leverages the fact that, while $h_{uu}$ and $ h^\mathrm{S}_{uu}$ diverge along the worldline, the harmonic modes of these fields are finite, allowing one to \emph{regularize} $h_{uu}$ on a mode-by-mode basis \cite{BaraOri00, HeffOtteWard14},
\begin{align} \label{eqn:modesum}
    h^\mathrm{R}_{uu} = \sum_{\ell=0}^\infty \left( h^{\ell}_{uu} - h^{\mathrm{S},\ell}_{uu} \right),
\end{align}
where
\begin{subequations}  \label{eqn:huul}
    \begin{align}
    h^{\ell}_{uu} &= \sum_{m=-\ell}^{+\ell} h^{\ell m}_{uu}(t_p,r_p) Y_{\ell m}(z_p)e^{im\phi_p},
    \\
    h^{\mathrm{S},\ell}_{uu} &= \sum_{m=-\ell}^{+\ell} h^{\mathrm{S},\ell m}_{uu}(t_p,r_p) Y_{\ell m}(z_p)e^{im\phi_p}.
\end{align}
\end{subequations}
While $h^{\mathrm{S}, \ell}_{uu}$ is not known exactly, it can be expressed as an expansion,
\begin{align} \label{eqn:huuS-expansion}
    \sum _{\ell=0}^\infty h^{\mathrm{S}, \ell}_{uu} &= H^{[D]} + \sum _{\ell=0}^\infty H^{[0]} + \frac{H^{[1]}}{2\ell+1} 
    \\ \notag
    & \qquad + \sum_{n=1}^\infty \frac{H^{[2n]}}{\prod_{k=1}^n (2\ell+1-2k)(2\ell+1+2k)},
\end{align}
where the \emph{regularization parameters} $H^{[k]}$ can be analytically derived \cite{HeffOtteWard14}. For $n \geq 1$, 
\begin{align}
    \sum_{\ell=0}^\infty \left[\prod_{k=1}^n (2\ell+1-2k)(2\ell+1+2k)\right]^{-1} = 0.
\end{align}
Only the leading-order parameters $H^{[-1]}$, $H^{[0]}$, and $H^{[1]}$ need to be included in \eqref{eqn:modesum} to achieve a formally convergent result. Higher-order parameters simply improve the rate of convergence of the mode-sum.

Importantly, (despite dropping gauge notation) the singular structure of $h_{uu}$ is not gauge-invariant. Thus, these parameters still depend on our choice of gauge and the way we choose to extend our definition of the four-velocity away from the worldline. Traditionally, regularization parameters are derived in Lorenz gauge and with an extension that is compatible with the choice $\hat{u}^\alpha(t,r,z,\phi) = u^\alpha(t)$ \cite{BaraOri03a, HeffOtteWard14, Heff21, HeffZeno22}. This leads to the convenient simplification $H^{[-1]} = 0 = H^{[1]}$. 

While we do not consider Lorenz gauge perturbations in this work, the leading-order Lorenz parameters can be applied to other gauges, provided those gauges are \emph{locally-Lorenz}, meaning they match the leading-order singular structure of Lorenz gauge in the neighborhood of the worldline \cite{PounMerlBara14}. As shown in Refs.~\cite{PounMerlBara14, VandShah15}, $H^{[0]}_\mathrm{Lor}$ can be directly applied to the mode-sum of our no-string radiation gauge solutions. Considering that the SRG perturbations can be expressed as the average of the ORG and IRG perturbations, $H^{[0]}_\mathrm{Lor}$ can be used to regularize SRG data, as well.

The parameter $H^{[0]}_\mathrm{Lor}$ has been calculated for nonprecessing eccentric orbits in Kerr \cite{HeffOtteWard14}, but it has not been explicitly generalized to precessing orbits. However, as pointed out by Heffernan,\footnote{Private communication} in Lorenz gauge $h_{uu}$ shares the same leading-order singular structure as a scalar field sourced by a point-particle in Kerr. Thus we can make use of the scalar field regularization parameter derived for generic orbits in Ref.~\cite{Heff21}. The full expression is provided in Appendix \ref{app:reg} for completeness.

Replacing $h^\mathrm{S}_{uu}$ with $H^{[0]}$ in \eqref{eqn:modesum} produces a formally convergent mode-sum regularization scheme, but in practice, the mode-sum does not converge at a sufficiently rapid rate. In our numerical calculations we sum over a finite number of $\ell$-modes, which introduces a truncation error,
\begin{align} \label{eqn:eps-trunc}
    \epsilon^{\ell_\mathrm{max}}_\mathrm{trunc} = \sum_{l=\ell_\mathrm{max}+1}^\infty \left(h^{l}_{uu} - h^{\mathrm{S},l}_{uu} \right).
\end{align}
Furthermore, from \eqref{eqn:huuS-expansion}, we see that terminating the number of regularization parameters produces the regularization error,
\begin{align} \label{eqn:eps-reg}
    \epsilon^{\ell_\mathrm{max}}_\mathrm{reg} = -\sum_{\ell=0}^{\ell_\mathrm{max}}\sum_{n=1}^\infty \frac{H^{[2n]}}{\prod_{k=1}^n (2l+1-2k)(2l+1+2k)}.
\end{align}
We expect $\epsilon^{\ell_\mathrm{max}}_\mathrm{trunc}$ to be a subdominant source of error, since the regularized field converges exponentially. On the other hand, $\epsilon^{\ell_\mathrm{max}}_\mathrm{reg} \sim \ell_\mathrm{max}^{-1}$ at large $\ell$, which is typically too large an error to extract meaningful results.

To circumvent this issue, we use the large-$\ell$ behavior of $\left( h^{\ell}_{uu} - H^{[0]} \right)$ to estimate these higher-order regularization parameters. We adopt the same fitting procedure as outlined in Ref.~\cite{Nasi22}, which adapted the algorithm first proposed in Ref.~\cite{VandShah15}. Because the completion and gauge contributions are known analytically, we only need to regularize contributions from the reconstructed metric perturbations. Therefore, we relate our reconstructed metric data to the redshift by defining the intermediate quantities $\tilde{z}^{\mathrm{rec},\ell}_1 = - \frac{1}{2} \tilde{z}_0 h^{\mathrm{rec},\ell}_{uu}$, $\tilde{z}^{\mathrm{S}}_1 = - \frac{1}{2} \tilde{z}_0 H^\mathrm{S}$, and $\tilde{z}^{\mathrm{S}[n]}_1 = - \frac{1}{2} \tilde{z}_0 H^{[n]}$, along with the partial sums
\begin{subequations} \label{eqn:z1-reg}
    \begin{align}
    \langle \tilde{z}_1^\mathrm{rec} \rangle_t^{\ell_\mathrm{max}} &= \sum_{\ell=0}^{\ell_\mathrm{max}} \left( \langle \tilde{z}^{\mathrm{rec},\ell}_1 \rangle_t - \langle \tilde{z}^{\mathrm{S}[n]}_1 \rangle_t \right),
    \\
    &= \langle \tilde{z}_1 \rangle_t + \epsilon^{\ell_\mathrm{max}}_\mathrm{reg} + \epsilon^{\ell_\mathrm{max}}_\mathrm{trunc},
\end{align}
\end{subequations}
but with $H^{[2n]}$, $h^{\mathrm{rec},\ell}_{uu}$, and $h^{\mathrm{S},\ell}_{uu}$ replaced by $\langle \tilde{z}^{\mathrm{S}[2n]}_1 \rangle_t $, $\langle \tilde{z}^{\mathrm{rec},\ell}_1 \rangle_t$, and $\langle \tilde{z}^{\mathrm{S}}_1 \rangle_t$ in Eqs.~\eqref{eqn:eps-trunc} and \eqref{eqn:eps-reg}. By varying $\ell_\mathrm{max}$ in the range $\ell_\mathrm{max} \in [\ell_1, \ell_2]$, we produce a set of partial sums $\langle \tilde{z}_1 \rangle_t^{\ell_\mathrm{max}}$. Assuming that $\epsilon^{\ell_\mathrm{max}}_\mathrm{trunc}$ is small enough to be ignored, we use this set of partial sums to perform a least-squares fit for the values of $\langle \tilde{z}^{\mathrm{S}[2n]}_1 \rangle_t$ (up to some $n=n_\mathrm{max}$ value) and $\langle \tilde{z}_1^\mathrm{rec} \rangle_t$. We then vary the values of $l_1$, $l_2$, and $n_\mathrm{max}$ to provide a set of fits. The mean of this set gives our final fit for $\langle \tilde{z}_1^\mathrm{rec} \rangle_t$, while two times the standard deviation of the set gives us an estimated extrapolation or fitting error. Using our analytic expressions for the gauge and completion pieces, we calculate the additional contributions $\langle \tilde{z}_1^\mathrm{comp} \rangle_t$ and $\langle \tilde{z}_1^\mathrm{gauge} \rangle_t$, leading to the full redshift value
\begin{align}
    \langle \tilde{z}_1\rangle_t = \langle \tilde{z}_1^\mathrm{rec} \rangle_t + \langle \tilde{z}_1^\mathrm{comp} \rangle_t + \langle \tilde{z}_1^\mathrm{gauge} \rangle_t.
\end{align}
Finally, we highlight that we can construct $h^{\mathrm{rec},\ell}_{uu}$, and thus $\langle \tilde{z}^{\mathrm{rec},\ell}_1 \rangle_t$, from our no-string metric perturbations by either taking the infinity-side limit $r\rightarrow r_p^+$, horizon-side limit $r\rightarrow r_p^-$, or the average of these two limits, when evaluating the metric perturbation along the worldline. While this will lead to slight differences in the values of $\langle \tilde{z}^{\mathrm{rec},\ell}_1 \rangle_t$ at the level of modes, it does not formally affect the regularized value of $\langle \tilde{z}_1\rangle_t$. In practice, however, different solutions have slightly different numerical errors. This leads to different estimates for $\langle \tilde{z}^{\mathrm{rec},\ell}_1 \rangle_t$, though these estimates should still be self-consistent within the numerical errors of our fitting procedure. Such self-consistency checks help validate our results and are further considered in Sec.~\ref{sec:redshift-results}.

\subsection{Reprojection and extension from the worldline}
\label{sec:reprojection}

To make use of the mode-sum regularization in \eqref{eqn:modesum}, we must also decompose the reconstructed components of $h_{uu}$ in Eq.~\eqref{eqn:huuModes} according to \eqref{eqn:huul}. First we expand the spin-weighted spherical harmonics in terms of the scalar harmonics $Y_{\ell m} = Y_{0lm}$ via
\begin{equation} \label{eqn:YslmExpansion}
		Y_{s\ell m}(z)
		= (1-z^2)^{-|s|/2} \sum_{j=|m|}^\infty \mathcal{A}^j_{s\ell m} 
		Y_{jm}(z),
\end{equation}
where the coefficients $\mathcal{A}^j_{s\ell m}$ are given by
\begin{subequations}
    \begin{align} \notag
	\mathcal{A}^j_{s\ell m} &= 
	(-1)^{m+s(1+\mathrm{sgn}(s))/2} \mathcal{C}_{|s|\ell j} \qquad \qquad
 \\
 &\qquad \qquad \times
		\left(
			\begin{array}{ccc}
				|s| & \ell & j
				\\
				0 & m & -m
			\end{array}
		\right)
		\left(
			\begin{array}{ccc}
				|s| & \ell & j
				\\
				s & -s & 0
			\end{array}
		\right),
  \\
  \mathcal{C}_{s\ell j}  &= \sqrt{\frac{4^{s} (s!)^2 (2\ell+1)(2j+1)}{(2s)!}}.
\end{align}
\end{subequations}
Due to the selection rules of the Wigner $3j$-symbols, ${}_s Y_{\ell m}$ only couples to $Y_{\ell-|s|,m}, Y_{\ell-|s|+1,m}, \dots, Y_{\ell+|s|,m}$, resulting in
\begin{align} \label{eqn:hYlm}
    h^{\mathcal{G}, \mathcal{J}}_{(\pm 2)\alpha\beta} &=
    \sum_{\ell m\omega} {\mathcal{H}^{\mathcal{G}, \mathcal{J}}_{(\pm2\ell m\omega)\alpha\beta}(r,z)} Y_{\ell m}(z) e^{i(m\phi - \omega t)}  +\mathrm{c.c.},
\end{align}   
with
\begin{align}
    {\mathcal{H}^{\mathcal{G}, \mathcal{J}}_{(\pm2\ell m\omega)\alpha\beta}}  &= \sum_{s_n=0}^2 \sum_{\ell'=\ell'_\mathrm{min}}^{\ell'_\mathrm{max}} \mathcal{A}^{\ell}_{\pm s_n\ell'm} \frac{h^{(s_n)\mathcal{G}, \mathcal{J}}_{(\pm 2\ell'm\omega)\alpha\beta}(r,z)}{(1-z^2)^{s_n/2}}.
\end{align}   
The mode-sum in Eq.~\eqref{eqn:hYlm} is still not decomposed onto a scalar spherical harmonic basis, because ${\mathcal{H}^{\mathcal{G}, \mathcal{J}}_{(\pm2\ell m\omega)\alpha\beta}}$ depends on $z$. One can remove this dependence by further decomposing ${\mathcal{H}^{\mathcal{G}, \mathcal{J}}_{(\pm2\ell m\omega)\alpha\beta}}$, then leverage the triple product rule for scalar harmonics to reduce the full-mode-sum expression for $h^{\mathcal{G}, \mathcal{J}}_{(\pm 2)\alpha\beta}$ onto a basis of $Y_{\ell m}(z)$. 

However, as observed by previous authors (e.g., \cite{Warb15, Vand16, Vand18}), this approach is not practical. Due to factors of $z$ in the denominators of ${\mathcal{H}^{\mathcal{G}, \mathcal{J}}_{(\pm2\ell m\omega)\alpha\beta}}$, the spherical harmonic decompositions of these coefficients decay very slowly in $\ell$. To circumvent this issue, we follow the approach carried out by van de Meent and Shah \cite{VandShah15, Vand16, Vand18}, and expand ${\mathcal{H}^{\mathcal{G}, \mathcal{J}}_{(\pm2\ell m\omega)\alpha\beta}}$ as a Taylor series in $z$ about $z_p$. Truncating this series at some finite order preserves the value of $h_{uu}$ at $x_p$, but changes how we extend the field $h_{uu}$ away from the worldline. As mentioned in the previous section, our mode-sum regularization scheme is sensitive to our choice of extension. Consequently, we must keep enough terms in our series expansion so that our modified choice of extension is still compatible with the Lorenz gauge regularization parameter $H^{[0]}$. For $h_{uu}$, $O(z-z_p)$ corrections to our extension only affect the higher-order regularization parameters. Since we do not include these higher-order terms in our mode-sum regularization, our reprojected field simply reduces to
\begin{align} \label{eqn:hYlmZp}
    \hat{h}^{\mathcal{G}, \mathcal{J}}_{(\pm 2)uu} &=
    \sum_{\ell m\omega} {\mathcal{H}^{\mathcal{G}, \mathcal{J}}_{\pm2\ell m\omega}(t,r)} Y_{\ell m}(z) e^{i(m\phi - \omega t)}  + \mathrm{c.c.},
\end{align}   
where ${\mathcal{H}^{\mathcal{G}, \mathcal{J}}_{\pm2\ell m\omega}(t,r)} = {\mathcal{H}^{\mathcal{G}, \mathcal{J}}_{(\pm2\ell m\omega)\alpha\beta}[r,z_p(t)]}u^\alpha(t) u^\beta(t)$. Note that, for a given $(\ell,m)$-mode, the complex conjugate terms will contribute to the $(\ell,-m)$-mode.

\subsection{Simplified mode-sum along the worldline}
\label{sec:huuWorldline}

When evaluated along the worldline, the summand in \eqref{eqn:hYlmZp} reduces to a function of the Mino phase variables $q_r$ and $q_z$, such that
\begin{subequations}
    \begin{align}
    {\mathcal{H}}^{(p)\mathcal{G}, \mathcal{J}}_{\pm2\ell m\omega}(q_r,q_z) &= \left. {\mathcal{H}^{\mathcal{G}, \mathcal{J}}_{\pm2\ell m\omega}(t,r)} Y_{\ell m}(z) e^{i(m\phi - \omega t)} \right\vert_{x=x_p},
    \\
    &= {\mathcal{H}^{\mathcal{G}, \mathcal{J}}_{(\pm2\ell m\omega)\alpha\beta}[r_p(q_r),z_p(q_z)]} 
    \\ \notag
    & \qquad \qquad \times u^\alpha(q_r,q_z)u^\beta(q_r,q_z)
    \\ \notag
    & \qquad \qquad \qquad \times Y_{\ell m}[z_p(q_z)] e^{i\phi_{m\omega}(q_r,q_z)},
\end{align}
\end{subequations}
where
\begin{subequations}
    \begin{align}
    e^{i(m \phi_p-\omega t_p)} &= e^{i[m\Delta\phi^{(z)}(q_z) - \omega \Delta t^{(z)}(q_z)]} 
    \\ \notag
    &\qquad \times e^{i[m\Delta\phi^{(r)}(q_r)-\omega \Delta t^{(r)}(q_r)]}e^{-ik q_z}e^{-inq_r},
    \\
    &= e^{i\phi_{m\omega}(q_r,q_z)},
\end{align}
\end{subequations}
Furthermore, as demonstrated in Appendix \ref{app:symmetries}, under complex conjugation we have
\begin{align} \label{eqn:paritySym}
    \bar{\mathcal{H}}^{(p)\mathcal{G}, \mathcal{J}}_{\pm2\ell m\omega}(q_r,q_z) = {\mathcal{H}}^{(p)\mathcal{G}, \mathcal{J}}_{\pm2\ell-m-\omega}(q_r,\pi+q_z).
\end{align}
Therefore, as mentioned in the previous section, the complex conjugate terms are equivalent to $(-m,-\omega)$ modes up to the parity transformation $q_z \rightarrow q_z+\pi$. Combining these symmetries, we re-express \eqref{eqn:hYlmZp} as
\begin{subequations} \label{eqn:mode-sum-huu-worldline}
    \begin{align}
    \hat{h}^{\mathcal{G}, \mathcal{J}}_{(\pm 2)uu} &=
    2\mathrm{Re}\sum_{\ell,m=0,\omega} {\mathcal{H}}^{(p)\mathcal{G}, \mathcal{J}}_{\pm2\ell m\omega}(q_r,q_z)
    \\ \notag
    &\qquad +
    2\mathrm{Re}\sum_{\ell,m>0,\omega} \Big[{\mathcal{H}}^{(p)\mathcal{G}, \mathcal{J}}_{\pm2\ell m\omega}(q_r,q_z) 
    \\ \notag
    & \qquad \qquad \qquad \qquad \qquad + {\mathcal{H}}^{(p)\mathcal{G}, \mathcal{J}}_{\pm2\ell m\omega}(q_r,\pi+q_z)\Big],
    \\
    &=
    \sum_{\ell=0}^\infty \hat{h}^{\mathcal{G}, \mathcal{J}, l}_{(\pm 2)uu}(q_r, q_z).
\end{align} 
\end{subequations}
With this reduced mode-sum, we only need to calculate and sum over all frequencies with $m\geq 0$.

\subsection{Time averages}

We now turn our attention to evaluating the infinite-time integrals in Eq.~\eqref{eqn:z}.
In the case of singly periodic orbits, \eqref{eqn:z} reduces to common expressions in the literature. More explicitly, if the small body's motion satisfies the periodic conditions
\begin{align}
    x_p^\alpha(\tau + \mathcal{T}) = \{t_p(\tau) + T, r_p(\tau), z_p(\tau), \phi_p(\tau) + \Phi\},
\end{align}
then we have
\begin{subequations} \label{eqn:z0z1}
\begin{align}
    \langle \tilde{z}_{(0)} \rangle_t &= \frac{\mathcal{T}}{T},
    &
    \langle \tilde{z}_{(1)} \rangle_t &= -\frac{1}{{2T}}\int_{0}^{T} \tilde{z}_{(0)}h^R_{uu} dt ,
    \\
    & & &= -\frac{1}{2}\langle \tilde{z}_{(0)} h_{uu}^\mathrm{R}\rangle_t.
\end{align}
\end{subequations}
For eccentric, nonprecessing (equatorial) orbits, one simply has $T \rightarrow T_r = 2\pi/\Omega_r$, while for noneccentric, precessing (spherical) orbits $T \rightarrow T_\theta = 2\pi/\Omega_\theta$. The numerical integrals can then be evaluated over a finite length of time.

For multiperiodic orbits, we transform to a phase-space parametrization with finite limits of integration. First, $dt/d\tau$ is re-expressed as a function of the orbital phases,
\begin{align}
    \frac{d\tau}{dt}(q_r,q_z) = \frac{d\tau}{dt}(q_r+2\pi,q_z) = \frac{d\tau}{dt}(q_r,q_z+2\pi).
\end{align}
The infinite time averages then translate into phase-space averages following Ref.~\cite{DrasHugh04},
\begin{align}
    \left\langle f(t)\frac{d\tau}{dt} \right\rangle_t = \frac{1}{\Upsilon_t} \int_0^{2\pi} \int_0^{2\pi} \Sigma_p(q_r,q_z) f(q_r,q_z) \frac{dq_r}{2\pi} \frac{dq_z}{2\pi},
\end{align}
where $\Sigma_p(q_r,q_z) = r_p^2(q_r) + a^2 z^2_p(q_z)$. More explicitly,
\begin{subequations}
    \begin{align}
    \langle \tilde{z}_{(0)} \rangle_t &= \frac{1}{\Upsilon_t} \int_0^{2\pi} \int_0^{2\pi} \Sigma_p \frac{dq_r}{2\pi} \frac{dq_z}{2\pi},
    \\
    \langle \tilde{z}_{(0)}h_{uu}^\mathrm{R} \rangle_t &= \frac{1}{\Upsilon_t} \int_0^{2\pi} \int_0^{2\pi} \Sigma_p h_{uu}^\mathrm{R} \frac{dq_r}{2\pi} \frac{dq_z}{2\pi},
\end{align}
\end{subequations}
with the integrals well suited to spectral integration methods \cite{HoppETC15, NasiOsbuEvan19}.

\section{Gauge comparisons}
\label{sec:gauge}

We compare the metric perturbations across IRG, ORG, SRG, and ARG.\footnote{Recall that we reconstruct the ARG metric perturbation from the radiative modes of the Hertz potentials. Therefore, the ARG perturbations are missing static and linear and time modes that we do not consider in this work.} In particular, we focus on the reconstructed metric perturbation, $h^\mathrm{G,\pm}_{\alpha\beta}$ in \eqref{eqn:h1pm}, since the other terms are either known analytically (e.g., $h^\mathrm{comp\pm}_{\alpha\beta}$ and the gauge-fixed $\pounds_{\xi^\pm}g_{\alpha\beta}$) or are not considered in this work (e.g., the remaining singular terms in the corrector tensor $x_{\alpha\beta})$.

The exact motion of the bound source (provided it is not plunging) will not affect the asymptotics of the metric perturbation at infinity and the horizon, nor will it alter the strength of the retarded field's divergence along the worldline. Therefore, to simplify comparisons between the different gauges, in this section we only focus on metric perturbations sourced by point-particles on noneccentric, nonprecessing (i.e., circular, equatorial) Kerr geodesics. In Sec.~\ref{sec:asymptotics} we compare the asymptotic behavior of the field in the four different gauges, while in Sec.~\ref{sec:singular} we compare the singular behavior of the different gauges near the worldline. Finally, in Sec.~\ref{sec:mode-sum-converge} we analyze the convergence of the mode-sum across the radial domain. These analyses will inform future second-order calculations, which require a first-order metric perturbation that is known globally.

\subsection{Asymptotics}
\label{sec:asymptotics}

To study the behavior at the horizon, we first transform to advanced time $v = t+r_*$ and shifted azimuthal angle
\begin{align}
    \tilde{\phi} = \phi - \frac{a}{2M\kappa} \ln\left(\frac{r-r_-}{r-r_+}\right).
\end{align}
Then, for constant slices of $v$, $z$, and $\tilde{\phi}$, the nonvanishing metric components have the asymptotic behaviors as $r\rightarrow r_+$,
\begin{subequations} \label{eqn:asympHor}
\begin{align}
    h^\mathrm{IRG}_{nn} &\sim \mathcal{B}^\mathrm{IRG}_{nn} \Delta^2,
    &
    h^\mathrm{ORG}_{ll} &\sim \mathcal{B}^\mathrm{ORG}_{ll}\Delta^{-2},
    \\
    h^\mathrm{IRG}_{n\bar{m}} &\sim \mathcal{B}^\mathrm{IRG}_{nn} \Delta,
    &
    h^\mathrm{ORG}_{l{m}} &\sim \mathcal{B}^\mathrm{IRG}_{n\bar{m}} \Delta^{-1},
    \\
    h^\mathrm{IRG}_{\bar{m}\bar{m}} &\sim \mathcal{B}^\mathrm{IRG}_{\bar{m}\bar{m}},
    &
    h^\mathrm{ORG}_{{m}{m}}&\sim \mathcal{B}^\mathrm{ORG}_{{m}{m}},
    \\
    h^\mathrm{SRG}_{nn}&\sim \mathcal{B}^\mathrm{SRG}_{nn} \Delta^2,
    &
    h^\mathrm{SRG}_{ll}&\sim \mathcal{B}^\mathrm{SRG}_{ll}\Delta^{-2},
    \\
    h^\mathrm{SRG}_{n\bar{m}}&\sim \mathcal{B}^\mathrm{SRG}_{n\bar{m}}\Delta,
    &
    h^\mathrm{SRG}_{l{m}}&\sim \mathcal{B}^\mathrm{SRG}_{l{m}}\Delta^{-1},
    \\
    h^\mathrm{SRG}_{\bar{m}\bar{m}}&\sim \mathcal{B}^\mathrm{SRG}_{\bar{m}\bar{m}},
    &
    h^\mathrm{SRG}_{{m}{m}}&\sim \mathcal{B}^\mathrm{SRG}_{{m}{m}},
\end{align}
\end{subequations}
with constant amplitudes $\mathcal{B}^\mathrm{\mathcal{G}RG}_{ab}$. The ARG perturbations share the same behavior as the SRG perturbations.

To examine the behavior at infinity, we instead transform to retarded time $u = t-r_*$. Then, for constant slices of $u$, $z$, and ${\phi}$ (or $\tilde{\phi}$), in the limit $r\rightarrow \infty$ the nonvanishing metric components go as
\begin{subequations} \label{eqn:asympInf}
\begin{align}
    h^\mathrm{IRG}_{nn}&\sim \mathcal{C}^\mathrm{IRG}_{nn} r,
    &
    h^\mathrm{ORG}_{ll}&\sim \mathcal{C}^\mathrm{ORG}_{ll} r^{-3},
    \\
    h^\mathrm{IRG}_{n\bar{m}}&\sim \mathcal{C}^\mathrm{IRG}_{n\bar{m}}r,
    &
    h^\mathrm{ORG}_{l{m}}&\sim \mathcal{C}^\mathrm{ORG}_{l{m}}r^{-2},
    \\
    h^\mathrm{IRG}_{\bar{m}\bar{m}}&\sim \mathcal{C}^\mathrm{IRG}_{\bar{m}\bar{m}},
    &
    h^\mathrm{ORG}_{{m}{m}}&\sim \mathcal{C}^\mathrm{ORG}_{{m}{m}}r^{-1},
    \\
    h^\mathrm{SRG}_{nn}&\sim \mathcal{C}^\mathrm{SRG}_{nn}r,
    &
    h^\mathrm{SRG}_{ll}&\sim \mathcal{C}^\mathrm{SRG}_{ll}r^{-3},
    \\
    h^\mathrm{SRG}_{n\bar{m}}&\sim \mathcal{C}^\mathrm{SRG}_{n\bar{m}}r,
    &
    h^\mathrm{SRG}_{l{m}}&\sim \mathcal{C}^\mathrm{SRG}_{l{m}}r^{-2},
    \\
    h^\mathrm{SRG}_{\bar{m}\bar{m}}&\sim \mathcal{C}^\mathrm{SRG}_{\bar{m}\bar{m}},
    &
    h^\mathrm{SRG}_{{m}{m}}&\sim \mathcal{C}^\mathrm{SRG}_{{m}{m}},
\end{align}
\end{subequations}
with constant amplitudes $\mathcal{C}^\mathrm{\mathcal{G}RG}_{ab}$.
Once again, the ARG perturbations share the same behavior as the SRG perturbations.

\begin{figure}[th!]
    \centering
    \includegraphics[width=0.99\linewidth]{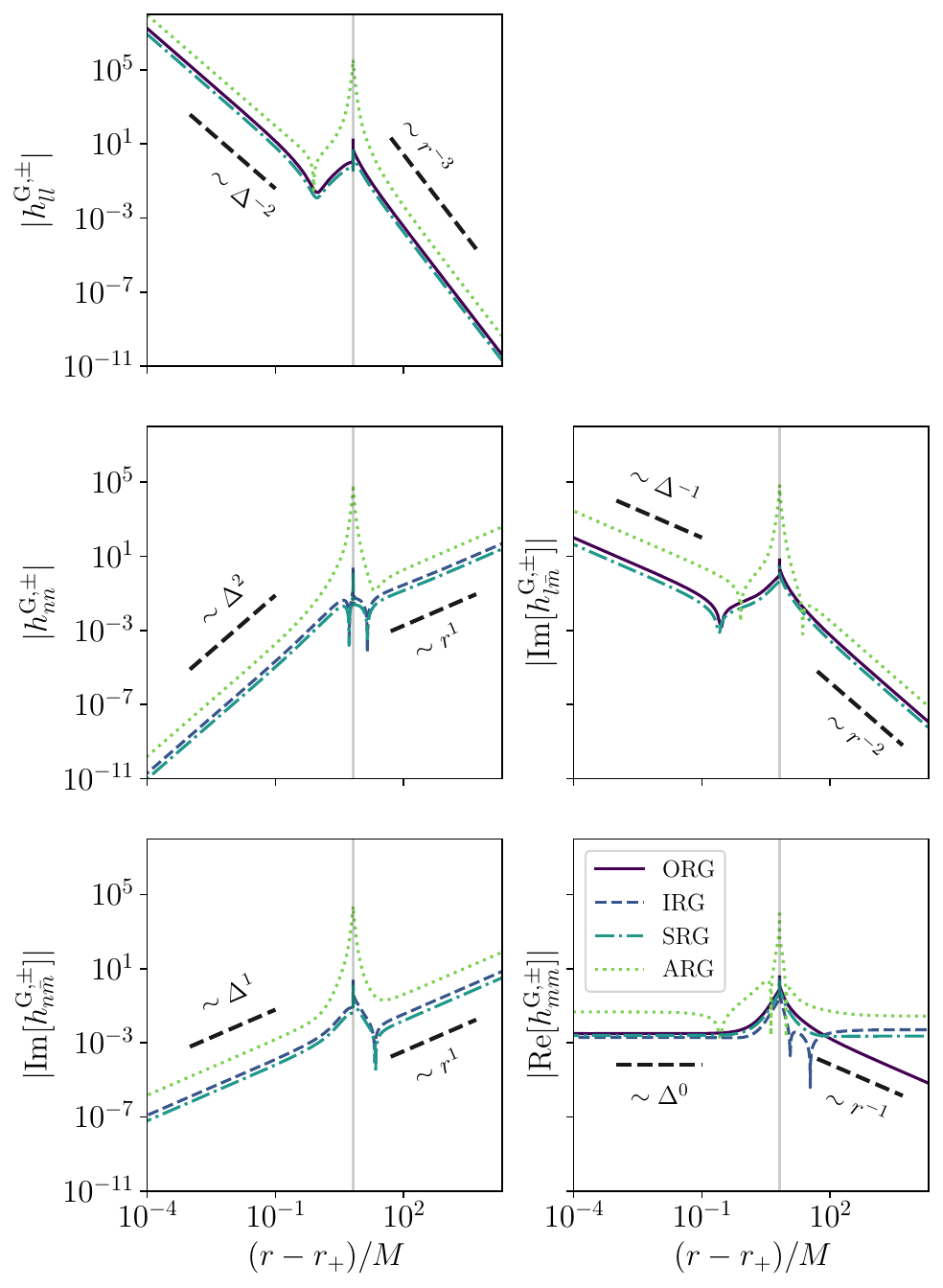}
    \caption{The nonvanishing, tetrad-projected components of the metric perturbation produced by a point-particle on a geodesic with parameters $(\hat{a},p,e,x)=(0.9, 8, 0,1)$. We compare the asymptotic behavior toward the horizon $r_+$ and infinity across the four different gauges. We give the real and imaginary pieces for the complex tetrad projections of the metric components. In the case of this circular orbit, we have $\mathrm{Re}[h_{lm}]=\mathrm{Re}[h_{nm}]=\mathrm{Im}[h_{mm}]=0$ for all four gauges.}
    \label{fig:asymp}
\end{figure}

\begin{figure*}[th!]
    \centering
    \includegraphics[width=0.95\linewidth]{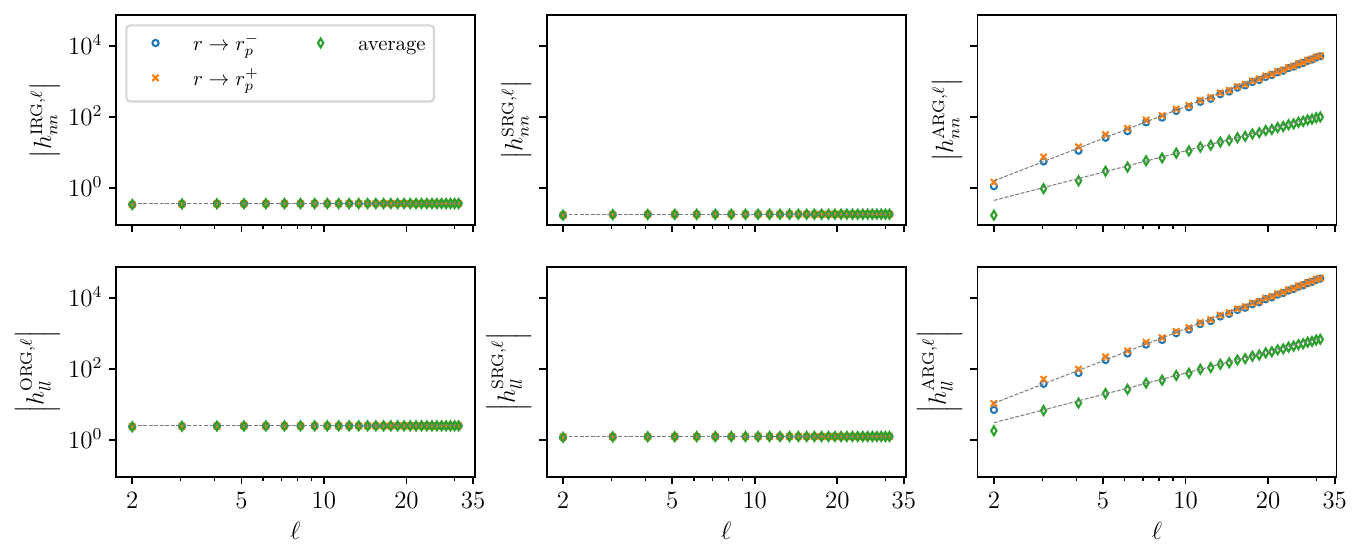}
    \caption{The $\ell$-mode contributions $h^{\mathrm{G}, \ell}_{\alpha\beta}$ to the reconstructed metric perturbation sourced by a small body on a circular geodesic with $(\hat{a},p,e,x)=(0.9,8,0,1)$. Mode contributions are evaluated along the worldline either by taking the limit from the horizon $r\rightarrow r_p^-$ (blue circles), from infinity $r\rightarrow r_p^+$ (orange $\times$), or from an average of the two limits (green diamonds). The leftmost column plots the metric components of the $\ell$-modes in IRG and ORG, the middle column in SRG, and the right in ARG.  The dashed lines provide references for the scaling of the modes with $\ell$. In the left and middle columns the lines scale as $\ell^{0}$, while those in the right column $\sim \ell^{2}$ and $\sim \ell^{3}$. We only include a subset of metric components, but they are representative of the behavior observed in every component of the metric perturbation in each gauge.}
    \label{fig:singular-convergence}
\end{figure*}

In Figure \ref{fig:asymp}, we highlight these asymptotic properties by plotting the metric perturbation produced by a point-particle on a noneccentric, nonprecessing (circular, equatorial) geodesic with parameters $(\hat{a},p,e,x)=(0.9,10,0,1)$. (Recall that $\hat{a}=a/M$.) The metric perturbations are constructed via Eq.~\eqref{eqn:hYlm} up to $\ell_\mathrm{max}=30$, but we do not include the static $m=0$ modes when reconstructing the ARG perturbation. To capture the behavior both at the horizon and infinity, we evaluate the metric perturbation along sharp time slices $T = 0$, with
\begin{align}
    T = \begin{cases}
        v, & r < r_0,
        \\
        t, & r = r_0,
        \\
        u, & r > r_0,
    \end{cases}
\end{align}
$r_0 = p M$ and at $z = 0$, $\tilde{\phi} = a/(2M\kappa) \ln[(r_0 - r_-)/(r_0-r_+)]$. As expected, the metric perturbations match the asymptotics in Eqs.~\eqref{eqn:asympHor} and \eqref{eqn:asympInf} with ORG perturbations regular at infinity and IRG perturbations regular at the horizon. Interestingly, the SRG and ARG perturbations are singular at both boundaries for the $mm$- and $\bar{m}\bar{m}$-components, inheriting the ``worst" behavior of their ORG and IRG counterparts. We also note that, for circular orbits, $h_{lm}$ and $h_{nm}$ are completely imaginary, while $h_{mm}$ is real across all of our gauges.

\subsection{Singular behavior}
\label{sec:singular}

As demonstrated by Pound \emph{et al.}~\cite{PounMerlBara14}, the no-string radiation gauge perturbations possess the singular behavior $h_{\alpha\beta} = O(s^{-1})$, where $s$ is a measure of the geodesic distance from the worldline. Consequently, when we decompose perturbation components in terms of $\ell$-modes $h^{\pm,\ell}_{\alpha\beta}$ [in a similar fashion to Eq.~\eqref{eqn:huul}], then we expect $h^{\pm,\ell}_{\alpha\beta} \sim \ell^{0}$ when evaluated at the location of the point-particle source. 

In Fig.~\ref{fig:singular-convergence} we plot the $\ell$-modes of the reconstructed metric perturbations sourced by a point-particle on an orbit defined by $(\hat{a},p,e,x)=(0.9, 8,0,1)$. We only include a subset of metric components, but they are representative of the behavior observed in every component of the metric perturbation in each gauge. The perturbations are evaluated along the worldline by either taking the radial limit by approaching from the horizon (blue circles), from infinity (orange $\times$), or from the average of these limits (green diamonds). We observe that the nonvanishing components of the IRG and ORG metric perturbations (in the left-most panels) both possess the expected behavior of $h_{\alpha\beta} = O(s^{-1})$ and that the SRG components (in the middle panels) match this singular structure. This is to be expected. From \eqref{eqn:reconstruct}, the SRG perturbation effectively reduces to the average of the ORG and IRG perturbations; consequently, it has the same singular behavior as the no-string radiation gauges.

On the other hand, we observe that the ARG components (right-most panels) are far more singular, with their $\ell$-modes diverging as $\ell^3$ when approaching the worldline from the horizon or infinity (blue circles and orange $\times$, respectively), suggesting that the no-string ARG perturbations possess the singular behavior $h^{\mathrm{ARG}\pm}_{\alpha\beta} = O(s^{-4})$. This highly singular structure reflects our reconstruction procedures outlined in Eqs.~\eqref{eqn:reconstruct} and \eqref{eqn:hertzInversion}. The reconstruction of the ARG perturbation primarily deviates from the other procedures when computing the Hertz potentials. While the Hertz potentials for IRG, ORG, and SRG are related to the Weyl scalars through fourth-order operators, the ARG potentials are related through a single time derivative. Because derivatives typically increase the singular behavior of a field by one order, we would then expect the ARG Hertz potentials, and likewise the metric perturbation, to be three orders more singular than the IRG, ORG, and SRG solutions, if they are to produce the same Weyl scalars. This is exactly what we observe in Fig.~\ref{fig:singular-convergence}.

\begin{figure*}[th!]
    \centering
    \includegraphics[width=0.8\linewidth]{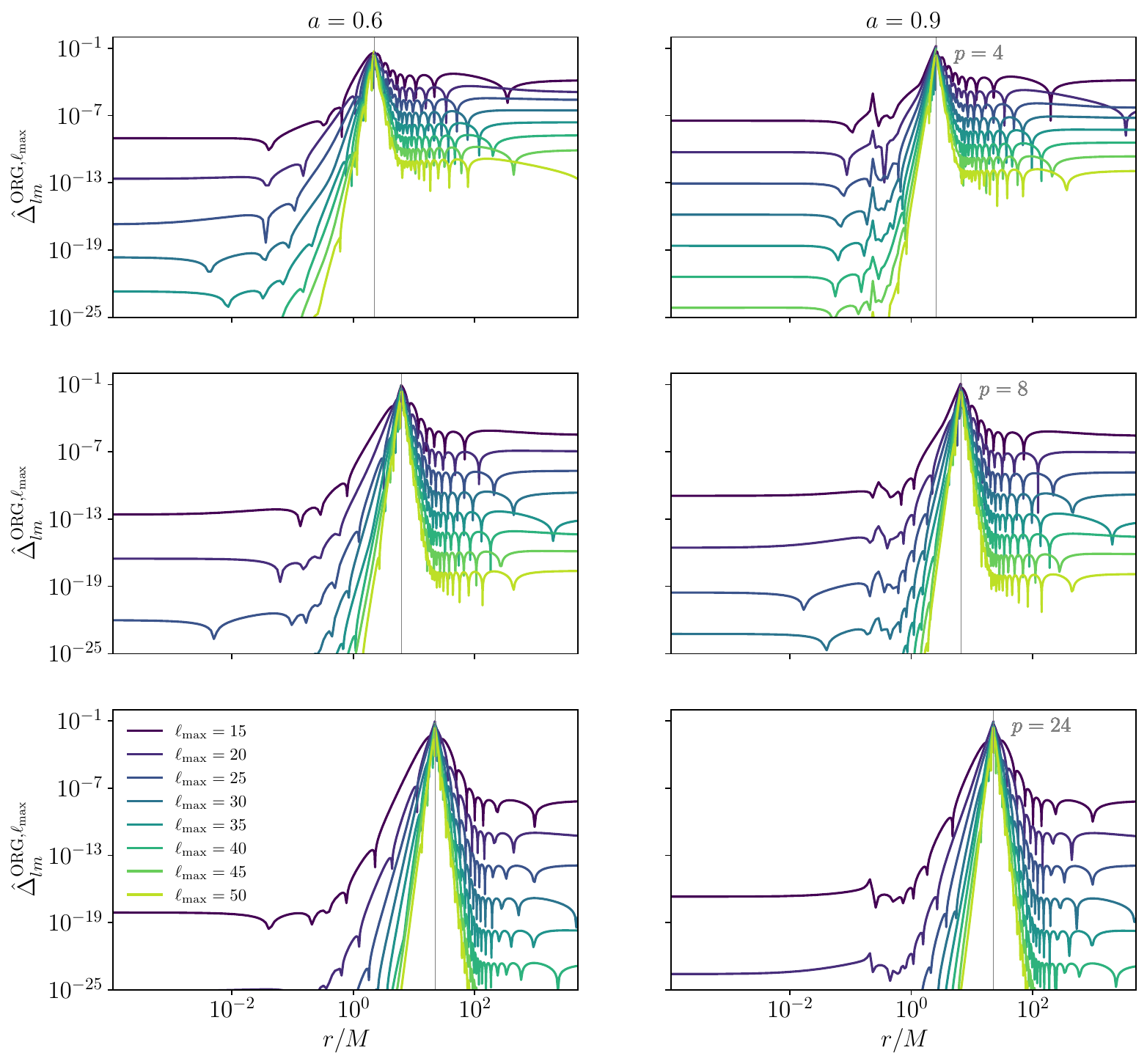}
    \caption{The relative contribution $\hat{\Delta}^{\mathrm{ORG},\ell_\mathrm{max}}_{lm}$ as a function of $(r-r_+)/M$ for fixed coordinates $T=z=\tilde{\phi}=0$. Each panel plots metric perturbation data for a different circular, equatorial geodesic source. Results in the top row are for a source with $p = 4$, the second for $p = 8$, and bottom for $p = 24$. Panels on the left plot perturbations of Kerr with $\hat{a}=0.6$, while the right panels are for $\hat{a}=0.9$.}
    \label{fig:lmode-convergence-log}
\end{figure*}

We also observe that there is a directional dependence in the divergence of the ARG solutions. Therefore, one can slightly mitigate their singular behavior by leveraging this additional structure and forming a no-string solution by taking the average of the two regular half-solutions (green diamonds). In this construction, the $\ell$-modes only diverge as $\ell^2$, suggesting $h^{\mathrm{ARG}}_{\alpha\beta} = O(s^{-3})$. Nonetheless, the ARG perturbation remains significantly more singular than the other gauges.

\subsection{Mode-sum convergence in vacuum}
\label{sec:mode-sum-converge}

While the $\ell$-mode-sum diverges on the worldline, it converges everywhere else, since the domain is vacuum away from the particle. To study this convergence, we define the quantity
\begin{align}
    \hat{\Delta}^{\mathrm{G},\ell_\mathrm{max}}_{\alpha\beta} = \left\vert h^{\mathrm{G}, \ell_\mathrm{max}}_{\alpha\beta} \times \left[\sum_{\ell=0}^{\ell_\mathrm{max}} h^{\mathrm{G}, \ell}_{\alpha\beta} \right]^{-1} \right\vert,
\end{align}
which measures the relative contribution of the final $\ell$-mode to the truncated sum. Since the full mode-sum converges exponentially off the worldline, this quantity estimates the relative truncation error. We find that convergence rates are broadly consistent across different gauges and components of the metric perturbation. Instead, the main factor influencing convergence is the nature of the source.

{\renewcommand{\arraystretch}{1.25}
\begin{table*}[ht!]
    \centering
    \setlength\tabcolsep{0pt}
    \caption{Redshift values $\langle \tilde{z}_1 \rangle_t$ for eccentric, nonprecessing orbits investigated in Ref.~\cite{VandShah15}. The columns $\langle \tilde{z}_1^\mathrm{ORG} \rangle_t$, $\langle \tilde{z}_1^\mathrm{IRG} \rangle_t$, and $\langle \tilde{z}_1^\mathrm{SRG} \rangle_t$ report the redshift calculated from metric perturbations in ORG, IRG, and SRG, respectively. The final column reports the maximum absolute difference between these three values and the redshift values given in Ref.~\cite{VandShah15}.}
    \label{tab:redshift-comp}
    \begin{tabular*}{\linewidth}{@{\extracolsep{\fill}} c c c c c c c}
        \hline
        \hline
        $\phantom{-}\hat{a}$ & $p\phantom{0}$ & $e\;$ & $\langle \tilde{z}_1^\mathrm{ORG} \rangle_t$ & $\langle \tilde{z}_1^\mathrm{IRG} \rangle_t$ & $\langle \tilde{z}_1^\mathrm{SRG} \rangle_t$ & abs. diff. \cite{VandShah15}  \\
        \hline
        $\phantom{-}0.9$ & $p_\mathrm{ISO} + 0.1\phantom{0}$ & $0.3\;$ & $0.11659(2)\phantom{0000}$ & $0.11659(2)\phantom{0000}$ & $0.11659(1)\phantom{0000}$ & $7.5\times 10^{-6}$ 
        \\
        $\phantom{-}0.9$ & $p_\mathrm{ISO} + 0.1\phantom{0}$ & $0.4\;$ & $0.11075(3)\phantom{0000}$ & $0.11075(2)\phantom{0000}$ & $0.11075(3)\phantom{0000}$ & $1.8\times 10^{-5}$ 
        \\
        $-0.9$ & $p_\mathrm{ISO} + 0.1\phantom{0}$ & $0.3\;$ & $0.0933206(5)\phantom{00}$ & $0.0933207(5)\phantom{00}$ & $0.0933208(6)\phantom{00}$ & $2.8\times10^{-7}$
        \\
        $-0.9$ & $p_\mathrm{ISO} + 0.1\phantom{0}$ & $0.4\;$ & $0.086501(1)\phantom{000}$ & $0.0865009(9)\phantom{00}$ & $0.0865008(9)\phantom{00}$ & $5.8\times 10^{-6}$ 
        \\
        $\phantom{-}0.9$ & $p_\mathrm{ISO} + 100\phantom{.}$ & $0.3\;$ & $0.008767529(8)$ & $0.008767529(8)$ & $0.00876753(2)\phantom{0}$ & $5.6\times 10^{-9}$ 
        \\
        $\phantom{-}0.9$ & $p_\mathrm{ISO} + 100\phantom{.}$ & $0.4\;$ & $0.00809051(8)\phantom{0}$ & $0.00809051(9)\phantom{0}$ & $0.0080905(2)\phantom{00}$ & $7.0\times 10^{-8}$ 
        \\
        $-0.9$ & $p_\mathrm{ISO} + 100\phantom{.}$ & $0.3\;$ & $0.00825732(3)\phantom{0}$ & $0.00825732(3)\phantom{0}$ & $0.00825733(5)\phantom{0}$ & $2.2\times10^{-8}$
        \\
        $-0.9$ & $p_\mathrm{ISO} + 100\phantom{.}$ & $0.4\;$ & $0.00760869(3)\phantom{0}$ & $0.00760869(3)\phantom{0}$ & $0.00760869(2)\phantom{0}$ & $2.2\times 10^{-9}$ 
        \\
        \hline
        \hline
    \end{tabular*}
\end{table*}}

Figure~\ref{fig:lmode-convergence-log} plots $\hat{\Delta}^{\mathrm{ORG},\ell_\mathrm{max}}_{lm}$ as a function of $(r-r_+)/M$, holding $T = z = \tilde{\phi} = 0$ fixed, for six noneccentric, nonprecessing sources with parameters $(\hat{a}, p, e=0, x=1)$ and various $\ell_\mathrm{max} \in [15, 50]$. The left panels show results for $\hat{a} = 0.6$, the right for $\hat{a} = 0.9$, and rows correspond to $p = \{4, 8, 24\}$ from top to bottom. In all cases, $\hat{\Delta}^{\mathrm{ORG},\ell_\mathrm{max}}_{l m}$ decays rapidly away from the particle location $r_0 = p M$, independent of $\ell_\mathrm{max}$. The dimensionless spin $\hat{a}$ has little effect on the decay rate with $|r - r_0|/M$, while the orbital radius $r_0/M$ strongly influences it. For smaller $r_0/M$, the decay is faster, but $\hat{\Delta}^{\mathrm{ORG},\ell_\mathrm{max}}_{l m}$ reaches a plateau at large $r$, oscillating around a fixed value. This behavior is clearest for $p=4$, where the error levels off around $r \gtrsim 8M$, regardless of $\ell_\mathrm{max}$. For example, with $\ell_\mathrm{max} = 51$, relative accuracy saturates near $10^{-12}$, while $\ell_\mathrm{max} = 25$ yields only $\sim 10^{-6}$. Improving accuracy at large $r$ thus requires including more modes. For larger orbits (e.g., $p=8$), the decay is slower, but the plateau occurs farther out and at lower values, e.g., $r \gtrsim 24M$.

Near the horizon, the behavior mirrors that at infinity: rapid initial decay followed by a plateau around $r - r_+ \lesssim M$. However, the plateau near the horizon occurs at smaller values rather than at large $r$. Thus, for a fixed $\ell_\mathrm{max}$, the metric perturbation is typically more accurate near the horizon than at spatial infinity.

\section{Redshift results}
\label{sec:redshift-results}

We present calculations of the first-order redshift correction $\langle \tilde{z}_1 \rangle_t$ sourced by precessing orbits for the first time. To validate our methods and results, we first construct redshift data for eccentric nonprecessing orbits in Sec.~\ref{sec:redshift-eq} and verify these results against previous calculations in the literature. We then extend these calculations to precessing orbits in Sec.~\ref{sec:redshift-gen}. Combined with \eqref{eqn:H6D}, these redshift results also provide the first numerical calculation of the conservative Hamiltonian to first-order in the mass-ratio for precessing binaries.

\subsection{Nonprecessing (equatorial) orbits}
\label{sec:redshift-eq}

We first validate our methods and calculations by comparing to published values in Table V of Ref.~\cite{VandShah15}. In particular, we focus on the sources that are closest to the innermost stable orbit $p_\mathrm{ISO}$ (i.e, $p = p_\mathrm{ISO} + 1$ in \cite{VandShah15}) and that possess the highest eccentricities ($e=\{0.3, 0.4\}$ in \cite{VandShah15}).\footnote{Methods for computing $p_\mathrm{ISO}$ are provided in Ref.~\cite{SteiWarb20}. In this work, we use the \textsc{KerrGeoPy} Python package \cite{KERRGEOPY, ParkNasi24}, which implements these methods, to compute $p_\mathrm{ISO}$.} In frequency-domain calculations, these are the most challenging sources to accurately resolve due to the slow convergence of the mode-sum in \eqref{eqn:mode-sum-huu-worldline}. Therefore, comparisons in these regions of parameter space provide the most stringent tests of our calculations. We consider both prograde orbits with spin $\hat{a}=0.9$ and retrograde orbits with $\hat{a}=-0.9$. We also compare with sources at large separations (i.e., $p = p_\mathrm{ISO} + 100$) to demonstrate that our code also has extensive reach across the parameter space. 

To obtain the redshift, we first compute the $\ell$-modes of $h_{uu}$ from the reconstructed (and unregularized) metric perturbation in ORG, IRG, and SRG using Eq.~\eqref{eqn:mode-sum-huu-worldline}. To obtain each $\ell$-mode, we truncate the sums over $m$ and $\omega$ in \eqref{eqn:mode-sum-huu-worldline} according to some precision threshold $\epsilon_\mathrm{threshold}$. More specifically, we terminate the mode summation when at least three consecutive terms in the sum pass the convergence condition,
\begin{align} \label{eqn:convergence-criteria}
    \frac{\left\langle{\tilde{z}_0\mathcal{H}}^{(p)\mathcal{G}, \mathcal{J}}_{\pm2\ell m\omega}\right\rangle_t}{\langle \tilde{z}_0H^{[0]}_\mathrm{Lor} \rangle_t} < \epsilon_\mathrm{threshold}.
\end{align}
For these calculations, we set $\epsilon_\mathrm{threshold}=10^{-6}$.

\begin{figure*}[!tbp]
    \centering
    \includegraphics[width=0.9\linewidth]{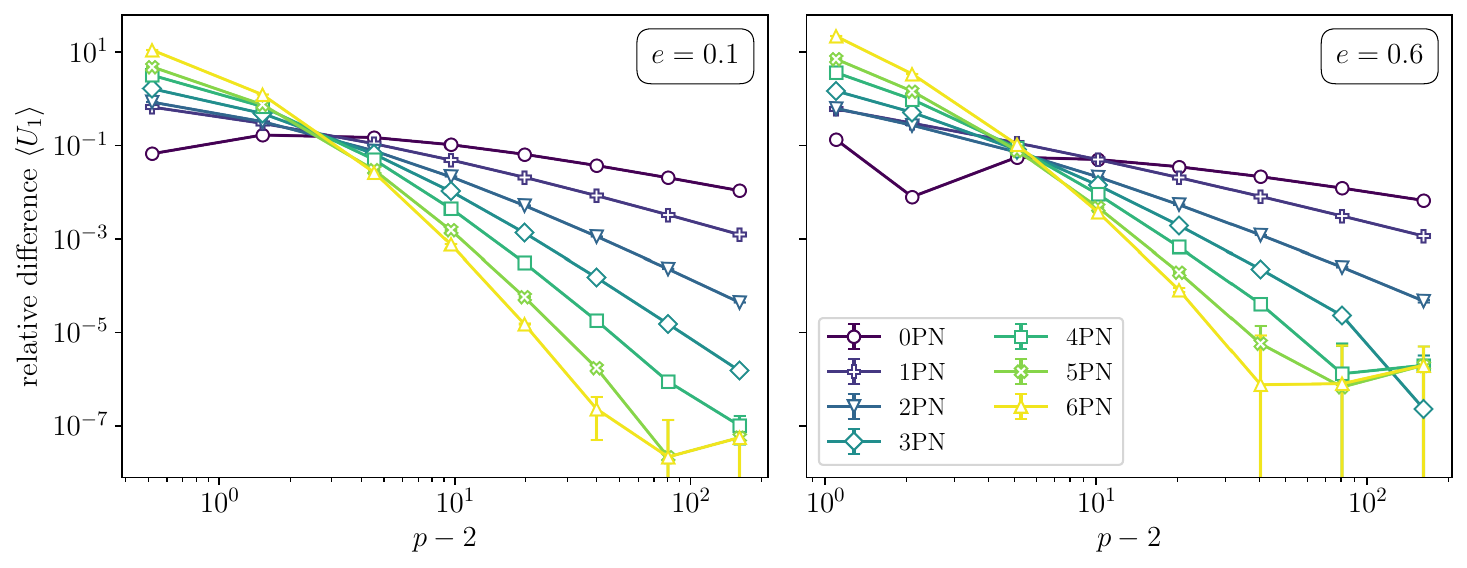}
    \caption{Relative difference between numerical calculations of $U_1$ and the analytical expansion of Ref.~\cite{Munn23} provided in the Black Hole Perturbation Toolkit \cite{BHPTK18}, truncated at various post-Newtonian (PN) orders up to 6PN. (Higher orders are considered in Fig.~\ref{fig:comp8PN}.) For example, diamonds denote the relative difference between our numerical results and the series truncated at 3PN (relative) order. This difference is plotted as a function of $p$ at fixed $(\hat{a}, e)=(0.99, 0.1)$ (left) and $(\hat{a}, e)=(0.99, 0.6)$ (right). Error bars represent estimated error in the numerical data.}
    \label{fig:comp6PN}
\end{figure*}

This procedure leads to three values for the the first-order redshift correction: $\langle \tilde{z}_1^\mathrm{ORG} \rangle_t$, $\langle \tilde{z}_1^\mathrm{IRG} \rangle_t$, and $\langle \tilde{z}_1^\mathrm{SRG} \rangle_t$. These values are reported in Table \ref{tab:redshift-comp}, along with the maximum error between these three values and the redshift values reported in Ref.~\cite{VandShah15}. With each redshift value, we also report the estimated error in the last digit (provided in parentheses) due to our regularization fitting procedure described in Sec.~\ref{sec:regularization} (e.g., $0.11659(2) = 0.11659 \pm 2\times 10^{-5}$). 

We find that our redshift values are consistent with one another (within their estimated error) across all three gauges and all sources. Furthermore for all but one source in Table \ref{tab:redshift-comp}, our redshift values match those in Ref.~\cite{VandShah15} within our error bounds. However, we find slight disagreement for the high-eccentricity, retrograde source $(\hat{a},p,e,x)=(-0.9,p_\mathrm{ISO}+1,0.4,1)$. While we report errors $\sim 10^{-6}$, we deviate with Ref.~\cite{VandShah15} at the level $\sim 6\times 10^{-6}$. This may indicate that we are underestimating our extrapolation error. Alternatively, the calculation of the higher modes in Eq.~\eqref{eqn:mode-sum-huu-worldline} requires evaluating highly oscillatory integrals. The Teukolsky solver in \texttt{pybhpt} can fail to adequately resolve these integrals, introducing numerical errors that would contaminate our extrapolated value for $\langle \tilde{z}_1 \rangle_t$. (See Appendix \ref{app:teuk} for further discussion of this high-frequency issue.) Thus, this may form the dominant source of error in $\langle \tilde{z}_1 \rangle_t$ for $(\hat{a},p,e,x)=(-0.9,p_\mathrm{ISO}+1,0.4,1)$. If this is the case, these errors may also be present for the $(\hat{a},p,e,x)=(0.9,p_\mathrm{ISO}+1,0.4,1)$ source but still remain subdominant to the extrapolation error.

\begin{figure*}
    \centering
    \includegraphics[width=0.95\linewidth]{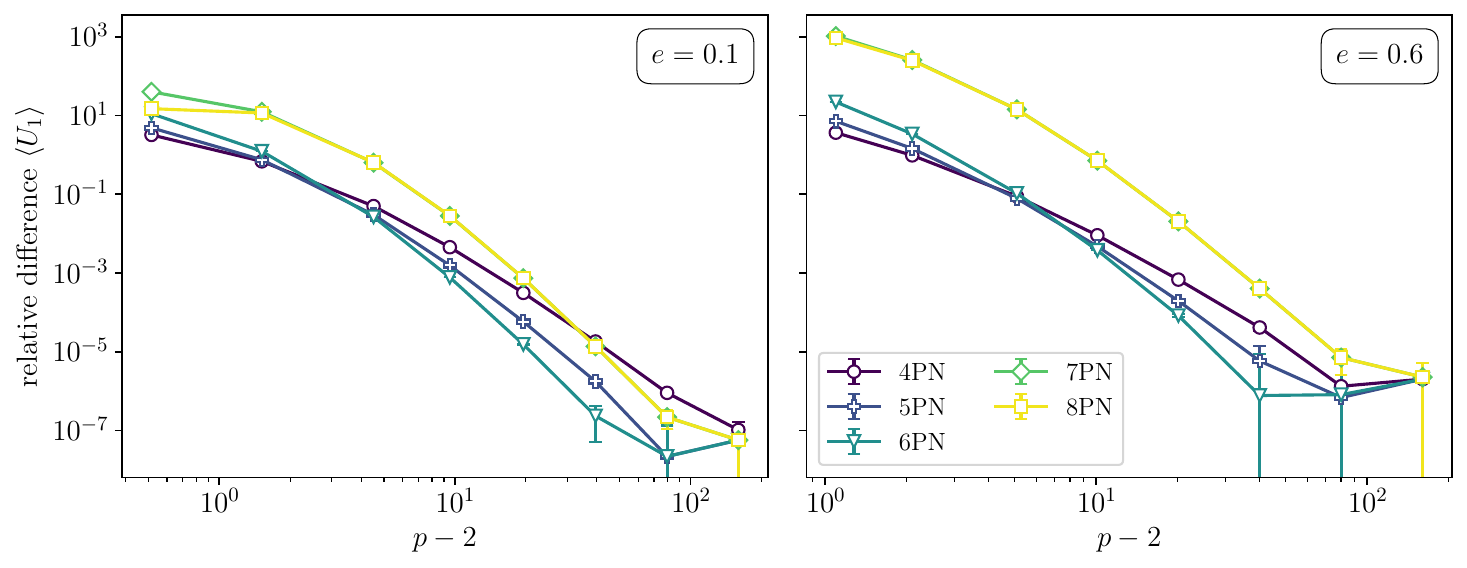}
    \caption{Similar to Fig.~\ref{fig:comp6PN}, the relative difference between numerical calculations of $U_1$ and the analytical expansion of Ref.~\cite{Munn23} truncated at 4, 5, 6, 7, and 8PN for $(\hat{a} = 0.99, p, e=0.1)$ and $(\hat{a} = 0.99, p, e=0.6)$. When including terms beyond 6PN, the agreement between the numerical and analytical results worsens. This behavior is present even when considering other spins and eccentricities.}
    \label{fig:comp8PN}
\end{figure*}

We also compare our numerical results to post-Newtonian (PN) expansions of the first-order redshift correction for eccentric, nonprecessing orbits in the Kerr spacetime \cite{BiniGera19a, Munn23}. Reference \cite{Munn23} computed the quantity $U_1 = - \langle \tilde{z}_1\rangle_t / \langle \tilde{z}_{(0)}\rangle_t^2$ up to 8PN order, and these expansions are publicly available via the \texttt{PostNewtonianSelfForce} \emph{Mathematica} package in the Black Hole Perturbation Toolkit \cite{BHPTK18}. Using these Toolkit results, Fig.~\ref{fig:comp6PN} evaluates our numerical data against the analytical expansions for orbits in a fixed background with spin $\hat{a}=0.99$. The relative difference is plotted as a function of $p$ for eccentricities $e=0.1$ (left panel) and $e=0.6$ (right panel). Each curve in Fig.~\ref{fig:comp6PN} represents a comparison with the PN series truncated at a different order up to 6PN; for instance, diamonds denote the relative difference between our numerical results and the series truncated at 3PN (relative) order. Error bars indicate the estimated uncertainty in our numerical results. 

As expected, the higher-order PN terms (up to 6PN) improve agreement with the numerical data for $p \gtrsim 7$, irrespective of eccentricity. At smaller values of $p$, the PN series begins to diverge, though even near the separatrix, the leading-order PN term maintains a relative agreement of $\lesssim 10^{-1}$. At large $p$, the comparison with higher-order PN expansions becomes limited by numerical error in our results, reaching a noise floor of approximately $10^{-8}$ for $e=0.1$ and $10^{-5}$ for $e=0.6$.

Notably, while the Toolkit provides $U_1$ up to 8PN order, the 7PN term appears to be inconsistent with our numerical findings. This is illustrated in Fig.~\ref{fig:comp8PN}, where the relative difference increases upon the inclusion of terms beyond 6PN, a trend that persists across various eccentricities and spin values. This discrepancy could be due to the eccentricity expansion being truncated at just $e^{10}$ rather than $e^{16}$ beyond 6PN, a mistake in our numerical evaluation of the Toolkit expressions, or some an in the reported higher-order analytical results. We leave the exact source of this discrepancy for future investigations.

\begin{figure}[b]
    \centering
    \includegraphics[width=0.95\linewidth]{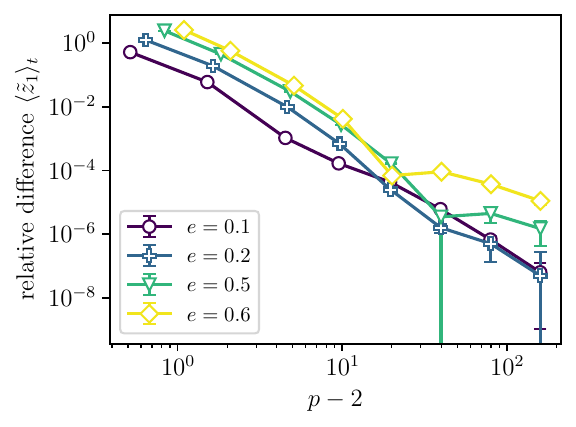}
    \caption{Relative difference between numerical calculations of $\langle \tilde{z}_1 \rangle_t$ and the analytical 3.5PN expansion of Ref.~\cite{BiniGera19a}. This difference is plotted as a function of $p$ at fixed $\hat{a} = 0.99$ for $e \in \{0.1, 0.25, 0.45, 0.6 \}$. Error bars represent estimated error in the numerical data.}
    \label{fig:comp3.5PN}
\end{figure}

Finally, motivated by the accuracy of the leading-order PN terms, we compare our results to those of Ref.~\cite{BiniGera19a}. This work provides analytical expansions for $\langle \tilde{z}_1\rangle_t$ up to orders $e^4$, $a^2$, and 3.5PN.\footnote{While Ref.~\cite{BiniGera19a} also presents $U_1$ up to 8.5PN, we restrict our comparison to their explicit lower-order expressions for $\langle \tilde{z}_1\rangle_t$.} Figure~\ref{fig:comp3.5PN} displays the relative difference between our numerical results and the expansions from Ref.~\cite{BiniGera19a} as a function of $p$ for $\hat{a}=0.99$ and $e \in \{0.1, 0.25, 0.45, 0.6\}$. We find that these lower-order expansions maintain strong agreement with our data even at higher eccentricities and spins for $p \gtrsim 7$, despite the omission of higher-order $e$ and $a$ corrections. These PN comparisons not only validate our results but demonstrate that even low-order expansions can provide leading-order estimates of conservative information deep into the strong-field.

\subsection{Precessing (inclined) orbits}
\label{sec:redshift-gen}

We repeat the redshift calculations in the previous section, but now extend them to precessing sources. 

\subsubsection{Validation}

\begin{figure}[t]
    \centering
    \includegraphics[width=0.9\linewidth]{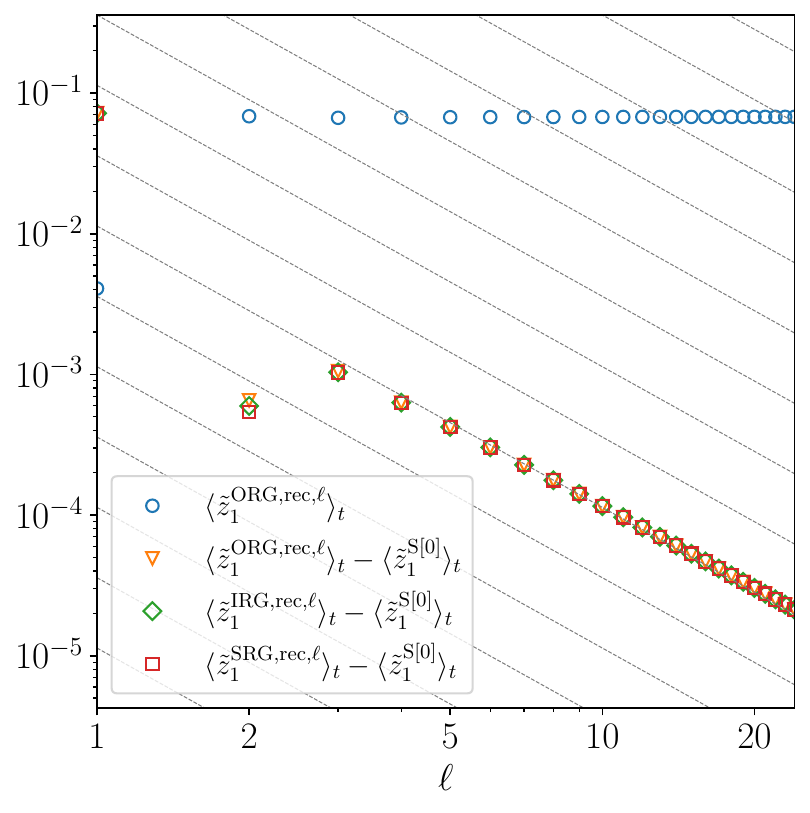}
    \caption{The $\ell$-mode contributions to $\langle \tilde{z}_1^{\mathrm{rec},\ell} \rangle_t$ for the orbital parameters $(\hat{a},p,e,x) = (0.5, 12.5,0.1,0.8)$, computed from the reconstructed metric in IRG, ORG, and SRG. The light gray reference lines are $\propto l^{-2}$.}
    \label{fig:convergence-validation}
\end{figure}

While we do not have external redshift values to compare against, we can validate our calculations through two consistency checks. First, we ensure that our generic regularization parameter, described in Sec.~\ref{sec:regularization}, properly regularizes the leading-order divergence in the $\ell$-mode-sum of the redshift. In Fig.~\ref{fig:convergence-validation}, we plot unregularized and regularized $\ell$-mode contributions to the redshift, $\langle \tilde{z}_1^{\mathrm{rec},\ell} \rangle_t$ and $\langle \tilde{z}_1^{\mathrm{rec},\ell} \rangle_t - \langle \tilde{z}_1^{\mathrm{S}[0]} \rangle_t$, respectively, [see Eq.~\eqref{eqn:z1-reg}] for a precessing, eccentric orbit $(\hat{a},p,e,x) = (0.5, 12.5, 0.1, 0.8)$. We construct the redshift using metric perturbations in no-strings ORG. As expected, $\langle \tilde{z}_1^{\mathrm{ORG,rec},l} \rangle_t$ (blue circles) asymptotes to a constant value at large $\ell$ (i.e., $\sim \ell^0$), while $\langle \tilde{z}_1^{\mathrm{ORG,rec},l} \rangle_t - \langle \tilde{z}_1^{\mathrm{S}[0]} \rangle_t$ (orange triangles) decays as $\sim \ell^{-2}$ (gray lines). We observe an identical fall-off when computing the regularized modes using (no-strings) IRG (green diamonds) and SRG (red squares) perturbations, as well, validating our regularization scheme across all three well-behaved gauges. This check provides a strong confirmation that our retarded mode-sum calculations faithfully reconstruct the local singular structure of the metric perturbation.

Second, we check that we obtain consistent values for $\langle \tilde{z}_1 \rangle_t$ when computing the redshift from the ORG, IRG, and SRG metric perturbations. Applying our regularization fitting scheme to the data in Fig.~\ref{fig:convergence-validation}, we find
\begin{subequations}
\begin{align}
    \langle \tilde{z}_1^\mathrm{ORG}\rangle_t &= 0.0654387(4),
    \\
    \langle \tilde{z}_1^\mathrm{IRG}\rangle_t &= 0.0654387(4),
    \\
    \langle \tilde{z}_1^\mathrm{SRG}\rangle_t &= 0.0654386(4).
\end{align}
\end{subequations}
As before, the number in parentheses refers to the estimated uncertainty in the last digit based on our fitting procedure. Mutual agreement between these redshift values provides further evidence that we have accurately constructed the regular contributions to the metric perturbations.

{\renewcommand{\arraystretch}{1.25}
\begin{table}[!t]
    \centering
    \setlength\tabcolsep{0pt}
    \caption{Redshift values $\langle \tilde{z}_1 \rangle_t$ for precessing orbits. Rows above the dividing line are noneccentric, while those below are eccentric. The numbers in parentheses refer to the estimated uncertainty in the last digit based on our fitting procedures.}
    \label{tab:redshift-precessing}
    \begin{tabular*}{\linewidth}{@{\extracolsep{\fill}} c c c c c}
        \hline
        \hline
        $\hat{a}$ & $p$ & $e$ & $x$ & $\langle \tilde{z}_1 \rangle_t$  \\
        \hline
        $0.9$ & $6.5$ & $0.0$ & $0.99$ & $\phantom{-}0.107222(1)\phantom{00}$
        \\
        $0.9$ & $10.5$ & $0.0$ & $0.99$ & $\phantom{-}0.07863001(5)$
        \\
        $0.9$ & $30.5$ & $0.0$ & $0.99$ & $\phantom{-}0.03114563(7)$
        \\
        $0.9$ & $50.5$ & $0.0$ & $0.99$ & $\phantom{-}0.01924561(3)$
        \\
        $0.9$ & $30.5$ & $0.0$ & $0.97$ & $\phantom{-}0.03107385(4)$
        \\
        $0.9$ & $30.5$ & $0.0$ & $0.95$ & $\phantom{-}0.03100323(4)$
        \\
        $0.9$ & $30.5$ & $0.0$ & $0.90$ & $\phantom{-}0.03083187(2)$
        \\
        $0.9$ & $100.5$ & $0.0$ & $0.99$ & $\phantom{-}0.00982077(2)$
        \\
        $0.9$ & $100.5$ & $0.0$ & $0.97$ & $\phantom{-0}0.009813797(7)$
        \\
        $0.9$ & $100.5$ & $0.0$ & $0.95$ & $\phantom{-0}0.009806954(5)$
        \\
        $0.9$ & $100.5$ & $0.0$ & $0.90$ & $\phantom{-0}0.009790422(4)$
        \\
        \hline
        $0.999$ & $p_\mathrm{ISO}+\phantom{0}0.1$ & $0.25$ & $0.7$ & $-0.1263(2)\phantom{0000}$
        \\
        $0.999$ & $p_\mathrm{ISO}+\phantom{0}0.2$ & $0.25$ & $0.7$ & $-0.0941(4)\phantom{0000}$
        \\
        $0.999$ & $p_\mathrm{ISO}+\phantom{0}1.0$ & $0.25$ & $0.7$ & $\phantom{-}0.02017(4)\phantom{000}$
        \\
        $0.999$ & $p_\mathrm{ISO}+10.1$ & $0.25$ & $0.7$ & $\phantom{-}0.059403(7)\phantom{00}$
        \\
        $0.999$ & $p_\mathrm{ISO}+30.1$ & $0.25$ & $0.7$ & $\phantom{-}0.026795(2)\phantom{00}$
        \\
        $0.999$ & $p_\mathrm{ISO}+30.1$ & $0.1$ & $0.7$ & $\phantom{-1}0.0282233(2)\phantom{00}$
        \\
        $0.9$ & $6.5$ & $0.2$ & $0.8$ & $\phantom{-}0.089638(4)\phantom{00}$
        \\
        $0.9$ & $6.5$ & $0.25$ & $\cos\left({49\pi}/{100}\right)$ & $\phantom{-}0.04316(2)\phantom{000}$
        \\
        $0.9$ & $6.5$ & $0.25$ & $0.3$ & $\phantom{-}0.05531(3)\phantom{000}$
        \\
        $0.9$ & $6.5$ & $0.25$ & $0.8$ & $\phantom{-}0.08792(2)\phantom{000}$
        \\
        $0.5$ & $8.0$ & $0.1$ & $0.8$ & $\phantom{-}0.087976(8)\phantom{00}$
        \\
        $0.5$ & $8.0$ & $0.4$ & $0.5$ & $\phantom{-}0.06344(4)\phantom{000}$
        \\
        $0.5$ & $8.0$ & $0.4$ & $0.8$ & $\phantom{-}0.07718(9)\phantom{000}$
        \\
        $0.5$ & $8.0$ & $0.6$ & $0.8$ & $\phantom{-}0.0616(4)\phantom{0000}$
        \\
        $0.5$ & $12.5$ & $0.1$ & $0.8$ & $\phantom{-}0.0654387(4)\phantom{0}$
        \\
        \hline
        \hline
    \end{tabular*}
\end{table}}

\subsubsection{Noneccentric, minimally precessing orbits}

Previous investigations of the redshift invariant have verified numerical calculations by both comparing their data to known post-Newtonian results, and they have used numerical data to extract higher-order terms in the post-Newtonian expansion of the redshift \cite{AkcaETC15, VandShah15}. However, to the best of our knowledge, current post-Newtonian expansions of the redshift are limited to nonprecessing systems, preventing a similar comparison in this work. Instead, in the top half of Table \ref{tab:redshift-precessing}, we present redshift values for a number of noneccentric, minimally precessing sources with a focus on orbits with $p > 10$ and $(1-x) < 0.1$. These values will facilitate future comparisons, once post-Newtonian expansions of the redshift for precessing orbits become available. Conceivably, one could use these results to estimate precession terms in the post-Newtonian expansion of the redshift---which we expect to be proportional to $a(1-x)$---but we save this potential investigation for future work.

\subsubsection{Eccentric, precessing orbits}

In the bottom half of Table \ref{tab:redshift-precessing}, we present $\langle \tilde{z}_1 \rangle_t$ for several eccentric, precessing sources. All values were computed from ORG perturbations. To construct each redshift estimate, we used up to $\ell_\mathrm{max}=30$ modes to regularize and fit for $\langle \tilde{z}_1 \rangle_t$. Furthermore, each $\ell$-mode was calculated using the mode-sum convergence criteria in Eq.~\eqref{eqn:convergence-criteria} with $\epsilon_\mathrm{threshold} \leq 10^{-5}$. We highlight that our results include sources with spins as high as $\hat{a}=0.999$, separations as close as $p-p_\mathrm{ISO} = 0.1$, eccentricities as large as $e=0.6$, and inclinations as great as $\cos(49\pi/100)\approx 0.0314108$.

\begin{figure}
    \centering
    \includegraphics[width=0.9\linewidth]{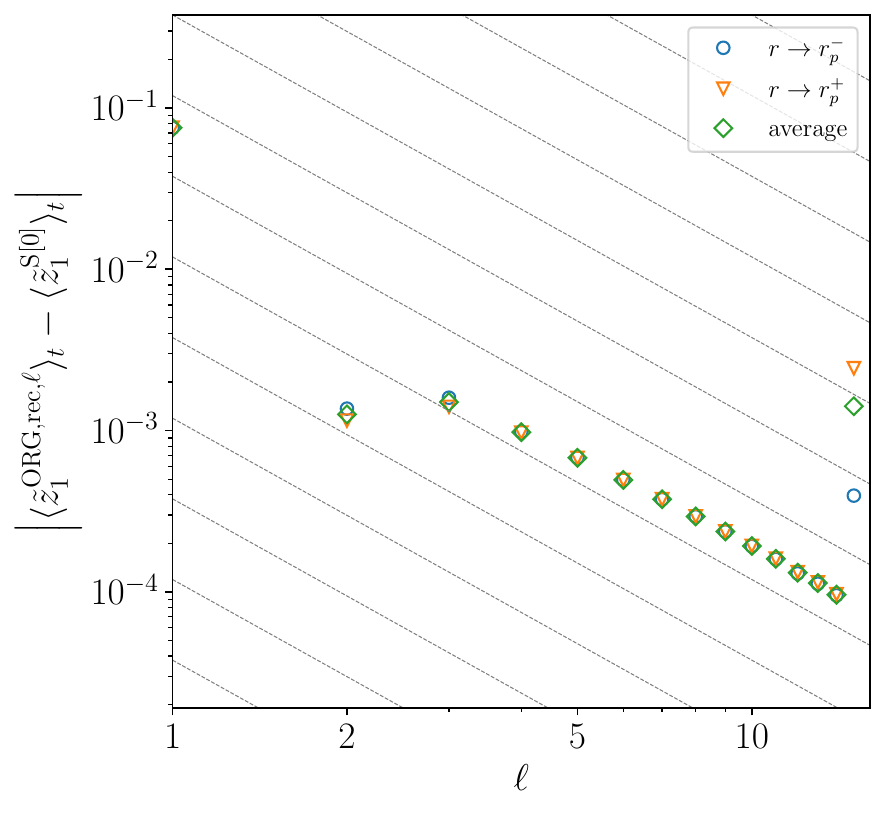}
    \caption{The $\ell$-mode contributions to $\langle \tilde{z}_1^{\mathrm{rec},\ell} \rangle_t$ for the orbital parameters $(\hat{a},p,e,x) = (0.5, 8, 0.6, 0.8)$, computed from the reconstructed metric in ORG using three different limits: approaching the particle from the horizon $r\rightarrow r_p^-$ (blue circles), infinity $r\rightarrow r_p^+$ (orange diamonds), and the average of these limits (green $\times$). The light gray reference lines are $\propto l^{-2}$. Numerical error dominates the final reported mode at $\ell = 16$.}
    \label{fig:highe-convergence}
\end{figure}

As we observed in Sec.~\ref{sec:redshift-eq}, we find it much more challenging to numerically resolve orbits with higher eccentricities and smaller separations, limiting our redshift results to just 2-3 digits of accuracy. For high eccentricities, the accuracy is not only limited by numerical errors in our computation of the Teukolsky amplitudes, as previously described in Sec.~\ref{sec:redshift-eq}, but also by catastrophic cancellation in our extended homogeneous solutions at high $\ell$-values (e.g., $\ell \gtrsim 16$), as discussed in Sec.~V A of Ref.~\cite{Vand16}. Thus, for $e=0.6$, we can only accurately calculate $\ell$-modes up to $\ell_\mathrm{max} \sim 15$. We highlight this behavior in Figure \ref{fig:highe-convergence}, where we plot the $\ell$-mode convergence of $\langle \tilde{z}_1^{\mathrm{rec},\ell} \rangle_t$ for the orbital parameters $(\hat{a},p,e,x) = (0.5, 8, 0.6, 0.8)$. The $\ell$-mode-sum converges until $\ell = 16$, where numerical errors due to catastrophic cancellations dominate. Truncating $\ell$-mode data around $\ell_\mathrm{max} \sim 15$ at higher eccentricities significantly diminishes our extrapolation of the large-$\ell$ behavior of the $\ell$-mode-sum, increasing the overall error of our regularization procedure (see Sec.~\ref{sec:regularization}).

For orbits closer to $p_\mathrm{ISO}$, particularly for those computed in the near-extremal Kerr case $\hat{a}=0.999$, the diminished accuracy results from slower convergence in the $\ell$-mode-sum due to beaming of the gravitational radiation as the small body orbits closer to the Kerr horizon. This also impacts the extrapolation of the large-$\ell$ behavior, producing more conservative error estimates from our regularization scheme. Inclination, on the other hand, has less of an impact on the accuracy of our results, with even our most inclined orbit possessing an estimated error $\sim 10^{-5}$ for $e \lesssim 0.4$.

\begin{figure}
    \centering
    \includegraphics[width=0.98\linewidth]{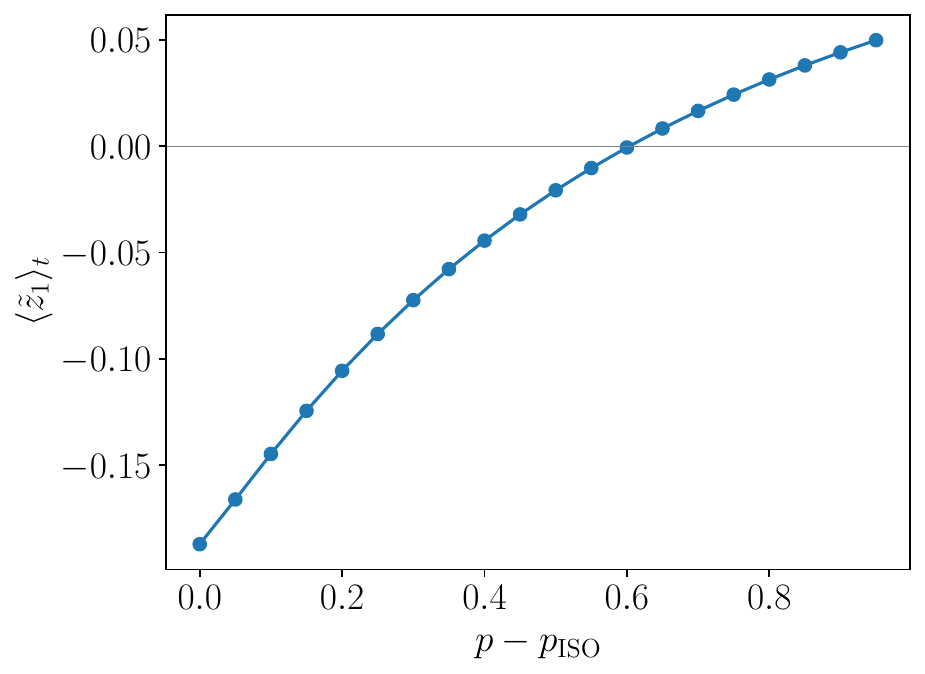}
    \caption{Plot of $\langle \tilde{z}_1 \rangle_t$ as a function of $p-p_\mathrm{ISO}$ for noneccentric, nonprecessing sources in a Kerr spacetime with $\hat{a}=0.999$ and $p_\mathrm{ISO} \approx 1.1817646$ [i.e., $(\hat{a},p,e,x)=(0.999,p,0,1)$].}
    \label{fig:z1-negative-circ}
\end{figure}

Interestingly, we observe for the first time negative values of the redshift correction $\langle \tilde{z}_1 \rangle_t$ for sources orbiting near the innermost stable orbit of a highly spinning Kerr black hole. Further investigation reveals that this feature is not exclusive to precessing orbits. In Fig.~\ref{fig:z1-negative-circ}, we plot $\langle \tilde{z}_1 \rangle_t$ as a function of $p$ for noneccentric, nonprecessing sources in a Kerr spacetime with $\hat{a}=0.999$. Even for these much more simple orbits, we find $\langle \tilde{z}_1 \rangle_t < 0$ for $p - p_\mathrm{ISO}\lesssim 0.6$. This suggests that, for highly spinning systems, there is a surface in $(p,e,x)$ along which $\langle \tilde{z}_1 \rangle_t$, and thus the interaction Hamiltonian, vanishes. Additionally, because $\langle \tilde{z}_1 \rangle_t$ decays with increasing $p$, $e$, and $1-x$, $\langle \tilde{z}_1 \rangle_t$ must also possess turning points where derivatives with respect to these orbital parameters vanish.

\section{Conclusion}
\label{sec:conclusion}

In this work, we studied first-order metric perturbations sourced by a massive point-particle on an eccentric, precessing orbit around a Kerr black hole. By employing four distinct reconstruction methods to obtain $h_{\alpha\beta}$ from the Weyl scalars $\psi_0$ and $\psi_4$, we calculated metric perturbations in IRG and ORG, along with two new gauges, which we refer to as SRG and ARG. We studied the similarities and differences in metric behavior across various gauges, highlighting that the SRG and ARG perturbations share the asymptotic behavior of the ORG and IRG perturbations. Furthermore, we found that our point-particle SRG perturbations share the same $O(s^{-1})$ singular structure near the particle as the IRG and ORG perturbations, while the ARG perturbations strongly diverge along the worldline as $\sim s^{-4}$, though this singular behavior can be reduced to $O(s^{-3})$ by leveraging the parity properties of the two half-string ARG solutions. Due to its highly singular nature, along with its breakdown for static modes, it remains unclear whether the ARG reconstruction procedure is a useful gauge for studying perturbations or if it serves best as an intermediate step in calculating the metric perturbation in more well-behaved gauges (e.g., Lorenz gauge \cite{DolaETC23}).

Using these methods, we produced the first computations of the generalized redshift (correction) $\langle \tilde{z}_1 \rangle_t$ along precessing orbits. We also provided new numerical results for eccentricities as high as $e=0.6$ and Kerr spins up to $a/M=0.999$---regions of parameter space that have not been investigated by previous redshift calculations in the self-force literature. By pushing to near-extremal spins, we observed for the first time that the redshift correction vanishes and then becomes negative near the innermost stable orbit. Furthermore, to make these calculations and parameter space investigations more accessible to the community, we created the open-source Python library \texttt{pybhpt}, which offers a numerical toolkit for future research in this domain. 

By combining our calculations with the Hamiltonian framework developed by Lewis \emph{et.~al} \cite{LewiETC25}, these redshift results provide a promising avenue for computing all conservative contributions to 1PA gravitational waveform models within the self-force framework. This Hamiltonian approach also has several other appealing properties. First of all, it only requires the calculation of a single perturbation field $h_{uu}$ to determine $\langle \tilde{z}_1 \rangle_t$, and thus the conservative Hamiltonian, rather than the four-components of the GSF $F_\alpha$ (up to first-order in the mass-ratio). Furthermore, $\langle \tilde{z}_1 \rangle_t$ is a quasi-invariant, unlike the highly gauge-dependent GSF, and is less singular than the unregularized self-force, since the GSF is constructed by regularizing coordinate derivatives of $h_{\alpha\beta}$. These characteristics make it much easier to accurately compute and validate $\langle \tilde{z}_1 \rangle_t$ compared to the GSF. Ultimately, the Hamiltonian approach exchanges coordinate derivatives of the metric perturbation for phase-space derivatives of the perturbed averaged redshift to drive the conservative dynamics. Therefore, future work will look to tabulate redshift data across the phase space in order to implement this Hamiltonian framework for waveform generation. 

There are also many avenues for future improvement of the underlying algorithms for computing the metric perturbation and redshift correction. These include constructing perturbation data along hyperboloidal slices and in compactified coordinates \cite{Zeng11, Mace20, MaceETC22}, implementing pseudospectral collocation methods for constructing the Weyl scalars and Hertz potentials \cite{MaceETC24}, incorporating the decomposition methods proposed by Spiers \cite{Spie24} for producing radiation gauge metric perturbations, and making our numerical routines more stable at higher eccentricities and smaller separations. Future work could also investigate synergies with other Hamiltonian approaches, such as effective-one-body and post-Newtonian models, to better understand the conservative dynamics of eccentric and precessing compact object binaries.

\begin{acknowledgments}
We thank Barry Wardell, Chris Kavanagh, and Adam Pound for invaluable feedback and discussions. This research was based on work originally supported by an appointment to the NASA Postdoctoral Program at the NASA Goddard Space Flight Center, administered by Oak Ridge Associated Universities under contract with NASA, and by NASA under award number 80GSFC21M0002. It was also supported from the ERC Consolidator/UKRI Frontier Research Grant GWModels (selected by the ERC and funded by UKRI [grant number EP/Y008251/1]). The authors also acknowledge the use of the IRIDIS High Performance Computing Facility, and associated support services at the University of Southampton, in the completion of this work.
\end{acknowledgments}

\section*{Data Availability}

The data that support the findings of this article are openly available \cite{PYBHPT, PYBHPTv092, Nasipak_metric-reconstruction-paper_2026, Nasipak2026-st}.

\appendix

\section{Null tetrads and GHP notation}
\label{app:GHP}

\subsection{General formalism}
\label{app:GHP-formalism}

The Geroch–Held–Penrose (GHP) formalism \cite{GeroHeldPenr73} provides a convenient set of tools for studying spacetimes with two principle null directions. We recommend Refs.~\cite{PounWard20, GreeHollZimm20, Spie24} for more extensive reviews of GHP notation in the context of black hole perturbation theory and its application to the Teukolsky equations and metric reconstruction in Kerr spacetime. In this appendix, we summarize the relevant GHP notation, coefficients, and operators, and how they are related to the metric reconstruction operators described in Sec.~\ref{sec:background}.

As before, we define a null tetrad $e^\alpha_\mathbf{a} \doteq (l^\alpha, n^\alpha, m^\alpha, \bar{m}^\alpha)$ such that $l^\alpha$ and $n^\alpha$ are aligned with the principal null directions of Kerr spacetime, and the tetrad is normalized such that $l^\alpha n_\alpha = -1 = - m_\alpha \bar{m}^\alpha$. In the GHP formalism, quantities are classified by their transformations under the rescaling and interchange of this tetrad. First of all, there is residual freedom to rescale the tetrad according to
\begin{subequations} \label{eqn:rescale-tetrad}
    \begin{align}
    l^\alpha &\rightarrow \lambda \bar{\lambda}l^\alpha,
    &
    n^\alpha &\rightarrow \lambda^{-1} \bar{\lambda}^{-1}n^\alpha,
    \\
    m^\alpha &\rightarrow {\lambda}\bar{\lambda}^{-1} m^\alpha,
    &
    \bar{m}^\alpha &\rightarrow \lambda^{-1}\bar{\lambda} \bar{m}^\alpha,
\end{align}
\end{subequations}
where $\lambda$ is a complex scalar. Additionally, one is free to exchange the principle null directions with which $l^\alpha$ and $n^\alpha$ are aligned.

Thus, a GHP quantity $f$ has GHP weights $\{p,q\}$, denoted as $f \mathring{=} \{p,q\}$, if $f \rightarrow \lambda^p \bar{\lambda}^q f$ under the transformation in Eq.~\eqref{eqn:rescale-tetrad}. Furthermore, one can define the GHP prime operation, which transforms GHP quantities under the exchange $l^\alpha \leftrightarrow n^\alpha$, $m^\alpha \leftrightarrow \bar{m}^\alpha$. From Eq.~\eqref{eqn:rescale-tetrad}, we see that this operation flips the sign of the GHP weights, i.e., $f' \mathring{=} \{-p,-q\}$, while complex conjugation switches the weights, i.e., $\bar{f} \mathring{=} \{q,p\}$. These GHP weights can also be related to the spin- and boost-weights via $s=(p-q)/2$ and $b=(p+q)/2$, respectively.

Akin to the Newman-Penrose formalism \cite{NewmPenr63}, one then defines the GHP scalars from covariant derivatives of the tetrad legs:
\begin{subequations}
    \begin{align}
        \kappa &= -l^\alpha m^\beta \nabla_\alpha l_\beta,
        \\
        \tau &= -n^\alpha m^\beta \nabla_\alpha l_\beta,
        \\
        \sigma &= -m^\alpha m^\beta \nabla_\alpha l_\beta,
        \\
        \rho &= -\bar{m}^\alpha m^\beta \nabla_\alpha l_\beta,
    \end{align} 
\end{subequations}
along with their primes $\kappa'$, $\tau'$, $\sigma'$, and $\rho'$. Similarly, there are the four GHP directional operators
\begin{subequations} \label{eqn:GHP-op}
    \begin{align}
        \thorn &= l^\alpha \nabla_\alpha - p \epsilon - q \bar{\epsilon},
        \\
        \eth &= m^\alpha \nabla_\alpha - p \beta + q \bar{\beta}',
        \\
        \thorn' &= n^\alpha \nabla_\alpha + p \varepsilon' + q \bar{\varepsilon}',
        \\
        \eth' &= \bar{m}^\alpha \nabla_\alpha + p \beta' - q \bar{\beta},
    \end{align}
\end{subequations}
where 
\begin{align}
    \beta &= \frac{1}{2}\left(m^\alpha \bar{m}^\beta \nabla_\alpha m_\beta - m^\alpha n^\beta \nabla_\alpha l_\beta \right),
    \\
    \varepsilon &= \frac{1}{2}\left(l^\alpha \bar{m}^\beta \nabla_\alpha m_\beta - l^\alpha n^\beta \nabla_\alpha l_\beta \right),
\end{align}
are scalars with no GHP weight. Following Wald's definition of adjoints \cite{Wald78}, the adjoints of \eqref{eqn:GHP-op} are then given by
\begin{subequations}
\begin{align}
    \thorn^\dagger &= -\thorn + \rho + \bar{\rho},
    \\
    \thorn'^\dagger &= -\thorn' + \rho' + \bar{\rho}',
    \\
    \eth^\dagger &= -\eth + \tau + \bar{\tau}',
    \\
    \eth'^\dagger &= -\eth' + \tau' + \bar{\tau}.
\end{align}
\end{subequations}

\subsection{Teukolsky and reconstruction operators}
\label{app:def-operators}

Using the GHP coefficients and operators summarized in Appendix \ref{app:GHP-formalism}, the operators $\mathcal{T}_{0,4}$ in Eq.~\eqref{eqn:Weyl} are given by
\begin{subequations} \label{eqn:T-op}
    \begin{align}
        2\mathcal{T}_0^{\alpha\beta} &= -(\edth - \bar{\tau}')^2 l^{\alpha} l^{\beta} - (\thorn - \bar{\rho})^2 {m}^{\alpha}{m}^{\beta}
        \\ \notag
        & \; + [(\thorn - \bar{\rho})(\edth - 2\bar{\tau}') + (\edth - \bar{\tau}')(\thorn - 2\bar{\rho})]l^{\alpha}{m}^{\beta},
        \\
        2\mathcal{T}_4^{\alpha\beta} &= -(\edth' - \bar{\tau})^2 n^{\alpha} n^{\beta} - (\thorn' - \bar{\rho}')^2 \bar{m}^{\alpha}\bar{m}^{\beta}
        \\ \notag
        & \; + [(\thorn' - \bar{\rho}')(\edth' - 2\bar{\tau}) + (\edth' - \bar{\tau})(\thorn' - 2\bar{\rho}')]n^{\alpha}\bar{m}^{\beta}.
    \end{align}
\end{subequations}
and their adjoints [see Eq.~\eqref{eqn:adjoint}] by
\begin{subequations} \label{eqn:Tdagger-op}
    \begin{align}
        2(\mathcal{T}^\dagger_0)_{\alpha\beta} &= - l_{\alpha} l_{\beta} (\edth - \tau)^2 - {m}_{\alpha} {m}_{\beta} (\thorn - \rho)^2
        \\ \notag
        & \qquad \qquad + l_{(\alpha}m_{\beta)}[(\thorn - \rho + \bar{\rho})(\edth - \tau) 
        \\ \notag
        & \qquad \qquad \qquad \qquad \quad + (\edth - \tau + \bar{\tau}')(\thorn - \rho)],
        \\
        2(\mathcal{T}^\dagger_4)_{\alpha\beta} &= - n_{\alpha} n_{\beta} (\edth' - \tau')^2 - \bar{m}_{\alpha} \bar{m}_{\beta} (\thorn' - \rho')^2
        \\ \notag
        & \qquad \qquad + n_{(\alpha}\bar{m}_{\beta)}[(\thorn' - \rho' + \bar{\rho}')(\edth' - \tau') 
        \\ \notag
        & \qquad \qquad \qquad \qquad \quad + (\edth' - \tau' + \bar{\tau})(\thorn' - \rho')].
    \end{align}
\end{subequations}
Likewise, $\mathcal{S}_{0,4}^{\alpha\beta}$ in Eq.~\eqref{eqn:source} are given by
\begin{subequations} \label{eqn:Sop}
    \begin{align}
        2\mathcal{S}_0^{\alpha\beta} &= -(\edth - \bar{\tau}' - 4\tau)(\edth - \bar{\tau}') l^{\alpha} l^{\beta} 
        \\ \notag
        & \qquad - (\thorn -4 \rho - \bar{\rho})(\thorn - \bar{\rho}) {m}^{\alpha}{m}^{\beta}
        \\ \notag
        & \qquad \qquad + [(\thorn - 4\rho - \bar{\rho})(\edth - 2\bar{\tau}') 
        \\ \notag
        & \qquad \qquad \qquad
        + (\edth - \bar{\tau}' - 4\tau)(\thorn - 2\bar{\rho})]l^{\alpha}{m}^{\beta},
        \\
        2\mathcal{S}_4^{\alpha\beta} &= -(\edth' - \bar{\tau} - 4\tau')(\edth' - \bar{\tau}) n^{\alpha} n^{\beta} 
        \\ \notag
        & \qquad - (\thorn' -4 \rho' - \bar{\rho}')(\thorn' - \bar{\rho}') \bar{m}^{\alpha}\bar{m}^{\beta}
        \\ \notag
        & \qquad \qquad + [(\thorn' - 4\rho' - \bar{\rho}')(\edth' - 2\bar{\tau}) 
        \\ \notag
        & \qquad \qquad \qquad
        + (\edth' - \bar{\tau} - 4\tau')(\thorn' - 2\bar{\rho}')]n^{\alpha}\bar{m}^{\beta},
    \end{align}
\end{subequations}
and their adjoints by
\begin{subequations} \label{eqn:Sdagger-op}
    \begin{align}
        2(\mathcal{S}^\dagger_0)_{\alpha\beta} &= - l_{\alpha} l_{\beta} (\edth - \tau)(\edth + 3\tau) 
        \\ \notag
        & \qquad - {m}_{\alpha} {m}_{\beta} (\thorn - \rho)(\thorn + 3\rho)
        \\ \notag
        & \qquad \qquad + l_{(\alpha}m_{\beta)}[(\thorn - \rho + \bar{\rho})(\edth + 3\tau) 
        \\ \notag
        & \qquad \qquad \qquad \qquad \quad + (\edth - \tau + \bar{\tau}')(\thorn + 3\rho)],
        \\
        2(\mathcal{S}^\dagger_4)_{\alpha\beta} &= - n_{\alpha} n_{\beta} (\edth' - \tau')(\edth' + 3\tau') 
        \\ \notag
        & \qquad - \bar{m}_{\alpha} \bar{m}_{\beta} (\thorn' - \rho')(\thorn' + 3\rho')
        \\ \notag
        & \qquad \qquad + n_{(\alpha}\bar{m}_{\beta)}[(\thorn' - \rho' + \bar{\rho}')(\edth' + 3\tau') 
        \\ \notag
        & \qquad \qquad \qquad \qquad \quad + (\edth' - \tau' + \bar{\tau})(\thorn' + 3\rho')].
    \end{align}
\end{subequations}
The Teukolsky operator in Eq.~\eqref{eqn:teuk}, along with its GHP prime, take the forms
\begin{subequations}
    \begin{align}
    \mathcal{O} &= (\thorn - 2s\rho -\bar{\rho})(\thorn' - {\rho}')
    \\ \notag
    & \qquad - (\eth -2s\tau -\bar{\tau}')(\eth' - \tau') - (1 - s) (1 - 2 s)\psi_2,
    \\
    \mathcal{O}' &= (\thorn' + 2s\rho' -\bar{\rho}')(\thorn - {\rho})
    \\ \notag
    & \qquad - (\eth' + 2s\tau' - \bar{\tau})(\eth - \tau) - (1 + s) (1 + 2 s)\psi_2,
\end{align}
\end{subequations}
while their adjoints are given by
\begin{subequations}
    \begin{align}
    \mathcal{O}^\dagger &= (\thorn' - \bar{\rho}')[\thorn - (1+2s)\rho]
    \\ \notag
    & \qquad - (\eth' - \bar{\tau})[\eth -(1+2s)\tau] - (1 + s) (1 + 2 s)\psi_2,
    \\
    &= \psi_2^{2s/3}\mathcal{O}'\psi_2^{-2s/3},
    \\
    \mathcal{O}'^\dagger &= (\thorn - \bar{\rho})[\thorn' - (1-2s)\rho']
    \\ \notag
    & \qquad - (\eth - \bar{\tau}')[\eth' -(1-2s)\tau'] - (1 - s) (1 - 2 s)\psi_2,
    \\
    &= \psi_2^{-2s/3}\mathcal{O}\psi_2^{2s/3}.
\end{align}
\end{subequations}
The second and fourth equalities are derived from the relations
\begin{subequations}
    \begin{align}
    \psi_2^{\alpha/3}\thorn\psi_2^{-\alpha/3} &= \thorn - \alpha \rho,
    \\
    \psi_2^{\alpha/3}\eth\psi_2^{-\alpha/3} &= \eth - \alpha \tau,
\end{align}
\end{subequations}
along with their primed versions. 

The Teukolsky-Starobinsky identities are then expressed as \cite{Torr94, PounWard20}
\begin{subequations} \label{eqn:teuk-starobinsky}
    \begin{align}
        \thorn^4\, \zeta^4 \psi_4 &= \edth'^4 \zeta^4 \psi_0 - 3 M \pounds_\xi \bar{\psi}_0,
        \\
        \thorn'^4 \zeta^4 \psi_0 &= \edth^4\, \zeta^4 \psi_4 + 3 M \pounds_\xi \bar{\psi}_4.
    \end{align}
\end{subequations}

\subsection{Kinnersley tetrad and coordinate expressions}
\label{app:kinnersley}

In this work, we make use of the Kinnersley tetrad. It is given in Boyer-Lindquist coordinates by
\begin{subequations}
\begin{align}
    (l^\mu) &\doteq \frac{1}{\Delta}(r^2+a^2, \Delta, 0 ,a),
    \\
    (n^\mu) &\doteq \frac{1}{2\zeta \bar{\zeta}}(r^2+a^2, -\Delta, 0 ,a),
    \\
    (m^\mu) &\doteq \frac{1}{\bar{\zeta}}\sqrt{\frac{1-z^2}{2}}\left(ia, 0, -1, \frac{i}{1-z^2}\right),
\end{align}
\end{subequations}
and $\zeta = (r-ia\cos\theta)$. The corresponding covectors are
\begin{subequations}
    \begin{align}
        (l_\mu) &\doteq \left(-1, \frac{\zeta \bar{\zeta}}{\Delta}, 0 ,a(1-z^2)\right),
        \\
        (n_\mu) &\doteq \frac{\Delta}{2\zeta \bar{\zeta}}\left(-1, -\frac{\zeta \bar{\zeta}}{\Delta}, 0 ,a(1-z^2)\right),
        \\
        (m_\mu) &\doteq -\frac{1}{\bar{\zeta}}\sqrt{\frac{1-z^2}{2}}\left(ia, 0, \frac{\Sigma}{1-z^2}, -i(r^2+a^2)\right).
    \end{align}
\end{subequations}
The GHP coefficients are given by
\begin{subequations}
    \begin{align}
    \tau &= -\frac{ia\sqrt{1-z^2}}{\sqrt{2} \zeta \bar{\zeta}},
    &
    \tau' &= -\frac{ia\sqrt{1-z^2}}{\sqrt{2} \zeta^2},
    \\
    \rho &= -\frac{1}{\zeta},
    &
    \rho' &= \frac{\Delta}{2\zeta^2\bar{\zeta}},
\end{align}
\end{subequations}
with $\kappa$, $\kappa'$, $\sigma$, and $\sigma'$ all vanishing. Furthermore,
\begin{subequations}
    \begin{align}
    \beta\phantom{'} &= \frac{z}{2\sqrt{2(1-z^2)}\bar{\zeta}},
    \\
    \beta' &= \frac{1}{\sqrt{2(1-z^2)}{\zeta}}\left(\frac{z}{2} - \frac{ia(1-z^2)}{\zeta}\right),
    \\
    \varepsilon\phantom{'} &= 0,
    \\
    \varepsilon' &= \frac{1}{2\zeta \bar{\zeta}}\left(\frac{\Delta}{\zeta} -r + M \right).
\end{align}
\end{subequations}

\section{Teukolsky source}
\label{app:source}

Recall that the full Teukolsky source is formed from the stress-energy via Eqs.~\eqref{eqn:source} and \eqref{eqn:Sop}. Transforming to the frequency-domain, we decompose the Teukolsky source into the mode-sums
\begin{subequations}
    \begin{align}
        -16\pi\zeta\bar{\zeta}T_0 =  \sum_{jm\omega} T_{+2jm\omega}(r) S_{+2jm\omega}(z)e^{i(m\phi-\omega t)},
        \\
        -16\pi\zeta\bar{\zeta}^5T_4 = \sum_{jm\omega} T_{-2jm\omega}(r) S_{+2jm\omega}(z)e^{i(m\phi-\omega t)},
    \end{align}
\end{subequations}
where $T_{\pm 2jm\omega}(r)$ is the radial Teukolsky source in Eq.~\eqref{eqn:teuk-radial}. 
Following the notation and methods of Refs.~\cite{SasaTago03, DrasHugh06}, the radial Teukolsky source then takes the general form
\begin{align}
    T_{\pm2jm\omega}(r) &= \epsilon \int_{-\infty}^\infty \frac{dt}{u^t}\, \Delta^{\mp 2}J_{\pm 2jm\omega}(t,r) e^{i[\omega t - m\phi_p(t)]},
\end{align}
where
\begin{align}
    J_{\pm 2jm\omega}(t,r) &= J^{(0)}_{\pm 2jm\omega}(t,r)\delta(r-r_p) 
    \\ \notag
    & \quad \quad \quad+ \partial_r [J^{(1)}_{\pm2jm\omega}(t,r)\delta(r-r_p)]
    \\ \notag
    & \quad \quad \quad \quad \quad \quad + \partial^2_r [J^{(2)}_{\pm2jm\omega}(t,r)\delta(r-r_p)],
\end{align}
with $\partial_r = \partial/\partial_r$. The $J^{(i)}_{\pm2jm\omega}(t,r)$ then further separate into the terms
\begin{subequations}
    \begin{align}
    J^{(i)}_{-2j m \omega}(t, r) &= A_{j m \omega}^{22(i)}(t,r)u_2(t)u_2(t)
    \\ \notag
    & \quad \quad \quad
    + A_{j m \omega}^{24(i)}(t,r)u_2(t)u_4(t) 
    \\ \notag
    & \quad \quad \quad \quad \quad \quad
    + A_{j m \omega}^{44(i)}(t,r)u_4(t)u_4(t),
    \\
    J^{(i)}_{+2j m \omega}(t, r) &= A_{j m \omega}^{11(i)}(t,r)u_1(t)u_1(t) 
    \\ \notag
    & \quad \quad \quad
    + A_{j m \omega}^{13(i)}(t,r)u_1(t)u_3(t) 
    \\ \notag
    & \quad \quad \quad \quad \quad \quad
    + A_{j m \omega}^{33(i)}(t,r)u_3(t)u_3(t),
    \end{align}
\end{subequations}
where $u_a = g_{\alpha\beta} u^\alpha e^\beta_a$ are the tetrad projections of the four-velocity $u^\alpha$. The nonvanishing terms $A^{ab(i)}_{j m \omega}(t,r) = \tilde{A}^{ab(i)}_{j m \omega}(r, \theta_p(t))$ are then explicitly given by
\begin{subequations}
    \begin{align}
  \tilde{A}_{j m \omega}^{11(0)}(r,\theta) &= -2\rho^{-2}\bar\rho^2\Delta^2 \hat{L}_{1}\left[\rho^{2} \hat{L}_{2}S_{+2j m \omega}\right],
    \\
 \tilde{A}_{j m \omega}^{22(0)}(r,\theta) &= \frac{-2\rho^{-4}}{\Delta^2}\hat{L}^{\dagger}_1\left[\rho^{2} \hat{L}^{\dagger}_2 S_{-2j m \omega}\right],
    \\
    \tilde{A}_{j m \omega}^{13(0)}(r,\theta) &= -4\sqrt{2} \bar\rho\Delta^2 \Big[ 
    \left(\rho+\bar{\rho}+i Q\right)\hat{L}_2 
    \\ \notag
    & \qquad \qquad \quad - a\sqrt{1-z^2} Q(\rho-\bar{\rho}) \Big]S_{+2j m \omega},
    \\
    \tilde{A}_{j m \omega}^{24(0)}(r,\theta) &= \frac{2\sqrt{2}\rho^{-2}\bar\rho}{\Delta} \Big[\left(\rho + \bar{\rho} - i P \right)\hat{L}^{\dagger}_2 
    \\ \notag
    & \qquad \qquad \quad + a \sqrt{1-z^2} P({\rho} - \bar{\rho}) \Big]S_{-2j m \omega},
    \\
    \tilde{A}_{j m \omega}^{33(0)}(r,\theta) &= 4\Delta^2 \left[Q^2 -i \partial_r Q -  2i \rho Q \right]S_{+2j m \omega},
    \\
    \tilde{A}_{j m \omega}^{44(0)}(r,\theta) &= \rho^{-2} \bar{\rho}^2\left[ P^2 + i \partial_r P  + 2i \rho P \right]S_{-2j m \omega} ,
    \\
    \tilde{A}_{j m \omega}^{13(1)}(r,\theta) &= 4\sqrt{2} \rho^{-1}\Delta^2 \hat{L}_2\left[\rho\bar{\rho} S_{+2j m \omega}\right],
    \\
    \tilde{A}_{j m \omega}^{24(1)}(r,\theta) &= -\frac{2\sqrt{2}\rho^{-3}}{\Delta}\hat{L}^{\dagger}_2 \left[\rho \bar\rho S_{-2j m \omega}\right],
    \\
    \tilde{A}_{j m \omega}^{33(1)}(r,\theta) &= 8\Delta^2\left[i Q + \rho \right]S_{+2j m \omega},
    \\
    \tilde{A}_{j m \omega}^{44(1)}(r,\theta) &= -2\rho^{-2}\bar{\rho}^2\left[i P  - \rho \right]S_{-2j m \omega},
    \\
    \tilde{A}_{j m \omega}^{33(2)}(r,\theta) &=  -4\Delta^2 S_{+2j m \omega},
    \\
    \tilde{A}_{j m \omega}^{44(2)}(r,\theta) &=-\rho^{-2}\bar{\rho}^2 S_{-2j m \omega},
\end{align}
\end{subequations}
where, analogous to Eq.~\eqref{eqn:chandraOps}, we have introduced the operators
\begin{align*}
\hat{L}_{\pm 2} &= -\frac{1}{\sqrt{1-z^2}}\left[(1-z^2) \partial_z +{m} - a\omega({1-z^2})\pm 2 z \right], \\
\hat{L}^\dagger_{\pm 2} &= -\frac{1}{\sqrt{1-z^2}}\left[(1-z^2) \partial_z -{m}+a\omega({1-z^2})\pm 2 z \right],
\end{align*}
along with the symbols $P=K/\Delta$ and $Q = P-2i\Delta^{-1}{\partial_r \Delta}$ to condense notation. Recall that $\rho = - \zeta^{-1} = -(r-iaz)^{-1}$.

\section{Metric completion}
\label{app:completion}

The completion piece of the reconstructed metric in Eq.~\eqref{eqn:hcomp} can also be expressed as
\begin{align}
    h^\mathrm{comp\pm}_{\alpha\beta} = c^\pm_M h_{\alpha\beta}^{(\delta M)} +  c^\pm_J h_{\alpha\beta}^{(\delta J)}
\end{align}
with
\begin{subequations}
\begin{align}
h_{tt}^{(\delta M)} &= \frac{2r \left( r^2 + 3a^2 z^2 \right)}{\Sigma^2}, \\
h_{t\phi}^{(\delta M)} &= -\frac{4 r a^3 z^2 (1 - z^2)}{\Sigma^2}, \\
h_{rr}^{(\delta M)} &= \frac{2r \left[M r^2 + a^2 \left( r - (r-3 M)z^2\right) \right]}{M\Delta^2}, \\
h_{zz}^{(\delta M)} &= -\frac{2a^2 z^2}{M(1-z^2)}, \\
h_{\phi\phi}^{(\delta M)} &= -\frac{2 a^2 (1 - z^2) \left[ \Sigma^2 + M r (r^2 - a^2 z^2)(1 - z^2) \right]}{M\Sigma^2},
\end{align}
\end{subequations}
and
\begin{subequations}
    \begin{align}
h_{tt}^{(\delta J)} &= -\frac{4a r z^2}{\Sigma^2}, \\
h_{t\phi}^{(\delta J)} &= -\frac{2 r (1 - z^2) \left( r^2 - a^2 z^2 \right)}{\Sigma^2}, \\
h_{rr}^{(\delta J)} &= -\frac{2a r \left[ r - (r-2M)z^2 \right]}{M\Delta^2}, \\
h_{zz}^{(\delta J)} &= \frac{2a z^2}{M(1-z^2)}, \\
h_{\phi\phi}^{(\delta J)} &= \frac{2 a (1 - z^2) \left[ \Sigma^2 + 2 Mr^3 (1 - z^2) \right]}{M\Sigma^2}.
\end{align}
\end{subequations}
Note that our explicit expression for $h^\mathrm{comp\pm}_{\alpha\beta}$ does not satisfy the radiation gauge conditions. Therefore, the completed metric perturbation is no longer in radiation gauge; only the piece that is reconstructed from the Hertz potentials satisfies the radiation gauge conditions (e.g., is traceless). While we could perform a gauge transformation to put the completion terms in radiation gauge, we do not do so in this work, since we are focused on calculating the redshift invariant, which will not depend on this additional gauge vector. 

\section{Reflection symmetries}
\label{app:symmetries}
We highlight the behavior of the various coefficients and mode functions that make up the Hertz potential under the transformation $(m,\omega) \rightarrow (-m,-\omega)$. For the constants,
\begin{subequations}
    \begin{align}
    \bar{Z}^\mathcal{J}_{\pm 2jm\omega} &= (-1)^{j+k} {Z}^\mathcal{J}_{\pm 2j-m-\omega},
    \\
    {b}^l_{\pm 2jm\omega} &= (-1)^{l+j} {b}^l_{\pm 2j-m-\omega},
    \\
    \mathcal{A}^{\ell'}_{s\ell m} &= (-1)^{s+\ell+\ell'} \mathcal{A}^{\ell'}_{s\ell-m},
    \\
    &= \mathcal{A}^{\ell'}_{-s\ell-m}.
\end{align}
\end{subequations}
For the mode functions and coefficients,
\begin{subequations}
    \begin{align}
    \bar{R}^\mathcal{J}_{\pm 2jm\omega}(r) &= {R}^\mathcal{J}_{\pm 2j-m-\omega}(r),
    \\
    \bar{Y}_{s\ell m}(z) &= (-1)^{s+m} Y_{-s\ell-m}(z),
    \\
    {Y}_{s\ell m}(z) &= (-1)^{\ell+m} Y_{-s\ell m}(-z),
    \\ \label{eqn:coeffSym}
    \bar{h}_{(\pm2)\alpha\beta}^{(n_t,n_r,n_s,n_\phi)}(r,z) 
    &= (-1)^{n_s}{h}_{(\pm2)\alpha' \beta'}^{(n_t,n_r,n_s,n_\phi)}(r,-z),
\end{align}
\end{subequations}
where the primed coordinates in \eqref{eqn:coeffSym} are given by $\alpha' =\{t,r,-z,\phi\}$. Note that we have also taken into account that $\mathcal{A}^{\ell'}_{\pm 2\ell m}$ is real and ${b}^\ell_{\pm 2jm\omega}$ is real for real frequencies. Combining these symmetries with $u^z(q_z) = - u^z(\pi+q_z)$ and $\phi_{m\omega}(q_r,q_z) = -\phi_{-m-\omega}(q_r,q_z) = \phi_{m\omega}(q_r,\pi+q_z) - k\pi$, we then arrive at Eq.~\eqref{eqn:paritySym}.
 
\section{Mode expressions for the radial Teukolsky-Starobinsky identities}
\label{app:TS}

In Boyer-Lindquist coordinates, the Teukolsky-Starobinsky identities in \eqref{eqn:teuk-starobinsky} reduce to differential identities for the homogeneous radial Teukolsky solutions,
\begin{subequations} \label{eqn:TS}
   \begin{align} 
    M^4\hat{D}_0 \left[{R}^\mathcal{J}_{-2jm\omega} e^{i(m\phi - \omega t)}\right] &= \alpha^{\mathrm{TS},\mathcal{J}}_{+2jm\omega} {R}^\mathcal{J}_{+2jm\omega} e^{i(m\phi - \omega t)},
    \\
    16\hat{D}_4 \left[{R}^\mathcal{J}_{+2jm\omega}e^{i(m\phi - \omega t)}\right] &= \alpha^{\mathrm{TS},\mathcal{J}}_{-2jm\omega} {R}^\mathcal{J}_{-2jm\omega}e^{i(m\phi - \omega t)},
\end{align} 
\end{subequations}
where $\hat{D}_0$ and $\hat{D}_4$ are provided in \eqref{eqn:chandraOps} and the solutions $\hat{R}^\mathcal{J}_{\pm2jm\omega}$ are normalized according to Eq.~\eqref{eqn:Rasymp}. For $|\omega| > 0$, the Teukolsky-Starobinsky amplitudes are given by
\begin{subequations}
    \begin{align}
    {\alpha}^\mathrm{TS,\mathcal{I}}_{-2jm\omega} & = 16M^4 \omega_{mkn}^4,
    &
    {\alpha}^\mathrm{TS,\mathcal{H}}_{+2jm\omega} & = 16 \kappa^4(iw - 2)_4,
    \\
    {\alpha}^\mathrm{TS,\mathcal{I}}_{+2jm\omega} & = \frac{\left\vert C^\mathrm{TS}_{jm\omega}\right\vert^2}{{\alpha}^\mathrm{TS,\mathcal{I}}_{-2jm\omega}},
    &
    {\alpha}^\mathrm{TS,\mathcal{H}}_{-2jm\omega} & = \frac{\left\vert C^\mathrm{TS}_{jm\omega}\right\vert^2}{{\alpha}^\mathrm{TS,\mathcal{H}}_{+2jm\omega}},
\end{align}
\end{subequations}
where $\kappa = \sqrt{1-\hat{a}^2}$, $w = (2M\omega(1 + \kappa) - m\hat{a})/\kappa$,
\begin{align} \label{eqn:CTS}
    \left\vert C^\mathrm{TS}_{jm\omega}\right\vert^2 &= \left[\lambda_{|\pm2|jm\omega}^2+4ma\omega-4a^2\omega^2 
	\right] 
    \\ \notag
    & \qquad \times \left[(\lambda_{|\pm2|jm\omega}-2)^2+36ma\omega-36a^2\omega^2 \right]
    \\ \notag
    & \qquad \qquad + (2\lambda_{|\pm2|jm\omega}-1)(96a^2\omega^2-48ma\omega)
    \\ \notag
    & \qquad \qquad \qquad +144\omega^2(M^2-a^2),
\end{align}
with
\begin{subequations} \label{eqn:CTS-ReIm}
    \begin{align}
    \mathrm{Im}\left[C^\mathrm{TS}_{jm\omega} \right] &= (-1)^{j+m+k}12M\omega,
    \\
    \mathrm{Re}\left[C^\mathrm{TS}_{jm\omega} \right] &= \sqrt{\left\vert C^\mathrm{TS}_{jm\omega}\right\vert^2 - \mathrm{Im}\left[C^\mathrm{TS}_{jm\omega} \right]^2},
\end{align}
\end{subequations}
and $\lambda_{|\pm s|jm\omega} = \lambda_{sjm\omega} + s(s+1)$ is invariant under $s\rightarrow -s$.\footnote{Note that Chandrasekhar defined a similarly invariant eigenvalue $\lambda_{|\pm s|jm\omega}' = \lambda_{sjm\omega} + s + |s|$, which is also used in the literature.} Likewise, $\left\vert C^\mathrm{TS}_{jm\omega}\right\vert^2$ is invariant under $(m, \omega) \rightarrow (-m, -\omega)$ and ${\alpha}^\mathrm{TS,\mathcal{J}}_{\pm 2jm\omega} = \bar{\alpha}^\mathrm{TS,\mathcal{J}}_{\pm 2j-m-\omega}$.
For $\omega=0$ modes,
\begin{subequations}
    \begin{align}
    {\alpha}^\mathrm{TS,\mathcal{I}}_{-2jm0} & = (j-2)_4,
    &
    {\alpha}^\mathrm{TS,\mathcal{H}}_{+2jm0} & = 1,
    \\
    {\alpha}^\mathrm{TS,\mathcal{I}}_{+2jm0} & = \frac{\left\vert C^\mathrm{TS}_{jm0}\right\vert^2}{\alpha^\mathrm{TS,\mathcal{I}}_{-2jm0}},
    &
    {\alpha}^\mathrm{TS,\mathcal{H}}_{-2jm0} & = \frac{\left\vert C^\mathrm{TS}_{jm0}\right\vert^2}{{\alpha}^\mathrm{TS,\mathcal{H}}_{+2jm0}}.
\end{align}
\end{subequations}
Furthermore, we have the simplification,
\begin{align}
    C^\mathrm{TS}_{jm0} = \lambda_{|\pm 2|jm0}(\lambda_{|\pm 2|jm0} - 2) = (j-2)_4.
\end{align}

% \section{Hertz amplitudes}
% \label{app:Hertz}

% To determine $\Psi^{\mathcal{G},\mathcal{J}}_{\pm 2jm\omega}$, we insert \eqref{eqn:Hertzmodesum} into \eqref{eqn:hertzEqns}. For the ORG and IRG amplitudes, one finds
% \begin{subequations}
%     \begin{align} \notag
%         \hat{D}_0 {\bar{\Psi}_{-2}^{\mathrm{I},\mathcal{J}}} &= \sum_{jm\omega} {\bar{\Psi}^{\mathrm{I},\mathcal{J}}_{- 2 jm\omega}}{\bar{S}_{-2jm\omega}} \hat{D}_0 \left[{\bar{R}^\mathcal{J}_{-2jm\omega}} e^{-i(m\phi - \omega t)}\right],
%         \\
%         &= \sum_{jm\omega} (-1)^{m} \bar{\Psi}^{\mathrm{I},\mathcal{J}}_{-2j-m-\omega} {S}_{+2jm\omega}
%         \\ \notag
%         & \qquad \qquad \qquad \quad \times \alpha^\mathcal{J}_{+2jm\omega} {R}^\mathcal{J}_{+2jm\omega} e^{i(m\phi - \omega t)},
%         \\ \notag
%         \hat{D}_4 \bar{\Psi}_{+2}^{\mathrm{O},\mathcal{J}} &= \sum_{jm\omega} \bar{\Psi}^{\mathrm{O},\mathcal{J}}_{+2jm\omega}\bar{S}_{+2jm\omega} \hat{D}_4 \left[\bar{R}^\mathcal{J}_{+2jm\omega}e^{-i(m\phi - \omega t)}\right],
%         \\
%         &= \frac{1}{16} \sum_{jm\omega} (-1)^{m} \bar{\Psi}^\mathcal{J}_{+2j-m-\omega} {S}_{-2jm\omega}
%         \\ \notag
%         & \qquad \qquad \qquad \quad \times \alpha^{\mathcal{J}}_{-2jm\omega} {R}^\mathcal{J}_{-2jm\omega}e^{i(m\phi - \omega t)},
%     \end{align}
% \end{subequations}

\section{Metric coefficients}
\label{app:metricCoeffs}

% [DOUBLE CHECK EXPRESSIONS TO MAKE SURE THEY MATCH CODE!]

First we consider metric coefficients produced by the reconstruction operator $\mathcal{S}^\dagger_4$. The $ll$-components are given by
\begin{subequations}
\begin{align}
    h_{(+2)ll}^{(0,0,1,0)} &= -\frac{i a \sqrt{1-z^2} \zeta}{M^2},
    \\
    h_{(+2)ll}^{(0,0,2,0)} &= -\frac{\zeta ^2}{2M^2},
    \\
    h_{(+2)ll}^{(1,0,0,0)} &= \frac{a^2 \left(1-z^2\right) \zeta}{M^2},
    \\
    h_{(+2)ll}^{(1,0,1,0)} &= -\frac{i a \sqrt{1-z^2} \zeta ^2}{M^2},
    \\
    h_{(+2)ll}^{(2,0,0,0)} &= \frac{a^2 \left(1-z^2\right) \zeta ^2}{2M^2},
\end{align}
\end{subequations}
the $lm$-components by
{\allowdisplaybreaks
\begin{subequations}
\begin{align}
    h_{(+2)lm}^{(0,0,0,0)} &= \frac{2 a^2 (r-M) z \sqrt{2-2 z^2} \zeta }{M^2\bar{\zeta }^2},
    \\
    h_{(+2)lm}^{(0,0,0,1)} &= -\frac{a^3 z \sqrt{1-z^2} \zeta }{\sqrt{2} M^2\bar{\zeta }^2},
    \\
    h_{(+2)lm}^{(0,0,1,0)} &= -\frac{\left[r^3-a^2 \left(r - 2 (r-M) z^2\right)\right] \zeta }{\sqrt{2}M^2 \bar{\zeta }^2},
    \\
    h_{(+2)lm}^{(0,0,1,1)} &= \frac{a \zeta ^2}{2 \sqrt{2}M^2 \bar{\zeta }},
    \\
    h_{(+2)lm}^{(0,1,0,0)} &= \frac{a^2 z \sqrt{1- z^2} \Delta  \zeta }{\sqrt{2} M^2\bar{\zeta }^2},
    \\
    h_{(+2)lm}^{(0,1,1,0)} &= -\frac{\Delta  \zeta ^2}{2 \sqrt{2} M^2\bar{\zeta }},
    \\
    h_{(+2)lm}^{(1,0,0,0)} &= \frac{ia \sqrt{1-z^2} \zeta }{\sqrt{2}M^2 \bar{\zeta }^2}
    \\ \notag
    & \qquad \qquad \times \left[2 a^2 (r - (r-M) z^2)-(r^2+a^2)\zeta \right],
    \\
    h_{(+2)lm}^{(1,0,0,1)} &= \frac{i a^2 \sqrt{1-z^2} \zeta ^2}{2 \sqrt{2} M^2\bar{\zeta }},
    \\
    h_{(+2)lm}^{(1,0,1,0)} &= \frac{\left(r^2+a^2\right) \zeta ^2}{2 \sqrt{2} M^2\bar{\zeta }},
    \\
    h_{(+2)lm}^{(1,1,0,0)} &= -\frac{i a \sqrt{1-z^2} \Delta  \zeta ^2}{2 \sqrt{2}M^2 \bar{\zeta }},
    \\
    h_{(+2)lm}^{(2,0,0,0)} &= \frac{i a \left(r^2+a^2\right) \sqrt{1- z^2} \zeta ^2}{2 \sqrt{2} M^2\bar{\zeta }},
\end{align}
\end{subequations}
}
and the $mm$-components by
\begin{widetext}
{\allowdisplaybreaks
\begin{subequations}
\begin{align}
    h_{(+2)mm}^{(0,0,0,0)} &= \frac{\left[2 M^2 r+a^2 r-2 M a^2-r^3+\left(2M^2+a^2+3 (r-2M) r\right)ia z\right] \zeta }{\bar{\zeta }^2},
    \\
    h_{(+2)mm}^{(0,0,0,1)} &= -\frac{a \left(a^2-(2 r-M) r+3 (r-M) iaz\right) \zeta }{2 M^2\bar{\zeta }^2},
    \\
    h_{(+2)mm}^{(0,0,0,2)} &= -\frac{a^2 \zeta ^2}{4 M^2\bar{\zeta }^2},
    \\
    h_{(+2)mm}^{(0,1,0,0)} &= \frac{\left(a^2-(3 r-2M) r+4(r-M) i a z\right) \Delta  \zeta }{2 M^2\bar{\zeta }^2},
    \\
    h_{(+2)mm}^{(0,1,0,1)} &= \frac{a\Delta \zeta ^2}{2 M^2\bar{\zeta }^2},
    \\
    \\
    h_{(+2)mm}^{(0,2,0,0)} &= -\frac{\Delta ^2 \zeta ^2}{4 M^2\bar{\zeta }^2},
    \\
    h_{(+2)mm}^{(1,0,0,0)} &= -\frac{\left(a^4-a^2 (2 r-M) r-3 (r-M) r^3+r^2 (4 r-5M)ia z+a^2 (4 r-4M)ia z\right) \zeta }{2 M^2\bar{\zeta }^2},
    \\
    h_{(+2)mm}^{(1,0,0,1)} &= -\frac{a \left(r^2+a^2\right) \zeta ^2}{2 M^2\bar{\zeta }^2},
    \\
    h_{(+2)mm}^{(1,1,0,0)} &= \frac{\left(r^2+a^2\right) \Delta  \zeta ^2}{2 M^2\bar{\zeta }^2},
    \\
    h_{(+2)mm}^{(2,0,0,0)} &= -\frac{\left(r^2+a^2\right)^2 \zeta ^2}{4 M^2\bar{\zeta }^2}
\end{align}
\end{subequations}
}
\end{widetext}
Next we consider metric coefficients produced by the reconstruction operator $\mathcal{S}^\dagger_0$. The $nn$-components are given by
{\allowdisplaybreaks
\begin{subequations}
\begin{align}
    h_{(-2)nn}^{(0,0,1,0)} &= -\frac{i a M^2 \sqrt{1-z^2}}{\zeta  \bar{\zeta }^2},
    \\
    h_{(-2)nn}^{(0,0,2,0)} &= -\frac{M^2}{2 \bar{\zeta }^2},
    \\
    h_{(-2)nn}^{(1,0,0,0)} &= -\frac{a^2 M^2 \left(1-z^2\right)}{\zeta  \bar{\zeta }^2},
    \\
    h_{(-2)nn}^{(1,0,1,0)} &= \frac{i a M^2 \sqrt{1-z^2}}{\bar{\zeta }^2},
    \\
    h_{(-2)nn}^{(2,0,0,0)} &= \frac{a^2 M^2 \left(1-z^2\right)}{2 \bar{\zeta }^2},
\end{align}
\end{subequations}
}
the $n\bar{m}$-components by
{\allowdisplaybreaks
\begin{subequations}
\begin{align}
    h_{(-2)n\bar{m}}^{(0,0,0,1)} &= -\frac{a^3 z M^2 \sqrt{2-2 z^2}}{\Delta  \zeta  \bar{\zeta }^2},
    \\
    h_{(-2)n\bar{m}}^{(0,0,1,0)} &= -\frac{\sqrt{2} rM^2}{\zeta  \bar{\zeta }^2},
    \\
    h_{(-2)n\bar{m}}^{(0,0,1,1)} &= \frac{a M^2}{\sqrt{2} \Delta  \bar{\zeta }},
    \\
    h_{(-2)n\bar{m}}^{(0,1,0,0)} &= -\frac{a^2 M^2 z \sqrt{2-2 z^2}}{\zeta  \bar{\zeta }^2},
    \\
    h_{(-2)n\bar{m}}^{(0,1,1,0)} &= \frac{M^2}{\sqrt{2} \bar{\zeta }},
    \\
    h_{(-2)n\bar{m}}^{(1,0,0,0)} &= \frac{a M^2 \sqrt{2-2 z^2} \left[\Delta r-\left(r^2+a^2\right) iaz \right]}{\Delta  \zeta  \bar{\zeta }^2},
    \\
    h_{(-2)n\bar{m}}^{(1,0,0,1)} &= -\frac{i a^2 M^2 \sqrt{2-2 z^2}}{2 \Delta  \bar{\zeta }},
    \\
    h_{(-2)n\bar{m}}^{(1,0,1,0)} &= \frac{M^2(r^2+a^2)}{\sqrt{2} \Delta  \bar{\zeta }},
    \\
    h_{(-2)n\bar{m}}^{(1,1,0,0)} &= -\frac{i a M^2 \sqrt{1- z^2}}{\sqrt{2} \bar{\zeta }},
    \\
    h_{(-2)n\bar{m}}^{(2,0,0,0)} &= -\frac{i a M^2 \left(r^2+a^2\right) \sqrt{1-z^2}}{\sqrt{2} \Delta  \bar{\zeta }},
\end{align}
\end{subequations}
}
and the $\bar{m}\bar{m}$ by
{\allowdisplaybreaks
\begin{subequations}
\begin{align}
    h_{(-2)\bar{m}\bar{m}}^{(0,0,0,1)} &= -\frac{2 a M^2}{\Delta ^2 \zeta }
    \\ \notag
    & \qquad \qquad \times\left[(2 r-3M)r - (r-M) ia z -a^2\right],
    \\
    h_{(-2)\bar{m}\bar{m}}^{(0,0,0,2)} &= -\frac{a^2M^2}{\Delta ^2},
    \\
    h_{(-2)\bar{m}\bar{m}}^{(0,1,0,0)} &= \frac{2M^2}{\zeta },
    \\
    h_{(-2)\bar{m}\bar{m}}^{(0,1,0,1)} &= -\frac{2 a M^2}{\Delta },
    \\
    h_{(-2)\bar{m}\bar{m}}^{(1,0,0,0)} &= \frac{2 M^2 \left[M \left(r^2-a^2\right)\zeta +{\left(r^2+a^2\right) \Delta }\right]}{\zeta\Delta ^2},
    \\
    h_{(-2)\bar{m}\bar{m}}^{(1,0,0,1)} &= -\frac{2 a M^2 \left(r^2+a^2\right)}{\Delta ^2},
    \\
    h_{(-2)\bar{m}\bar{m}}^{(1,1,0,0)} &= -2M^2-\frac{4 M^3 r}{\Delta},
    \\
    h_{(-2)\bar{m}\bar{m}}^{(2,0,0,0)} &= -\frac{\left(r^2+a^2\right)^2}{\Delta ^2}
\end{align}
\end{subequations}
}
All other terms either vanish or are given by the complex conjugates of the above coefficients.

\begin{widetext}

\section{Regularization parameter}
\label{app:reg}

In Lorenz gauge, the leading-order regularization parameter for $h_{uu}$ is given by
\begin{align}
    H^{[0]} = \frac{4}{\pi}\sqrt{\frac{\eta}{k}} K(k)
\end{align}
where
\begin{align*}
    \eta^2 & = \frac{a^4 (2 M r+\Sigma )^2+2 a^2 (2M r+\Sigma ) \left[M^2\Sigma 
   \left(\mathcal{L}_z^2-\mathcal{Q}\right)-a^2z^2
   \left(\mathcal{E}^2 \Sigma +2 Mr\right)\right]+\left[\Sigma 
   \left(\mathcal{L}_z^2+\mathcal{Q}\right)+a^2z^2 \left(\mathcal{E}^2 \Sigma
   +2 M r\right)\right]^2}{\Sigma^2 },
    \\
    \zeta &= \frac{2 M r \left(r^2+a^2\right)}{\Sigma }+a^2-r \left[\left(\mathcal{E}^2-2\right)
   r+2M\right]+\mathcal{E}^2 \Sigma +M^2(\mathcal{L}_z^2+\mathcal{Q}),
    \\
    k &= \frac{2\eta}{\eta + \zeta},
\end{align*}
and $K(k)$ is the complete elliptic integral of the first kind,
\begin{align}
    K(k) = \int_0^{\pi/2} \frac{1}{\sqrt{1-k \sin^2 x}}\, dx.
\end{align}

\end{widetext}

\section{Numerical implementation}
\label{app:pybhpt}

We implement the methods described in Secs.~\ref{sec:mode-sum-full-sec} and \ref{sec:mode-sum-redshift-full-sec} via the Python package \textsc{pybhpt}, which was built upon the scalar self-force \CC{} code employed in Ref.~\cite{Nasi22}. The \textsc{pybhpt} v1 library has the following modules:
\begin{enumerate}
    \item \emph{geo.py}: a module for computing bound geodesics and related quantities in Kerr spacetime;
    \item \emph{swsh.py}: a module for computing spin-weighted spheroidal harmonics;
    \item \emph{radial.py}: a module for solving homogeneous solutions of the radial Teukolsky equation, along with radial monodromy data (see Ref.~\cite{Nasi24} for more on monodromy);
    \item \emph{teuk.py}: a module for solving inhomogeneous solutions to the radial Teukolsky equation with a point-particle source;
    \item \emph{hertz.py}: a module for computing the mode functions of Hertz potentials;
    \item \emph{metric.py}: a module containing coefficients for reconstructing the metric from the Hertz potential(s);
    \item \emph{redshift.py}: a module containing coefficients for reconstructing the redshift invariant from the Hertz potential(s);
    \item \emph{flux.py}: a module for computing scalar and gravitational wave fluxes produced by a point-particle source.
\end{enumerate}
In the following subsections, we summarize the key numerical algorithms employed within each package.

\subsection{\emph{geo}}

Kerr geodesic quantities---such as the orbital functions in Eq.~\eqref{eqn:geo} and frequencies in Eq.~\eqref{eqn:frequencies}---are accessed through the \emph{pybhpt.geo} library. Numerical routines are implemented in \CC{} but are made accessible in Python via Cython. This library makes use of the same code and spectral integration routines described in Appendix B 1 of Ref.~\cite{Nasi22}.

\subsection{\emph{swsh}}

Spin-weighted spheroidal harmonics are computed with the \emph{pybhpt.swsh} library using the series expansions in Eqs.~\eqref{eqn:mixingCoeffs} and \eqref{eqn:YslmExpansion}. It uses the same numerical routines as described in Appendix B 4 of Ref.~\cite{Nasi22} to compute the mixing coefficients, but these algorithms are generalized to arbitrary spin-weight $s$. The module uses a combination of \CC{} code wrapped with Cython and pure Python implementations of the \CC{} routines.

\subsection{\emph{radial}}

Homogeneous radial Teukolsky solutions to Eq.~\eqref{eqn:teuk-radial} are produced with the \emph{pybhpt.radial} library. Again, the library makes use of the same \CC{} routines as described in Appendix B 5 of Ref.~\cite{Nasi22}, but now generalized to arbitrary $s$. Rather than directly solving Eq.~\eqref{eqn:teuk-radial}, we perform the transformation proposed by Zengino{\v{g}}lu \cite{Zeng11b} to reduce the oscillatory nature of the solution at the boundaries,
\begin{subequations} \label{eqn:hblTransform}
\begin{align}
    \Psi^\mathcal{H}_{sjm\omega}(r) &= r \Delta^{s} e^{-i(m\varphi- \omega r_*)}R^\mathcal{H}_{sjm\omega}(r),
    \\
    \Psi^\mathcal{I}_{sjm\omega}(r) &= r \Delta^{s} e^{-i(m\varphi + \omega r_*)}R^\mathcal{I}_{sjm\omega}(r),
\end{align}
\end{subequations}
with $\varphi(r) = \frac{a}{2M\kappa}\mathrm{ln}\frac{r-r_+}{r-r_-}$ and $\kappa =\sqrt{1-a^2/M^2}$. Solutions $\Psi^\mathcal{J}_{sjm\omega}(r)$ are obtained using an explicit embedded Runge-Kutta Prince-Dormand $(8, 9)$ method and an adaptive stepper from the GSL ODE library. As described in Ref.~\cite{Nasi22}, boundary conditions are computed using the confluent Heun form of the Teukolsky equation. The confluent Heun equation for the scalar ($s=0$) case is provided in Ref.~\cite{Nasi22}, while its generalization to arbitrary $s$ is given in Sec.~II A of Ref.~\cite{Nasi24}.

\begin{widetext}
To ensure the stability of the ODE solver, we always use the $s\leq 0$ ($s \geq 0$) Teukolsky equation to produce $R^\mathcal{I}_{sjm\omega}$ ($R^\mathcal{H}_{sjm\omega}$). We then obtain the $s \geq 0$ ($s \leq 0$) via the Teukolsky-Starobinsky identities in Eq.~\eqref{eqn:TS}. Rather than using the full fourth-order equations, we replace higher-order derivatives by iteratively applying the Teukolsky equation, resulting in the relations
\begin{align} \label{eqn:linearComboR}
    \alpha^\mathrm{TS,\mathcal{J}}_{-sjm\omega}R^\mathcal{J}_{-sjm\omega}(r) &= {M^{-2s}\Delta^{s}}\big[ f_{sjm\omega}(r) R^\mathcal{J}_{sjm\omega}(r) + g_{sjm\omega}(r)\partial_r R^\mathcal{J}_{sjm\omega}(r)\big].
\end{align}
For the $s=-2$ case, we have
\begin{subequations}
\begin{align}
    g_{-2jm\omega} & = \frac{4 i}{\Delta} \big\{[2 (r-M)^2-\Delta (\lambda_{-2jm\omega}+2)] K
    +2 K^3 + 2 \omega  \Delta \left[\Delta - r(r-M)\right]\big\},
   \\
   f_{-2jm\omega} &= 
   \frac{8 K^2}{\Delta^2} \left[(r-M)^2-\Delta (3 i \omega r + \lambda_{-2jm\omega} +1) + K^2\right] -\frac{4iK}{\Delta} \left[5 i\omega  \Delta - (r-M) (8 i\omega r +\lambda_{-2jm\omega})\right]
   \\ \notag
   & \qquad \qquad \qquad \qquad \qquad \qquad \qquad \qquad \qquad + \left[\lambda_{-2jm\omega} (\lambda_{-2jm\omega}+2)+12  \omega ^2 r^2+4 i \omega 
   (\lambda_{-2jm\omega} r+3M)\right],
\end{align}
\end{subequations}
while for $s=2$,
\begin{subequations}
\begin{align}
    g_{+2jm\omega}(r) & = \bar{g}_{-2jm\omega}(r)
    \\
    f_{+2jm\omega}(r) &= \bar{f}_{-2jm\omega}(r) + \frac{4(r-M)}{\Delta}\bar{g}_{-2jm\omega}(r).
\end{align}
\end{subequations}
\end{widetext}

The module also provides a routine for computing solutions via the Mano-Suzuki-Takasugi (MST) series expansions \cite{ManoSuzuTaka96b}, though this method is numerically unstable and considered deprecated. However, the module does include a relatively stable method for computing the renormalized angular momentum $\nu$ using the monodromy techniques described in Ref.~\cite{Nasi24}, which could be used not only to construct MST solutions but also scattering amplitudes. Regardless, these methods are not employed in this work.

\subsection{\emph{teuk}}
\label{app:teuk}

Inhomogeneous radial Teukolsky solutions in Eq.~\eqref{eqn:EHSmode}, in particular the Teukolsky amplitudes $Z_{\pm 2 jm\omega}^\mathcal{J}$ [as defined in Eq.~\eqref{eqn:Zteuk}], are computed using the \emph{pybhpt.teuk} library. Using the results in Appendix \ref{app:source}, this module extends the spectral source integration algorithm \cite{HoppETC15, NasiOsbuEvan19} described in Appendix B 6 of Ref.~\cite{Nasi22} to spin-weights $s=0,\pm 2$. 

These source integration routines produce values and numerical error estimates for $Z_{\pm 2 jm\omega}^\mathcal{J}$. As highlighted in Ref.~\cite{Nasi22}, the errors produced by the numerical integration can become quite significant for large values of $\ell$, $n$, and $k$. This is because the source integral reduces to phase space integrals of the form
\begin{align}
    Z_{s jm\omega_{mkn}}^\mathcal{J} \sim \int \int (\cdots) \times r^{\pm l}_p(q_r) e^{inq_r}e^{ikq_z}. 
\end{align}
At large values of $\ell$, $n$, and $k$, the integrand becomes highly oscillatory and can span many orders of magnitude, leading to catastrophic numerical cancellation. The estimated relative error is given by the ratio of $Z_{s jm\omega}^\mathcal{J}$ and the largest numerical value sampled in the integrand of Eq.~\eqref{eqn:Zteuk}. For large mode numbers, we observe errors $>10^{-2}$, indicating that almost all significant digits have been catastrophically canceled, making the numerically computed values for $Z_{s jm\omega}^\mathcal{J}$ highly untrustworthy. In these cases, we do not include these modes in our mode-sum calculations. For high eccentricities, we believe this is one of our most significant sources of error.

\subsection{\emph{hertz}}

We construct Hertz potential mode functions using the \emph{pybhpt.hertz} library. Given inhomogeneous Teukolsky solutions and a choice of gauge, this library computes the Hertz amplitudes $\mathcal{Z}^\mathcal{J}_{sjm\omega}$ via Eqs.~\eqref{eqn:hertzAmpRG1} and \eqref{eqn:hertzAmpRG2}, along with the mode functions $R^\mathcal{J}_{sjm\omega}$ and $S_{sjm\omega}$ and their first and second derivatives. As with the previous libraries, the methods are implemented in \CC{} but made accessible via a Cython/Python wrapper.

\subsection{\emph{metric}}

The metric coefficients described in Appendix \ref{app:metricCoeffs} are provided in the library \emph{pybhpt.metric}.

\subsection{\emph{redshift}}

The redshift coefficients, which are constructed by combining the results in Sec.~\ref{sec:huuWorldline} with those in Appendix \ref{app:metricCoeffs}, are also provided in the library \emph{pybhpt.redshift}.

\subsection{\emph{flux}}

While not the focus of this work, one can also calculate the contributions of each Teukolsky mode to the energy, angular momentum, and Carter constant ``fluxes" via the \emph{pybhpt.flux} library. Given an inhomogeneous Teukolsky mode produced by the \emph{pybhpt.teuk} library, the \emph{pybhpt.flux} module allows one to compute the associated fluxes radiated down the horizon and out to infinity.

\bibliography{parent}

\end{document}